\documentclass{aa}  

\usepackage[breaklinks=true,colorlinks,citecolor=blue]{hyperref}
\usepackage{graphicx,animate}
\usepackage{enumerate}
\usepackage{epsfig}
\usepackage{color}
\usepackage{txfonts}
\usepackage{natbib}
\usepackage{multirow}
\usepackage{ulem} 
\usepackage{amssymb}

\usepackage{movie15}

\usepackage{booktabs}
\usepackage{array}   

\usepackage{academicons}
\newcommand{\orcid}[1]{\href{https://orcid.org/#1}{\textcolor[HTML]{A6CE39}{\aiOrcid}}}
\usepackage{orcidlink}

\usepackage{xcolor}

\defcitealias{Mapping-model}{Paper I}
\defcitealias{Mapping-disk}{Paper II}
\defcitealias{Mapping-bulge}{Paper III}
\defcitealias{Mapping-extragalactic}{Paper IV}
\defcitealias{Mapping-sfh}{Paper V}

\definecolor{ao}{rgb}{0.0, 0.5, 0.0}

\newcommand{\kmps}{\rm km~s\ensuremath{^{-1} }\,}
\newcommand{\kmskpc}{km~s\ensuremath{^{-1}}~kpc\ensuremath{^{-1} }\,}
\newcommand{\Msun}{M\ensuremath{_\odot}}

\newcommand{\Msunkpcs}{M\ensuremath{_\odot}~kpc\ensuremath{^{-2} }\,}

\newcommand{\Gaia}{{\it Gaia}\,}

\newcommand{\aFe}{\ensuremath{\rm [\alpha/Fe]}\,}

\newcommand{\MgFe}{\ensuremath{\rm [Mg/Fe]}\,}
\newcommand{\FeH}{\ensuremath{\rm [Fe/H]}\,}

\newcommand{\sz}{\ensuremath{\rm \sigma_{z}}}
\newcommand{\sphi}{\ensuremath{\rm \sigma_{\phi}}}
\newcommand{\sr}{\ensuremath{\rm \sigma_{R}}}

\newcommand{\vz}{\ensuremath{\rm V_z}}
\newcommand{\vp}{\ensuremath{\rm V_\phi}}
\newcommand{\vr}{\ensuremath{\rm V_R}}

\newcommand{\papername}{Rediscovering the Milky Way with orbit superposition approach and APOGEE data}

\newcommand{\thankssdss}{Funding for the Sloan Digital Sky Survey IV has been provided by the Alfred P. Sloan Foundation, the U.S. Department of Energy Office of Science, and the Participating Institutions. SDSS acknowledges support and resources from the Center for High-Performance Computing at the University of Utah. The SDSS web site is www.sdss4.org. \\

SDSS is managed by the Astrophysical Research Consortium for the Participating Institutions of the SDSS Collaboration including the Brazilian Participation Group, the Carnegie Institution for Science, Carnegie Mellon University, Center for Astrophysics | Harvard \& Smithsonian (CfA), the Chilean Participation Group, the French Participation Group, Instituto de Astrofísica de Canarias, The Johns Hopkins University, Kavli Institute for the Physics and Mathematics of the Universe (IPMU) / University of Tokyo, the Korean Participation Group, Lawrence Berkeley National Laboratory, Leibniz Institut für Astrophysik Potsdam (AIP), Max-Planck-Institut für Astronomie (MPIA Heidelberg), Max-Planck-Institut für Astrophysik (MPA Garching), Max-Planck-Institut für Extraterrestrische Physik (MPE), National Astronomical Observatories of China, New Mexico State University, New York University, University of Notre Dame, Observatório Nacional / MCTI, The Ohio State University, Pennsylvania State University, Shanghai Astronomical Observatory, United Kingdom Participation Group, Universidad Nacional Autónoma de México, University of Arizona, University of Colorado Boulder, University of Oxford, University of Portsmouth, University of Utah, University of Virginia, University of Washington, University of Wisconsin, Vanderbilt University, and Yale University.}

\newcommand{\thanksgaia}{This work presents results from the European Space Agency (ESA) space mission Gaia. Gaia data are being processed by the Gaia Data Processing and Analysis Consortium (DPAC). Funding for the DPAC is provided by national institutions, in particular the institutions participating in the Gaia Multi-Lateral Agreement (MLA). The Gaia mission website is https://www.cosmos.esa.int/gaia. The Gaia Archive website is http://archives.esac.esa.int/gaia.}

\newcommand{\thanksmiapb}{This research was supported by the Munich Institute for Astro-, Particle and BioPhysics (MIAPbP) which is funded by the Deutsche Forschungsgemeinschaft (DFG, German Research Foundation) under Germany´s Excellence Strategy-EXC-2094-390783311}

\newcommand{\thanksleiden}{The authors thank the organizers of the ``A new dawn of dwarf galaxy research'' workshop and the Lorentz Center~(Leiden) for the organizational and financial support of the meeting, which provided an insightful atmosphere which stimulated the progress of this work.}

\defcitealias{Mapping-1}{Paper I}
\defcitealias{Mapping-2}{Paper II}
\defcitealias{Mapping-3}{Paper III}
\defcitealias{Mapping-4}{Paper IV}
\defcitealias{Mapping-5}{Paper V}

\begin{document}

\title{\papername\ II. Chrono-chemo-kinematics of the disc}

\titlerunning{Rediscovering the Milky Way disc}

\authorrunning{S. Khoperskov et al.}

\author{Sergey Khoperskov$^1$\thanks{sergey.khoperskov@gmail.com}\orcidlink{0000-0003-2105-0763}, 
Matthias Steinmetz$^1$\orcidlink{0000-0001-6516-7459}, 
Misha Haywood$^3$\orcidlink{0000-0003-0434-0400}, 
Glenn van de Ven$^2$\orcidlink{0000-0003-4546-7731}, 
Davor Krajnovic$^1$, \\
Bridget Ratcliffe$^1$\orcidlink{0000-0003-1124-7378}, 
Ivan Minchev$^1$\orcidlink{0000-0002-5627-0355}, 
Paola Di Matteo$^3$, 
Nikolay Kacharov$^1$\orcidlink{0000-0002-6072-6669}, \\
Léa Marques$^{1,4}$,
Marica Valentini$^1$\orcidlink{0000-0003-0974-4148}, 
Roelof S. de Jong$^1$\orcidlink{0000-0001-6982-4081} }

\institute{$^1$ Leibniz-Institut für Astrophysik Potsdam (AIP),
              An der Sternwarte 16, 14482 Potsdam, Germany \\
              $^2$ Department of Astrophysics, University of Vienna, Türkenschanzstraße 17, A-1180 Vienna, Austria \\
              $^3$ GEPI, Observatoire de Paris, PSL Research University, CNRS, Place Jules Janssen, 92195 Meudon, France \\
              $^4$ Universität Potsdam, Institut für Physik und Astronomie, Karl-Liebknecht-Str. 24-25, 14476 Potsdam, Germany}

\abstract{The stellar disc is the dominant luminous component of the Milky Way~(MW). Although our understanding of its structure is rapidly expanding due to advances in large-scale surveys of stellar populations across the Galaxy, our picture of the disc remains substantially obscured by selection functions and incomplete spatial coverage of observational data. In this work, we present the comprehensive chrono-chemo-kinematic structure of the MW disc, recovered using a novel orbit superposition approach combined with data from APOGEE DR 17.

We detect periodic azimuthal metallicity variations within $6-8$~kpc with an amplitude of $0.05-0.1$~dex peaking along the bar major axis. The radial metallicity profile of the MW also varies with azimuth, displaying a pattern typical among other disc galaxies: a decline outside the solar radius and an almost flat profile in the inner region, attributed to the presence of old, metal-poor high-$\alpha$ populations, which comprise $\approx 40\%$ of the total stellar mass. The geometrically defined thick disc and the high-$\alpha$ populations have comparable masses, with differences in their stellar population content, which we quantify using the reconstructed 3D MW structure.

The well-known $\rm [\alpha/Fe]$-bimodality in the MW disc, once weighted by stellar mass, is less pronounced at a given metallicity for the whole galaxy but distinctly visible in a narrow range of galactic radii~(5-9 kpc), explaining its relative lack of prominence in external galaxies and galaxy formation simulations. Analysing a more evident double age-abundance sequence, we construct a scenario for the MW disc formation, advocating for an inner/outer disc dichotomy genetically linked to the MW's evolutionary stages. In this picture, the extended solar vicinity is a transition zone that shares chemical properties of both the inner~(old age-metallicity sequence) and outer discs~(young age-metallicity sequence).
}


\maketitle

%-------------------------------------------------------------------

\section{Introduction}
Disc galaxies are extremely common in the local universe~\citep{2010ApJ...723...54K, 2010ApJ...716..942F, 2014MNRAS.444L..80L}, so it is not surprising that the MW is also a disc-dominated system. Furthermore, the mass distribution of galaxies peaks exactly at the MW mass,  making its type the most frequent system~\citep{2013ApJ...771L..35V}.
Given its current star formation rate, disc morphology, and bulge structure, it is reasonable to consider the MW as a typical galaxy. This makes the study of the MW and its disc particularly relevant for understanding not only our Galaxy but also the broader context of galaxy formation.

Since the works by \cite{1983MNRAS.202.1025G, 2003A&A...410..527B, 2004A&A...415..155B, 2004AN....325....3F}, based on the local solar vicinity data, the MW stellar disc has been considered to be a superposition of thin and thick components. The parameters of these components are debated in the literature~\citep{2001ApJ...553..184C, 2002ApJ...578..151S, 2008ApJ...673..864J, 2012MNRAS.420.1423F, 2016PASA...33...27D, 2016ARA&A..54..529B}, partially because the definitions of thick/thin components depend on the context \citep[see, e.g.][for the ambiguity of the thin and thick disc definitions]{2013A&A...560A.109H,2016ApJ...831..139M}. There is a tendency to name high-$\alpha$ populations as a thick disc, which is perhaps a reasonable assumption, but mainly inside the solar circle~\citep{2003A&A...397L...1F,2003MNRAS.340..304R,2005A&A...433..185B}. In the outer parts of the Galaxy, the thicker component is made of low-$\alpha$ stars as first seen in APOGEE \citep{2015ApJ...808..132H}, which is explained as the flaring of mono-age populations \citep{2015ApJ...804L...9M}. The geometric definition of thin and thick disc components is somewhat uncertain due to the presence of a boxy-peanut bulge in the centre~\citep{1977Natur.265..515O, 1978PASJ...30....1M, 1994ApJ...425L..81W} and the warp at the outskirts~\citep{1957AJ.....62...90B, 1957AJ.....62...93K, 1989ApJ...341L..13D}.

Thick stellar components appear to be detected in nearly all edge-on disc galaxies~\citep{comeron11,2002AJ....124.1328D,2006AJ....131..226Y,2018A&A...610A...5C}; however, their formation scenarios remain a matter of debate~\citep{2019A&A...623A..89C, 2016MNRAS.460L..89K,2019A&A...623A..19P}. Simulations suggest that the role of mergers in the formation of thick discs is important~\citep{1993ApJ...403...74Q,2008MNRAS.391.1806V}, whereas wet and dry mergers require special treatment, as they may result in different orbital compositions~\citep{2011A&A...525L...3D,2004ApJ...612..894B,2008MNRAS.391.1806V}. One problem with this scenario is that satellite bombardment results in strong disc flaring, not observed in external edge-on galaxies \citep{vanderkruit82,degrijs98}. Thick stellar populations can also form rapidly at high redshift during intense starbursts in a turbulent interstellar medium (ISM), which result in ab initio geometrically thick stellar components with the high velocity dispersion~\citep{2004ApJ...612..894B, 2009ApJ...707L...1B}; these, however, are typically confined to the inner disc since they form at high redshift. Alternatively, thick stellar discs have been proposed to result from secular evolution involving stellar radial migration~(\citealt{2009MNRAS.396..203S}; but see \citealt{2012A&A...548A.127M,vera-ciro14}, for counter-arguments); a number of works, however, have shown that in the presence of realistic disc formation in the cosmological context migration cools the outer disc, rather than heating it (e.g., \citealt{2014AA...572A..92M,2016MNRAS.459..199G}). Finally, thick discs can be explained with the nested flares of mono-age stellar populations, resulting from merger interactions with the host disc over cosmic time~\citep{2015ApJ...804L...9M, 2021MNRAS.501.5105G}.%; in this model, heating by mergers can produce thick discs that do not flare by considering the more realistic scenario of inside-out growing discs in a cosmological environment.

One of the notable features of the MW disc that has garnered significant attention from the Galactic community over the past decade is the chemical bimodality of its stellar populations. Specifically, two distinct \aFe sequences are visible in the $\aFe-\FeH$ plane in a broad range of \FeH for the disc stars~\citep{2014AJ....147..116H, 2014ApJ...796...38N, 2015ApJ...808..132H}. Although the exact shape and metallicity range of the bimodality varies slightly, most current datasets concur on the presence of this feature, which remains evident even without considering stellar mass along the two sequences~\citep{2019MNRAS.486.1167B,2021MNRAS.506..150B,2022MNRAS.513..232N,2022Natur.603..599X,2023ApJ...954..124I,2024A&A...682A...9G}. From a theoretical standpoint, there is a broad range of chemical evolution models~\citep{2014ApJ...796...38N,2017MNRAS.472.3637G,2018MNRAS.477.5072M,2020MNRAS.498.1710P,2021MNRAS.507.5882S,2021MNRAS.508.4484J,2023MNRAS.523.3791C, 2024A&A...690A.208S} and galaxy formation simulations~\citep{2018MNRAS.480L..38V, 2019MNRAS.484.3476C, 2018MNRAS.477.5072M, 2020MNRAS.491.5435B, 2021MNRAS.503.5846R, 2021MNRAS.501.5176K} qualitatively reproducing the MW-like $\aFe$-bimodality. However, it is unclear how to pick a model that is the most relevant for the MW, thus capturing the most adequate scenario of its evolution. Perhaps the community may benefit from a bimodality comparison project summarising the theoretical landscape, allowing us to critically confront the role of particular processes shaping the MW disc.

Moreover, the relevance of the Galactic community's efforts to understand the $\alpha$-bimodality of the MW disc remains unclear when applied to the study of external galaxies. So far, the MW is the only prominent case of the $\aFe$-bimodality, based on star counts that are heavily weighted towards the solar radius. Even the most massive nearby dwarf galaxies, such as the Large Magellanic Cloud~(LMC) and the Small Magellanic Cloud~(SMC), do not appear to host such a feature \citep{2012ApJ...761...33L, 2014AN....335...79M, 2020ApJ...895...88N, 2021ApJ...923..172H}. The Andromeda galaxy~(M31) is the only nearby giant external disc galaxy where the study of resolved stellar populations is currently feasible. Using JWST abundance data, \cite{2024AAS...24342805N, 2024IAUS..377..115N} showed that there is no $\alpha$-bimodality in M31; instead, the data can be explained by a single high star formation efficiency evolutionary track similar to what is seen in the MW bulge. Conversely, \cite{2022A&A...666A.109A} found evidence for bimodality in the disc of M31 using planetary nebulae \citep[see also][for the chemical evolution models and comparison with JWST data]{2023ApJ...956L..14K}. However, it is essential to note that it is still the early days of the M31 chemical abundance mapping, as the spatial coverage of the M31 disc and the number of stars with sufficiently precise chemical information are rather limited compared to the MW data and may not be representative of the entire galaxy.

High-resolution integral field spectroscopic observations of external galaxies also do not show convincing evidence for the existence of MW-like $\alpha$-bimodality, or chemically-distinct disc components in other systems~\citep{2008ApJ...683..707Y, 2019A&A...625A..95P, 2019A&A...623A..19P}. This may be due to the limitations of the full spectra fitting techniques, which may also act against discoveries of the extragalactic bimodalities~\citep{2021ApJ...913L..11S, 2023arXiv231018258W}. Additionally, there is no clear way to translate the $\alpha$-bimodality observed in the MW to an extragalactic perspective. The MW disc is not entirely covered by spectroscopic data, which may be biased by the survey selection function~\citep{2021ApJ...909...77G,2022MNRAS.513..232N}. In other words, we do not yet fully understand how much stellar mass (or light) corresponds to each $\alpha$-sequence population at different galactocentric radii and heights from the midplane.

The time evolution of radial metallicity gradients is a vital constraint for chemical evolution models. For instance, a negative radial metallicity gradient is believed to be a universal signature of the inside-out formation of disc galaxies~\citep{1976MNRAS.176...31L, 1989MNRAS.239..885M, 2000MNRAS.313..338P, 2003ceg..book.....M, 2021A&ARv..29....5M}. It is caused by more generations of stars that started to enrich the ISM in the centre earlier compared to the outer disc, where the enrichment is either less efficient or results from fewer stars releasing metals to the ISM. The ab initio gradients may change over time as the galactic disc expands and the star formation proceeds unevenly along the radius. It is known that secular processes, such as radial migration~\citep{2002MNRAS.336..785S}, also affect the observed present-day metallicity gradients. In agreement with such an idea, the MW radial metallicity gradient shows monotonic flattening with increasing stellar age~\citep[see, e.g.][]{2005A&A...433..127M, 2010ApJ...714.1096S, 2024MNRAS.535..392L, 2023ApJ...954..124I, 2023MNRAS.526.2141W, 2023A&A...669A.119M}. \cite{2017A&A...600A..70A} showed more complex behaviour of the MW metallicity variations with radius where the slope of the radial metallicity gradient was compatible with a flat distribution for older ages, then it steepens and finally flattens again, suggesting a two-phase disc formation history~\citep[see also][]{2015RAA....15.1209X, 2017ApJS..232....2X,2015RAA....15.1240H}. The most extreme result was presented by \cite{2023NatAs...7..951L}, who found that the stellar light-weighted metallicity profile of the MW has a broken shape, with a mildly positive gradient inside $7$~kpc and a steep negative gradient outside. A similar conclusion can be reached by looking at the mean metallicity maps shown in \cite{2021A&A...656A.156Q} and \cite{2019MNRAS.490.4740B}. The latter results may suggest that the MW might not have a typical metallicity distribution for a galaxy of its mass. However, does this imply that the MW was not formed inside-out, or does it simply reflect a peculiar gas infall or merger history? It is believed that the MW had a relatively quiet merger history~\citep{2020ARA&A..58..205H, 2024NewAR..9901706D}, in agreement with simulated MW analogues~\citep{2019MNRAS.484.4471F} with a relatively gentle impact of mergers on the disc structure~\citep{2019A&A...632A...4D,2020MNRAS.494.3880B}, which makes its enrichment history well reproducible with the closed box chemical evolution models, naturally producing negative radial metallicity gradients~\citep{1989MNRAS.239..885M, 2007A&A...462..943C, 2018MNRAS.481.2570G}.

At the same time, it is plausible that different subsamples of MW stars exhibit radically different radial metallicity profiles. Once their contributions are mixed together in various proportions, this can lead to apparent radial metallicity gradient variations. This variation can (partially) be attributed to hidden biases in sample selection. For instance, it is well-known that, beyond the radial metallicity gradient, there also exist vertical metallicity gradients \citep[e.g.,][]{2014A&A...572A..33M,2017MNRAS.471.3057M} and azimuthal metallicity variations \citep[e.g.,][]{2023MNRAS.525.3318H, 2024arXiv240518120H}. Therefore, accurately determining the representative metallicity profiles of the disc as a whole requires correcting observational data for the survey (or subsample) selection function \citep{2017A&A...606A..97N, 2021ApJ...909...77G, 2022MNRAS.513.4130L}. This is a complex task because these abundance gradients vary with position in the Galaxy, particularly in the inner region dominated by the bar and the boxy-peanut bulge, which (re)shape stellar populations depending on their kinematics~\citep{2015A&A...577A...1D, 2017MNRAS.469.1587D, 2018A&A...616A.180F}. 

While it may not be seen directly in the star counts, the MW disc morphological structure, e.g., the presence of spirals and bar, is still reflected in the kinematics of stellar populations~\citep{2012MNRAS.425.2335S, 2013MNRAS.436..101W, 2018A&A...616A..11G}. Using APOGEE DR 16 \cite{2021A&A...656A.156Q} presented the mean Galactocentric radial velocity XY-map revealing a prominent quadrupole pattern, known in external galaxies~\citep{2001ApJ...550..243B, 2021A&A...649A..30C, 2016A&A...588A..33Q} and predicted in  simulations~\citep{2001ApJ...550..243B, 2023A&A...674A..37G, 2024ApJ...968...86B, 2024MNRAS.528.3576V}. Simulations predict the velocity pattern to be aligned with the bar, resulting in a zero net mass flow along the bar. Although in \cite{2021A&A...656A.156Q}, the kinematic maps cover both sides of the bar; the mean velocity pattern is not aligned with the known bar orientation~\citep{1992ApJ...384...81W}. The authors measured the  bar orientation angle as of $20^\circ$, which is fairly low compared to recent estimates~\citep[see, e.g.][]{2015MNRAS.450.4050W, 2013MNRAS.435.1874W, 2016ARA&A..54..529B}. Although \cite{2019MNRAS.490.4740B} used the same APOGEE data, being more conservative regarding the sample selection, their data do not cover the entire bar, and it is hard to assess the alignment of the radial velocity quadrupole pattern with the bar orientation. With the arrival of the massive \Gaia DR3 sample, this tension becomes more evident, as shown by \cite{2023A&A...674A..37G}, the distance uncertainty obscures the observed orientation of the radial velocity quadrupole~\citep[see also][]{2024MNRAS.528.3576V}. We note that such an effect has been known for quite some time, from early works about the MW bar structure~\citep{1997ApJ...477..163S, 2013ApJ...776L..19D, 2020MNRAS.492.3128G} showing a visual alignment of the bar along the Sun-Galactic centre line, which is artificially stretched along the line-of-sight due to distance uncertainties. Therefore, in order to obtain realistic kinematic information about the MW disc, the distance uncertainties should also be taken into account. The latter, however, is not easy to implement due to natural limitations of the parallax measurements affecting the distance determination~\citep{2018AJ....156...58B, 2021AJ....161..147B}.

In light of upcoming data from the new generation spectroscopic surveys, like SDSS-V~\citep{2019BAAS...51g.274K} and 4MOST~\citep{2019Msngr.175....3D}, new approaches for the interpretation of the observational data are needed in order to fill in the gaps in our understanding of the composition of the MW disc and its projection into extragalactic perspective. In the first work of the series, we developed a novel orbit superposition method based on the mock APOGEE-like MW data. We have shown that a reasonable assumption about the 3D structure of the galactic potential, taking into account a rigidly rotating bar, allows us to calculate the weights of orbits of stars in the mock catalogue and recover not only the detailed mass distribution and kinematics but also the metallicity information across the entire galaxy. Hence, it can be used for all sorts of stellar parameters, evolutionarily-linked to the orbits of stars. In this work, we use this approach to analyse the kinematics, chemical composition and age structure of the MW disc stellar populations using the APOGEE DR 17 dataset. We need to mention that the approach developed in the present paper is partially inspired by the work of \cite{2021A&A...653A.143W}, who mapped the abundances of the APOGEE DR14 stars in the bulge region along the orbits. However, due to poor disc coverage and the lack of the orbit superposition approach, it was not possible to recover the complete abundance distribution for the MW. Compared to \cite{2021A&A...653A.143W}, we use an improved version of the MW potential, which now reproduces the mass distribution not only in the inner MW but also beyond the solar radius~\citep{2022MNRAS.514L...1S}.

The paper is structured as follows. In Section~\ref{sec2::method} we describe our selections from the APOGEE survey, the orbit superposition model ingredients and workflow of our approach. In Section~\ref{sec2::results_intro} we use the results of the orbit superposition to examine the structure of the MW disc stellar populations depending on their chemistry and ages. The kinematics analysis of the MW disc stellar populations is presented in Section~\ref{sec2::results_kinematics}. Section~\ref{sec2::results_abundances} describes the abundance structure of the Galactic disc. We present the metallicity gradients and discuss age-abundance MW disc composition in Sections~\ref{sec2::results_met_gradients} and \ref{sec2::results_AMR}, respectively. In Sections~\ref{sec2::discussion} we discuss the results of the paper in a broader context of the MW formation history. The paper summary is presented in Section~\ref{sec2::summary}.

\section{Data and Method}\label{sec2::method}

\begin{figure}
    \centering
    \includegraphics[width=1\hsize]{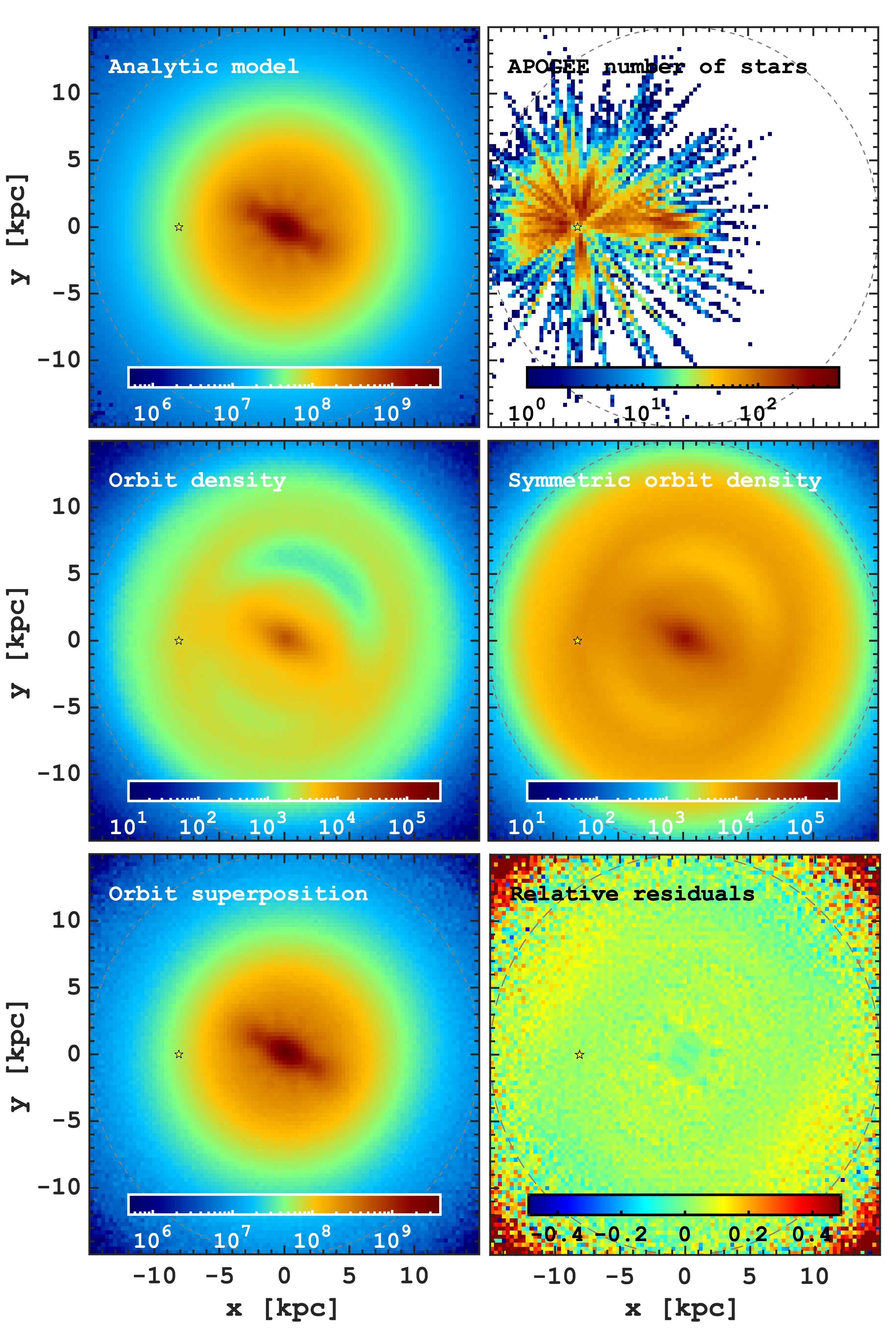}
    \caption{Recovery of the MW density structure using orbital superstition and APOGEE dataset. The top left panel shows the projected MW stellar density from the analytic model~\citep{2022MNRAS.514L...1S}. The top right panel shows the face-on distribution of stars in the APOGEE sample~(see Section~\ref{sec2::data}). The second row shows the raw density of orbits of the APOGEE stars before mirroring~(left) and after~(right). The bottom row shows the result of the orbit superposition, which is the weighted density of the orbits after mirroring~(left) and the difference~(right) between the top left and reconstructed density. The grey dashed circle corresponds to $15$~kpc inside which the orbit superposition solution was obtained. The yellow asterisk marks the position of the Sun.}
    \label{fig02::density_reconstruction}
\end{figure}

\begin{figure*}
    \centering
    \includegraphics[width=0.49\hsize]{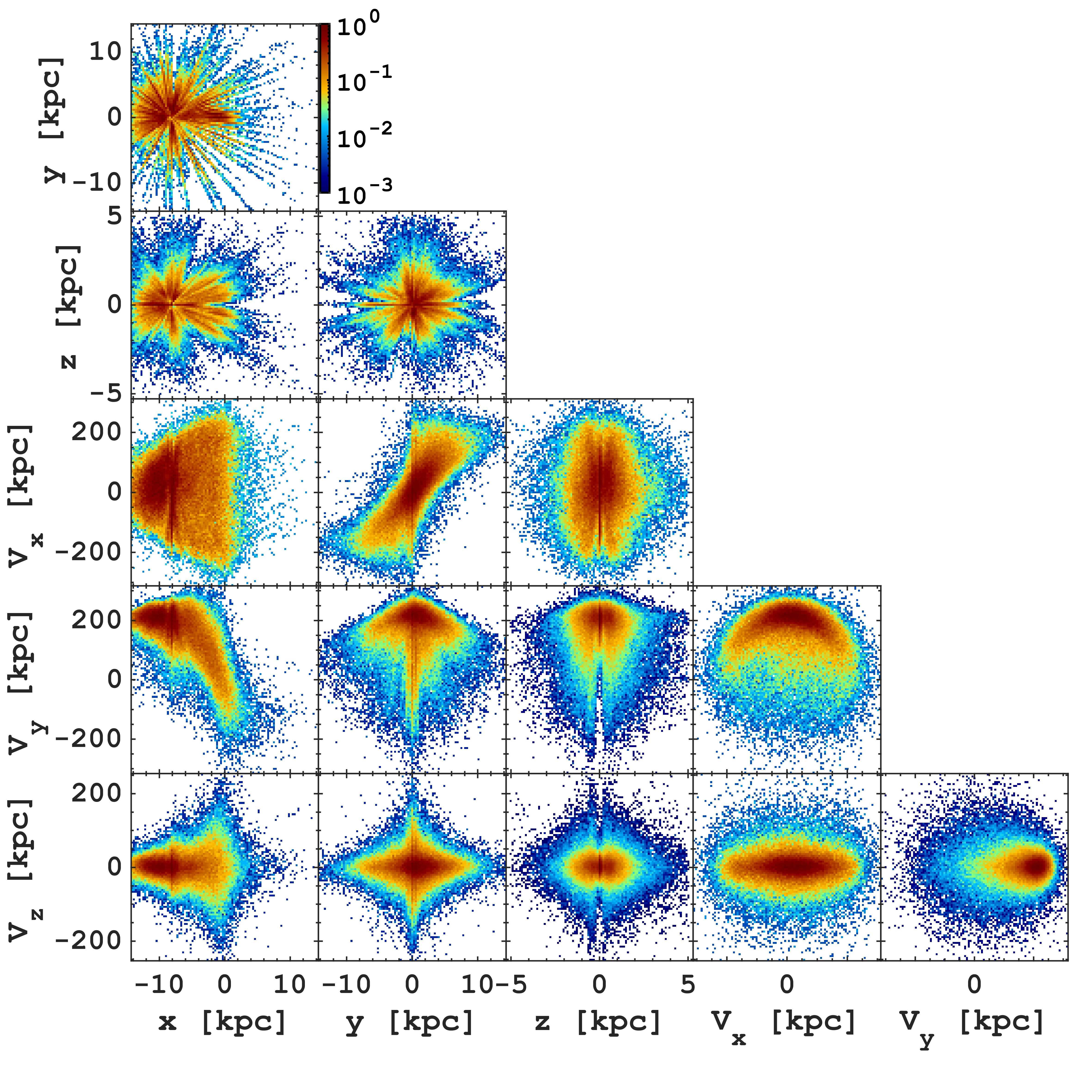}
    \includegraphics[width=0.49\hsize]{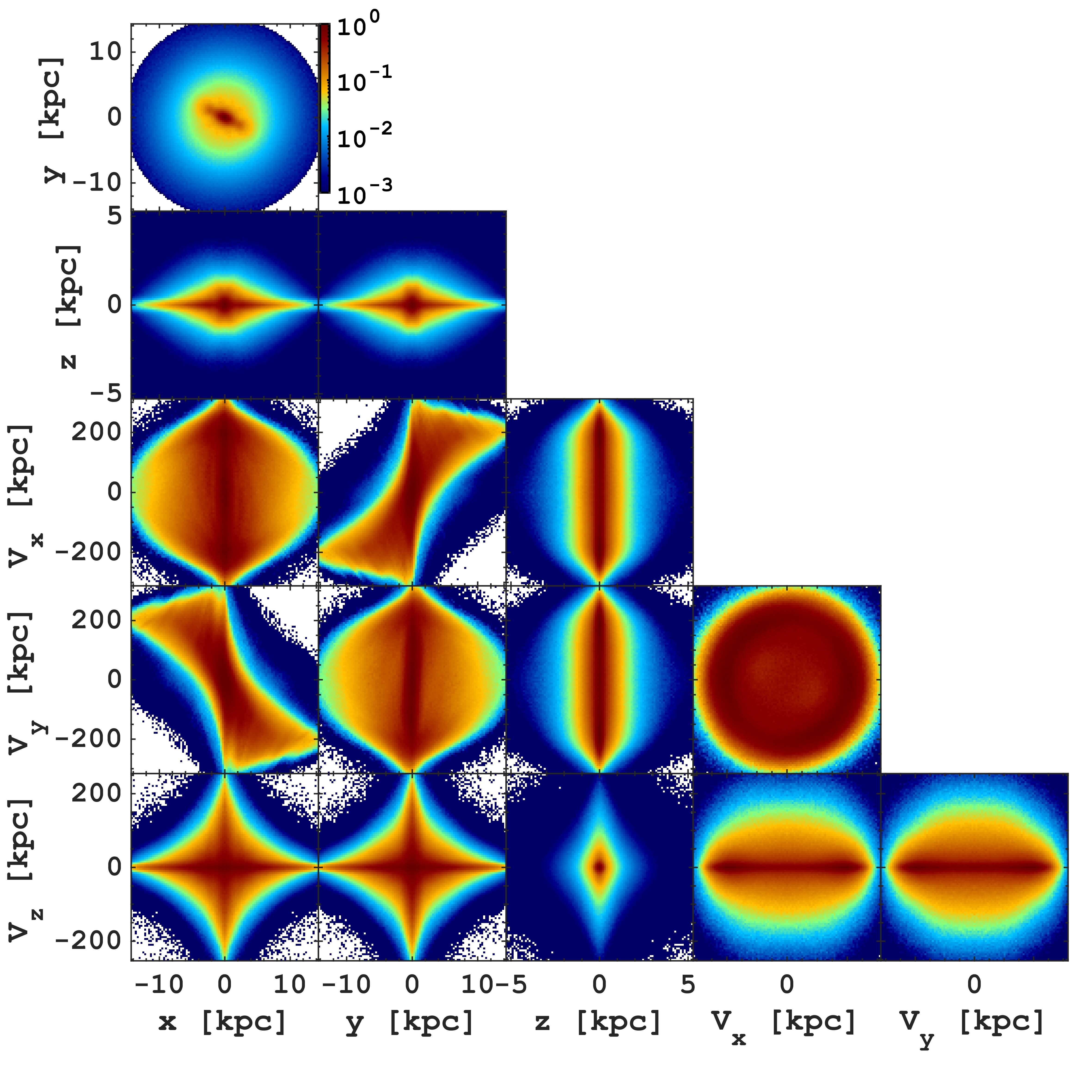}
    \caption{Comparison of distribution functions for APOGEE~(left) and orbit superposition reconstruction~(right) in Cartesian coordinates. In each panel, the density is normalized by the maximum value, which is shown in the log scale.}
    \label{fig02::DFs}
\end{figure*}

% \begin{figure}
%     \centering
%     \includegraphics[width=1\hsize]{figs02/disk_surface_densities_APOGEE-Dr17_0000.jpg}
%     \caption{Surface density profiles of low- and high-$\alpha$ populations~(left) and mono-age populations~(right, according to the colourbar). }
%     \label{fig02::surface_densities}
% \end{figure}

\subsection{APOGEE data}\label{sec2::data}
In this work, we use radial velocities, atmospheric parameters and stellar abundances~(\FeH and \MgFe) from APOGEE DR17~\citep{2022ApJS..259...35A} which were complemented by the proper motions from the \Gaia DR3  catalogue~\citep{2023A&A...674A...1G}, providing us with a sample of stars with full 6D phase-space information. We only use stars with radial velocity uncertainties better than $2$~\kmps, distance errors  $<20\%$ and proper motion errors better than $10\%$. In order to cover a larger area across the MW disc, we select giant stars with $\rm log g < 2.2$ and flagged them as $\rm ASPCAPFLAG=0$ and $\rm EXTRATARG=0$. For the analysis we adopted stellar ages from the \cite{2024AJ....167...73S} catalogue  and $\sigma_{age}<2$~Gyr. The age catalogue for the giant stars has been extensively analysed by \cite{2023ApJ...954..124I}, where its reasonable behaviour was demonstrated in many applications.

The choice of giant stars from the APOGEE catalogues, whose stellar parameters might not be so precise as the ones for dwarfs, is dictated by the need to cover a larger area of the MW~\citep[see the comparison in][]{2023ApJ...954..124I} which, as we have shown in our previous paper~\citep[][hereafter \citetalias{Mapping-model}]{Mapping-model}, is vital for proper implementation of the orbit superposition method. In the present work, we also follow a non-conservative approach by not removing stars with relatively high abundance uncertainties from our sample, as they are being used to propagate the chemical abundance information along the orbits of the stars.

\subsection{MW potential and orbit integration}\label{sec2::MW_potential}
The important ingredient of our method is the 3D gravitational potential of the MW. In this work, we use the potential from \cite{2022MNRAS.514L...1S}, which is an updated analytic model of the potential constructed by \cite{2017MNRAS.465.1621P}. This potential has been built by adjusting the mass distribution in the central region of the MW constrained by the kinematic data from the BRAVA and OGLE surveys and the entire bar region from the ARGOS survey and also including the red clump giant star counts~\citep{2013MNRAS.435.1874W,2015MNRAS.450.4050W}. The updated analytic model of \cite{2022MNRAS.514L...1S} takes into account the correct behaviour of the mass distribution outside the bar region and reproduces well the 3D N-body density of the bar, including the X-shape structure of the bulge, which we adopted from the AGAMA package~\citep{2019MNRAS.482.1525V}. To our knowledge, this is the most detailed and precise model of the MW potential whose analytic version is easy to implement in our approach using the AGAMA software.

For the adopted MW potential, the bar rotational frequency is set to be $37.5$~\kmskpc, which lies within the range of recent estimations~\citep{2008A&A...489..115R, 2022ApJ...925...71L, 2016ApJ...824...13L, 2019MNRAS.490.4740B, 2019MNRAS.488.4552S, 2022MNRAS.512.2171C}. More importantly, this value of the bar pattern speed is obtained by constructing the potential we use~\citep{2022MNRAS.514L...1S}.
In this case, we, however, neglect possible second-order time-dependent processes like the bar deceleration~\citep{2020A&A...638A.144K, 2021MNRAS.500.4710C}, possible short time-scale variations of its parameters caused by re-connection to spiral arms~\citep{2020MNRAS.497..933H,2024MNRAS.528.3576V} and possible evolution of the X-shaped component~\citep{2019A&A...622L...6K,2020MNRAS.494.5936F} mainly because their exact amplitudes are poorly constrained in the MW. We, therefore, integrate the orbits of the APOGEE stars in a rigid, rotating 3D X-shaped barred galactic potential over $5$ Gyr. The output of the orbit integration incorporates 500 data points~(positions and velocities) per orbit for each star. Additionally, we account for uncertainties in the abundances and age determination for each star by propagating them along the orbit, assuming a normal distribution within the specified uncertainty range for each star individually. By doing so, we assume that each star represents a sample of stars whose parameters~(abundances and ages) differ from each other within the uncertainty range for that star. The last component of our model is the determination of masses~(or weights) of the orbits, which we describe next.

\subsection{Orbit superposition}\label{sec2::orbits_superposition}
Once we obtain the orbits of the APOGEE stars, we have a library of natural orbits for the superposition and determination of their individual contribution. We follow the approach described in \citetalias{Mapping-model}, where the weights of the orbits, $w_i$, were obtained from the following equation:
\begin{equation}
    \rho_{s}({\bf r}) = \sum^N_i w_i \rho({\bf r})_i\,\label{eq2::model}
\end{equation}
where each orbit is characterised by a 3D density distribution $\rho({\bf r})_i$. The 3D density of the stellar component of the MW~($\rho_{s}({\bf r})$) is taken from the analytic approximation of the potential~\citep{2022MNRAS.514L...1S}, representing one of the potential components used to integrate the orbits. All the densities were discretized onto a Cartesian grid of $30\times30\times30$~kpc with 50 bins along each direction. To account for the asymmetry of the orbital library relative to the bar, we mirrored each orbit around all three axes. 

To address a potential degeneracy in the solution of Eq.~\ref{eq2::model}, we divided the orbital library into five random subsamples and calculated the weights for each subsample individually. We repeated this procedure ten times with different random combinations of orbits in each subsample. Consequently, we obtained ten realizations of weights for each orbit, and the mean values of these weights were used in the following analysis. As described in \citetalias{Mapping-model}, such an approach is not absolutely necessary, but it ensures the stability of the solution.

The results of the orbit superposition are presented in Fig.~\ref{fig02::density_reconstruction}, where, for simplicity, we show the stellar surface density~(or number/density of stars), while the solution was obtained using 3D volume stellar density. The top panels show the analytic stellar density of the MW~($\rho_{s}({\bf r})$, left) from \cite{2022MNRAS.514L...1S} and the input APOGEE sample~(right). The second row presents a stacked orbital library before mirroring~(left) and after~(left). These panels essentially assume equal weights for all the orbits, demonstrating the importance of mirroring the orbital library widely used in orbit superposition methods~\citep{2008MNRAS.385..647V, 2022ApJ...941..109T}. The stacked orbital library already represents some interesting features, such as the elongated inner region of the bar and the ring at the solar radius. The latter one is trivially caused by the overdensity of stars at the solar radius in the initial sample~(see APOGEE map in top right). Likely, such a feature is similar to the $5$~kpc ring found by \cite{2021A&A...653A.143W}, who mapped the unweighted orbits in the barred potential~\citep{2017A&A...598A..66P}, where their input APOGEE sample of stars has the maximum density at around $5$~kpc.

\begin{figure*}
    \centering
    \includegraphics[width=1\hsize]{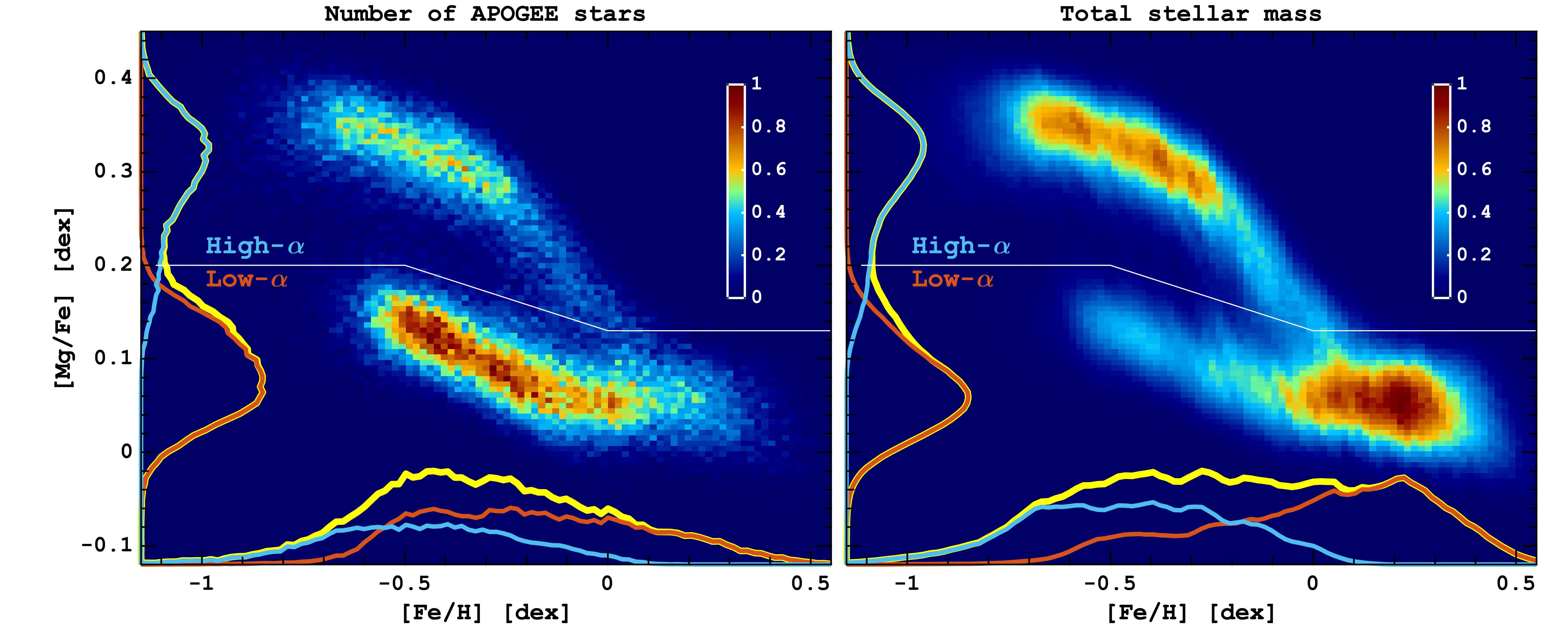}
    \caption{Chemical abundance structure of the MW disc in the $\MgFe-\FeH$ plane. The left panel shows the number of stars with abundances available in our APOGEE selection, while the right panel depicts the stellar mass-weighted distribution obtained using the orbit superposition method. In both panels, the maps are normalized by the maximum value. The yellow lines show the distributions of \MgFe and \FeH separately, where the contribution from high- and low-$\alpha$ populations is marked by blue and red, respectively. The solid white line is used to separate high- and low-$\alpha$ populations.}
    \label{fig02::afe_feh_all}
\end{figure*}

The left bottom panel of Fig.~\ref{fig02::density_reconstruction} presents the final result of the orbit superposition adopting the mean weights of the orbits, as described above. The bottom right panel shows the relative residuals between the analytic stellar density distribution and the orbit superposition solution. From the bottom panels, it is evident that our implementation of the orbit superposition provides us with a very good solution. It perfectly recovers the structure of the MW disc and the bar present in the analytic model, with some local deviations not exceeding $5-10\%$ within 15~kpc from the Galactic centre where the solution was obtained.

A more comprehensible comparison of the orbit superposition results with the APOGEE data is shown in Fig.~\ref{fig02::DFs}, where we present corresponding distribution functions. The figure shows a corner plot for the 6D phase-space data in Cartesian coordinates, demonstrating that our approach allows us not only to fill the gaps observed by the spectroscopic survey distribution function but also to extend to uncovered space. This figure demonstrates that our orbit superposition approach not only allows us to populate the MW stellar density distribution with the orbits of observed stars but also to augment the kinematic space. More detailed analysis of the kinematic data is presented in the following sections. 

\section{Mass-weighted MW disc stellar populations}\label{sec2::results_intro}

\subsection{Bimodality: global view}\label{sec2::results_intro_bimodality}
Before discussing the details of structural, kinematic, and abundance variations across the MW, we introduce its main components. One of the commonly used MW disc structural decompositions is based on the chemical abundances, particularly the $\aFe-\FeH$ plane. In Fig.~\ref{fig02::afe_feh_all} we show the $\MgFe-\FeH$ distributions of the number of stars from APOGEE~(left) and the stellar mass-weighted distributions obtained in the orbit superposition~(right). Since the distribution function units are different in the left and right panels, both maps are normalized by the maximum density value to contrast similarities and differences. We define high- and low-$\alpha$ sequences using the white line shown in both panels. This division is somewhat arbitrary~\citep[see,][for alternatives]{2019ApJ...874..102W, 2023ApJ...954..124I, 2024ApJ...974..227H} but will be used when referring to the disc populations.

The input APOGEE distribution on the left panel of Fig.~\ref{fig02::afe_feh_all} shows the well-known double $\alpha$-sequence~(see references in the introduction) traced well in the metallicity range from $\approx -0.7$ to $\approx -0.1$~dex. A faint `bridge' or a lower bend of the `knee' connects the high-$\alpha$ sequence to the super-solar metallicity tail of the low-$\alpha$ at higher metallicities. The bimodal distribution of the \MgFe ratio is very prominent along the vertical axis of the panel, while the \FeH distribution function~(DF) shows a broad distribution with a single maximum around $-0.5$~dex. Once we count the contribution from high- and low-$\alpha$ populations separately, marked by the white solid line, both components show rather flat \FeH-DFs.

The stellar mass-weighted distribution in the right panel of Fig.~\ref{fig02::afe_feh_all} looks slightly different compared to the input APOGEE sample. While the \MgFe-DF also shows a prominent bimodality with nearly the same fractional contribution of both components, the distribution of the low-$\alpha$ stars is more compact. However, the \FeH-DF between the two samples is very different; the \FeH-DF of the APOGEE sample is weighted towards the low-metallicity tail, while using the stellar mass-weighted sample it shows an excess of high-metallicity stars, significantly flattening the shape of the DF. This is quite easy to understand because the most metal-rich stars can be formed only in the very centre of the Galaxy, which is poorly covered by APOGEE. Therefore, these metal-rich stars are concentrated towards the centre or, more precisely, pass through the inner region and are assigned larger weights (or mass). Such redistribution of weights results in a less prominent low-metallicity tail of the low-$\alpha$ sequence. The bridge, or the knee, also appears more prominent once weighted by the stellar mass, likely because the corresponding phases of the MW enrichment happen in the inner MW.

Overall, we conclude that the stellar mass-weighed \MgFe-\FeH distribution preserves the main double sequence feature of the APOGEE sample; it, however, differs in terms of the fractional contribution of these two components, especially as a function of metallicity. The difference between the raw APOGEE and orbit superposition-based  \MgFe-\FeH maps highlights the effect of the selection function. Although APOGEE, being a near-infrared survey, penetrates close to the Galactic centre, populated by more metal-rich stars, it is still biased towards stars with more outer disc abundances. The orbit superposition method allows us to correct this bias by adding more mass to the metal-rich populations, amplifying their fractional contribution. For the first time here, we can show how the selection-function independent \MgFe-\FeH distribution should look for the entire MW, allowing for more direct comparison with simulations, where there is no need to mimic the APOGEE-like selection. In fact, the disc $\alpha$-bimodality looks less noticeable, as the metal-poor tail of the low-$\alpha$ sequence is much fainter compared to the APOGEE data.

\subsection{High-/low-$\alpha$: density structure}\label{sec2::results_intro_density}

\begin{figure*}
    \centering
    \includegraphics[width=1\hsize]{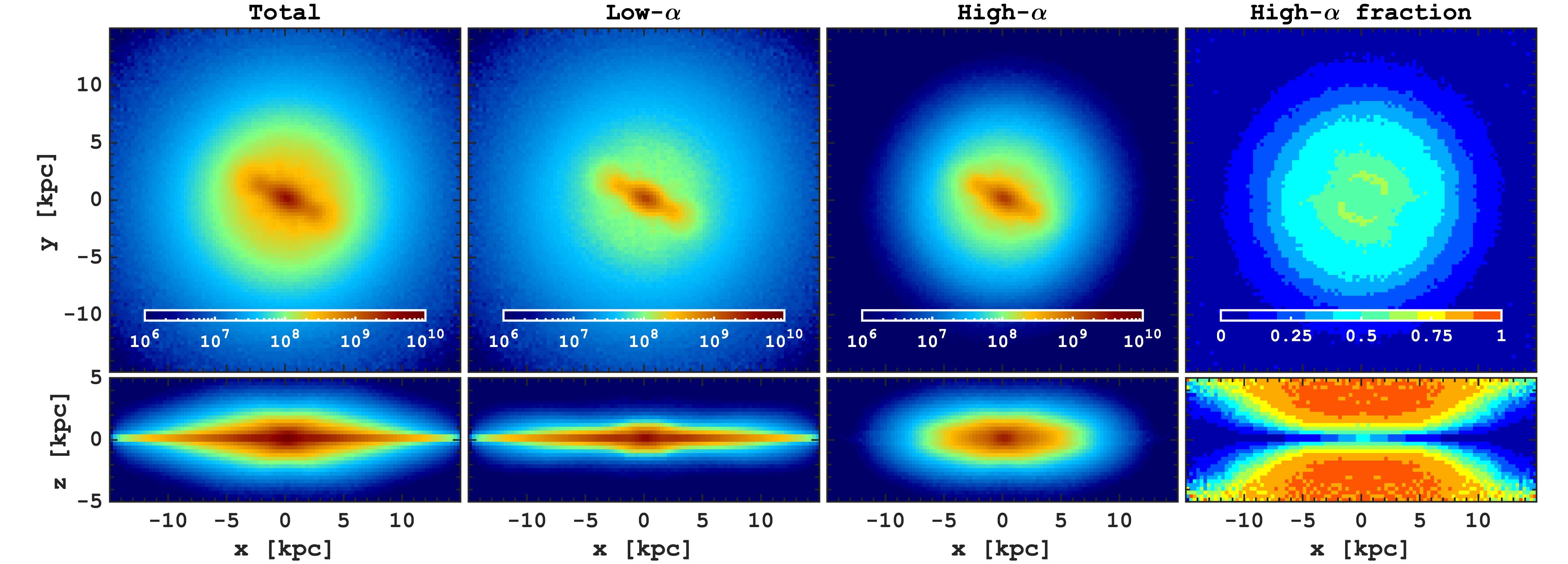}
    \caption{Stellar surface density~(\Msunkpcs) of low- and high-$\alpha$ populations in the MW disc. The projected mass fraction of the high-$\alpha$ populations is shown in the rightmost panel.}
    \label{fig02::disk_density_high_low}
\end{figure*}

After identifying the high- and low-$\alpha$ populations of the MW disc in the chemical abundance space, we proceed to examine their spatial properties. Figure~\ref{fig02::disk_density_high_low} presents the face-on and edge-on surface stellar density for all stars, as well as for the high- and low-$\alpha$ populations separately. The rightmost panel illustrates the projected stellar mass fraction of the high-$\alpha$ population. A notable feature in the density maps is the distinct difference in the large-scale morphology of the high- and low-$\alpha$ populations, previously described in literature~\citep[see, e.g.][]{2023A&A...674A..38G, 2023ApJ...954..124I}, but shown here, thanks to the orbit superposition implementation, in greater detail and across the whole MW disc. The low-$\alpha$ stars primarily form a radially extended, thinner disc component, whereas the high-$\alpha$ stars exhibit a thicker and more centrally concentrated component. The stellar mass fraction of the high-$\alpha$ population, approximately $0.5$, remains nearly constant within 5-6 kpc but becomes dominant at larger distances from the mid-plane. The U-shape of the high-$\alpha$ mass fraction in the bottom rightmost panel may imply the importance of the flaring of the low-$\alpha$ stars, but it can be simply the result of the geometric superposition of differently scaled high/low-$\alpha$ populations. We address this issue in the following sections.

\begin{figure*}
    \centering
    \includegraphics[width=1\hsize]{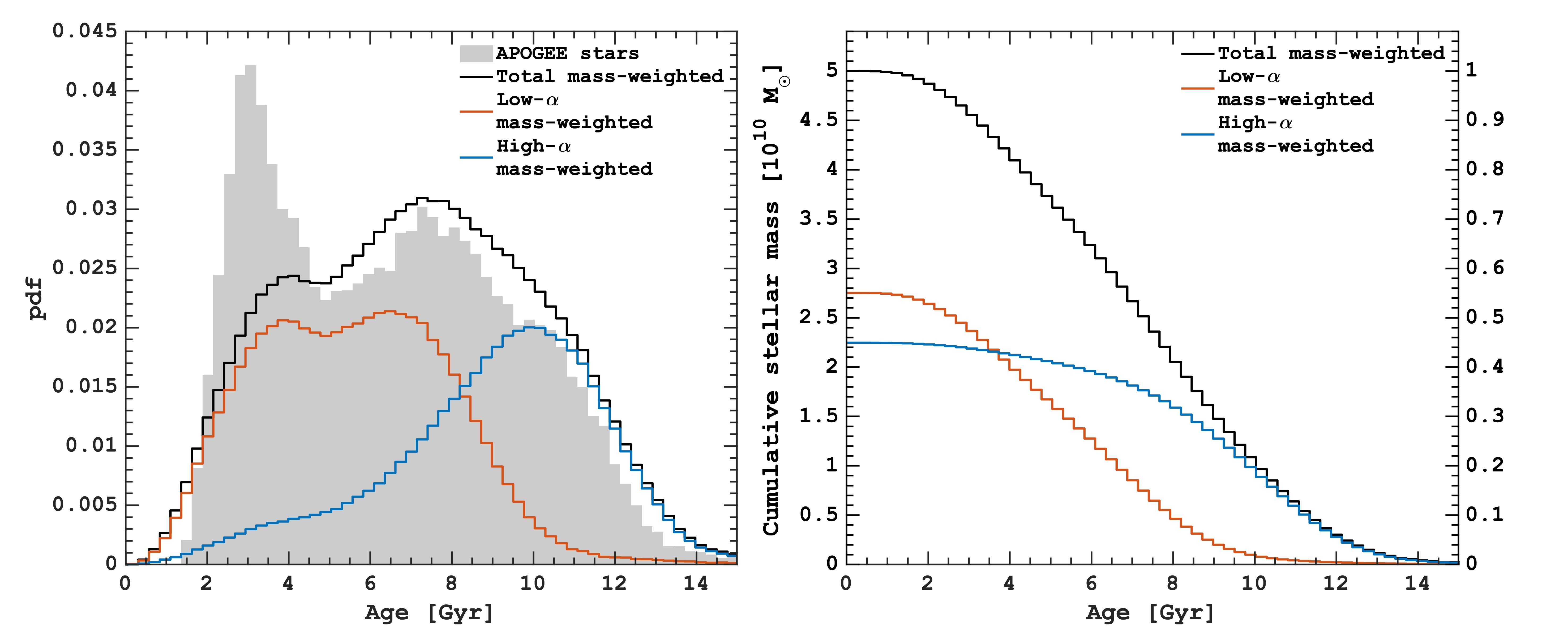}
    \caption{Left panel shows the age distribution for APOGEE stars~(grey shaded area) and the stellar mass-weighted age distributions for all stars in black and high-/low-$\alpha$ populations in blue and red, respectively. The right panel shows the cumulative mass distributions for the stellar mass mass-weighted components. The orbit superposition method results in the mass-weighted age distribution, which can be considered the star formation history of the entire Galaxy, as it gives direct information about the amount of stellar mass formed over time. Since we propagate the age uncertainties along the orbits, in some cases, we have ages larger than the age of the universe.  }
    \label{fig02::age_distr0}
\end{figure*}

\begin{figure*}
    \centering
    \includegraphics[width=1\hsize]{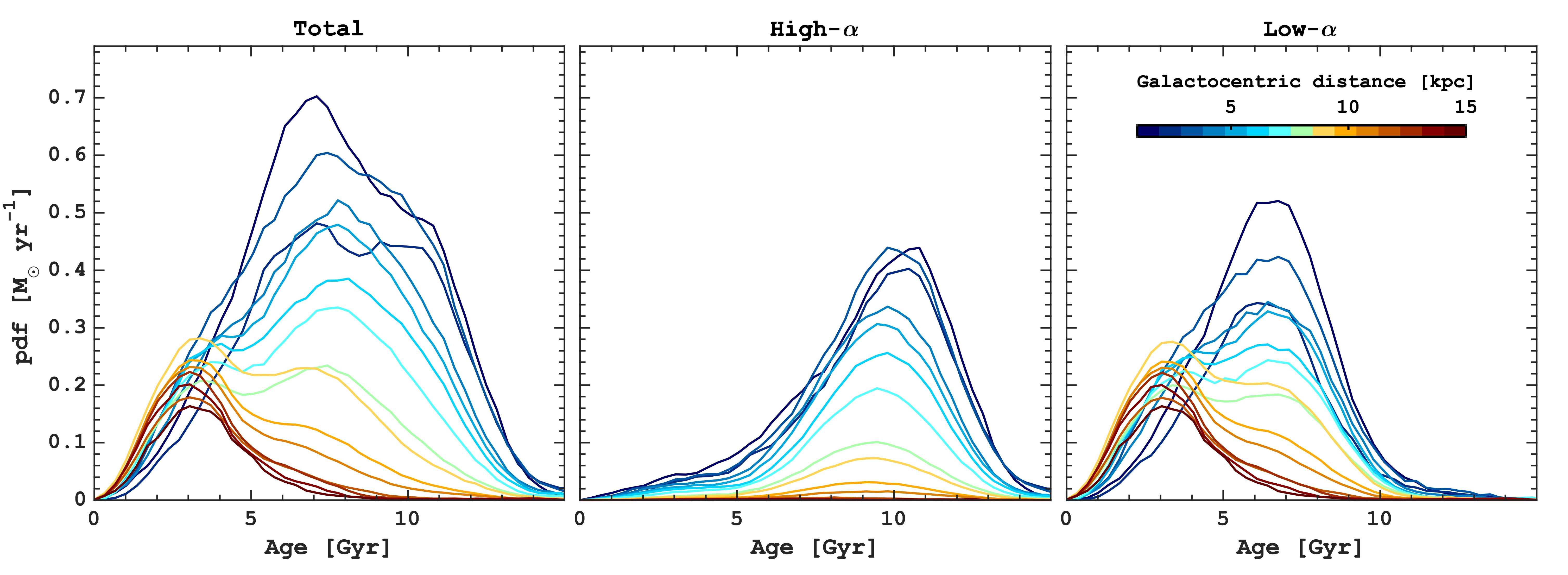}
    \caption{Stellar mass-weighted age distribution of the MW stellar populations at different Galactocentric radii for all~(left) and high/low-$\alpha$ populations in the middle and right panels, respectively. The lines of different colours correspond to different Galactocentric radii, as marked on the rightmost panel.}
    \label{fig02::age_radii_distibution}
\end{figure*}

Our orbit superposition approach is designed to capture the three-dimensional density of the MW disc and its primary asymmetric features—the bar and bulge structure. Although the edge-on projection we present here does not match the viewpoint from the solar position, the prominent boxy/peanut component is clearly visible (see bottom panels of Fig.~\ref{fig02::disk_density_high_low}), especially for the low-$\alpha$ populations and it is less evident for the high-$\alpha$ stars, in agreement with the current understanding of the bulge structure~\citep[see, e.g.][]{2017MNRAS.469.1587D,2018A&A...616A.180F}. An in-depth analysis of the MW bulge using the orbit superposition method is addressed in a follow-up paper of the series~\citetalias{Mapping-bulge}. Fig.~\ref{fig02::disk_density_high_low} demonstrates that both high- and low-$\alpha$ populations trace the bar; however, the bar associated with low-$\alpha$ stars is visibly longer, thinner, and more prominent than that of the high-$\alpha$ stars. Such kinematic behaviour was well explained in several studies, implying that the kinematics of stellar populations defines how they are shaped within the common gravitational potential~\citep[dubbed as kinematic fractionation][]{2017MNRAS.469.1587D}. In such a scenario, assuming that stellar discs heat rapidly as they form, both the in-plane and vertical random motions correlate with stellar age and chemistry, leading to different density distributions for kinematically cold and kinematically hot stars. However, the radially compact and vertically extended distribution of the pre-existing high-$\alpha$ populations affects the way how they are mapped into the bar and bulge region, highlighting the importance of the thicker component to understanding the MW inner region~\citep{2019A&A...628A..11D}.

\subsection{MW disc age structure}\label{sec2::results_age_structure}
Although the ages of stars are measured for relatively small samples compared to full 6D phase-space information, they play a crucial role in understanding the assembly of the MW components. The advancement of various machine-learning approaches, calibrated using highly precise asteroseismic data, has significantly improved age determinations in recent years~\citep{2021MNRAS.503.2814C,2019MNRAS.489.2079L,2023MNRAS.522.4577L}. Despite this progress, we are still far from having enough stars with measured ages to cover the entire Galaxy, which is now possible with our approach. Given that the kinematics of stars, and consequently their orbits, correlate with ages in complex ways~\citep[see, e.g.][]{2019ApJ...883..177N}, we transfer the age information from \cite{2024AJ....167...73S} along the orbits of stars; hence, in our orbit superposition method, we treat ages as tags for stars in the APOGEE catalogue. As mentioned earlier, we adopt a non-conservative approach: instead of excluding stars with high age uncertainties, we propagate these uncertainties along the orbits. This method ensures we retain as much information as possible while acknowledging and managing the uncertainties inherent in the age dataset. This, of course, may result in smoothing out some short time-scale features. However, it should not significantly impact our understanding of the overall structure and build-up of the bulk of the MW stellar disc.

In Fig.~\ref{fig02::age_distr0}, we present the age distribution for our selection of the APOGEE stars from \cite{2024AJ....167...73S}, alongside the stellar mass-weighted distribution for all stars and for the high- and low-$\alpha$ populations separately. It is important to underline that the raw age distribution~(grey shaded area in the left panel), unless corrected for observational biases and the survey selection function, does not directly correspond to the star formation history of the entire Galaxy. However, the stellar mass-weighted age distribution~(black line) obtained in our orbit superposition analysis does account for these factors. Nonetheless, the difference between the two distributions is not drastic. Both age distributions on the left closely follow each other up to around 6 Gyr. For younger ages, the APOGEE stars show a distinct peak at around 4 Gyr, which has almost vanished in the mass-weighted distribution. This suggests that the prominence of the peak is likely an artifact of the survey selection function rather than a real feature of the star formation history. We address this issue in detail in Section~\ref{sec2::results_AMR}, where we study the age-metallicity relation.

Although the exact timing and mechanism of the transition between the two $\alpha$-sequences are debated, it is well established from numerous studies that high-$\alpha$ stars are older than the relatively younger low-$\alpha$ populations, a view widely accepted today~\citep{2013A&A...560A.109H, 2014A&A...562A..71B, 2021A&A...645A..85M}. Figure~\ref{fig02::age_distr0}~(left) illustrates this, showing the mass distribution of the high-$\alpha$ populations peaking at the age of 10 Gyr, by which time about three-quarters of its mass was already formed. In contrast, the low-$\alpha$ sequence began to form later, reaching its peak around 6 Gyr ago. The star formation rate remained nearly constant for the most recent $2-3$~Gyr of evolution, only starting to decline approximately 4~Gyr ago. We note the presence of relatively young~($<6$ Gyr) high-$\alpha$ populations, known to exist in the MW~\citep{2016A&A...595A..60J, 2018ApJ...860...49M, 2018MNRAS.475.5487S, 2020ApJ...903...12S} originating from the thick disc stars being subjected to a mass increase by either mass transfer or collision and/or coalescence~\citep[see, e.g.][ and references therein]{2023A&A...671A..21J,2023A&A...676A.108C}. We estimate their fraction to be about $5\%$ among the total mass in our orbit superposition dataset, which is in agreement with other estimates~\citep{2015MNRAS.451.2230M}.

In the right panel of Figure~\ref{fig02::age_distr0} we show the cumulative mass growth of two disc components, which suggests that the masses of $\alpha$-sequences are very close to each other and, for instance, the mass of the high-$\alpha$ is about 40\% of the total stellar mass of the MW disc. In this analysis, we do not isolate the contribution of the bulge and the bar, as they are both mostly made of disc stars. Our estimate of the mass of the high-$\alpha$ populations is very much in agreement with \cite{2015A&A...578A..87S, 2022A&A...659A..64S}, who found their mass fraction of $\approx 44\%$ using chemical evolution models. We note, however, that it is not correct to assign all this mass to the geometric thick disc of the MW, because, as we showed in Fig.~\ref{fig02::disk_density_high_low}, the inner high-$\alpha$ stars are still concentrated close to the midplane and the low-$\alpha$ populations flare in the outer disc~\citep{2015ApJ...804L...9M}. Nevertheless, it was suggested that geometric thick discs make up a significant fraction of the baryonic content in external galaxies~\citep{2011ApJ...741...28C, 2017ApJ...847...14E}.

\begin{figure*}
    \centering
    \includegraphics[width=1\hsize]{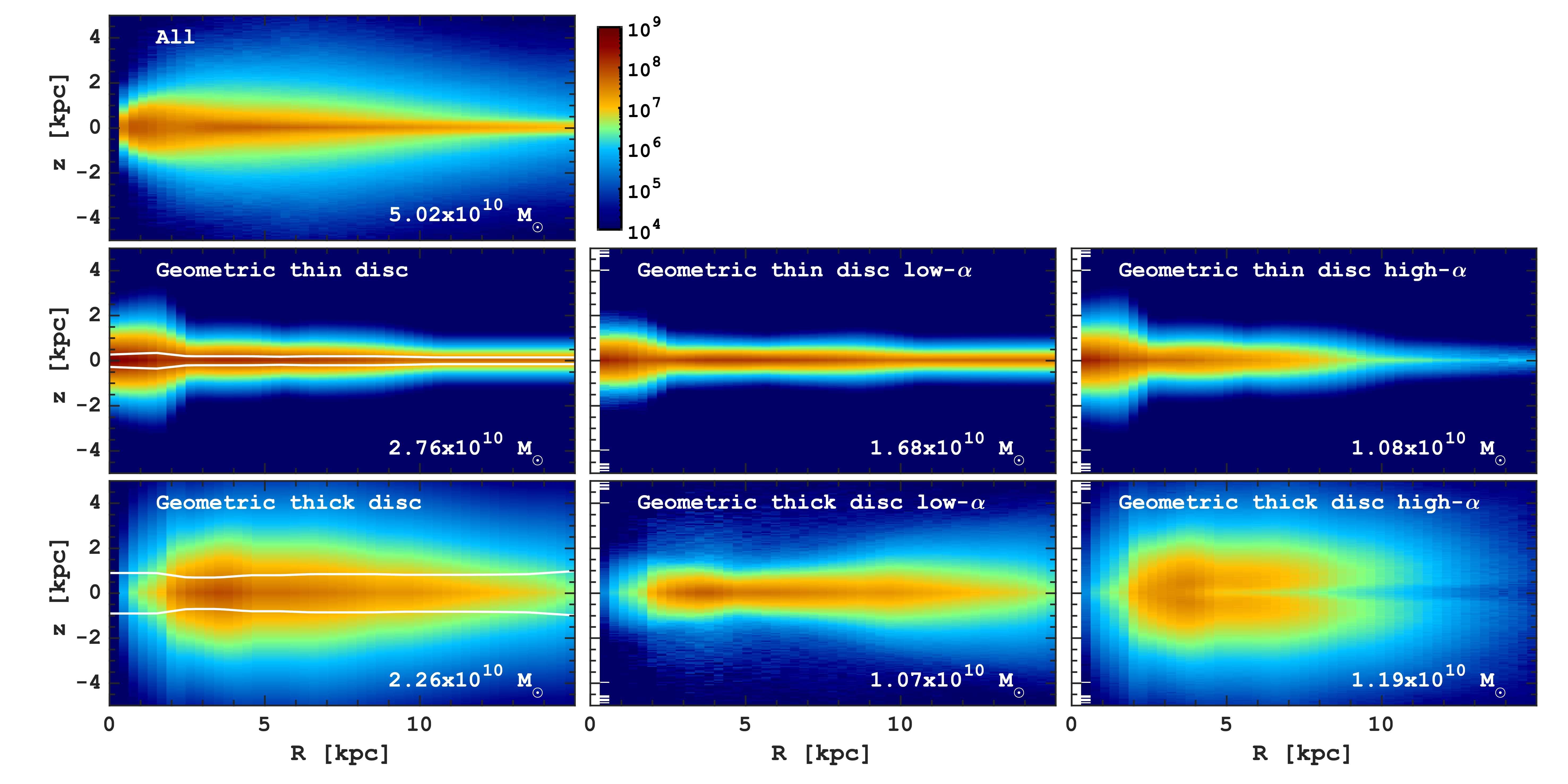}
    \caption{Decomposition of geometric thin and thick MW stellar discs. From top to bottom, the left panels show the stellar density distribution in $R-z$ coordinates for all stellar populations, as well as geometric thin and thick discs. Geometric disc components structure is obtained by fitting the vertical density profiles with double-exponential law. The middle and right panels show the structure of low- and high-$\alpha$ populations in geometric thin~(second row) and thick components~(bottom row). The purely chemical selection is shown in Fig.~\ref{fig02::disk_density_high_low}. The white lines in panels with geometric thin and thick discs show the corresponding exponential scale height. In each panel, the total stellar masses of corresponding components are given in the bottom right corner. The mass of the geometric thick disc is about $40\%$, which matches the mass of the high-$\alpha$ sequence. However, only half of the geometric thick MW disc is comprised of high-$\alpha$ populations.}
    \label{fig02::disk_decomposition}
\end{figure*}

\begin{figure*}
    \centering
    \includegraphics[width=1\hsize]{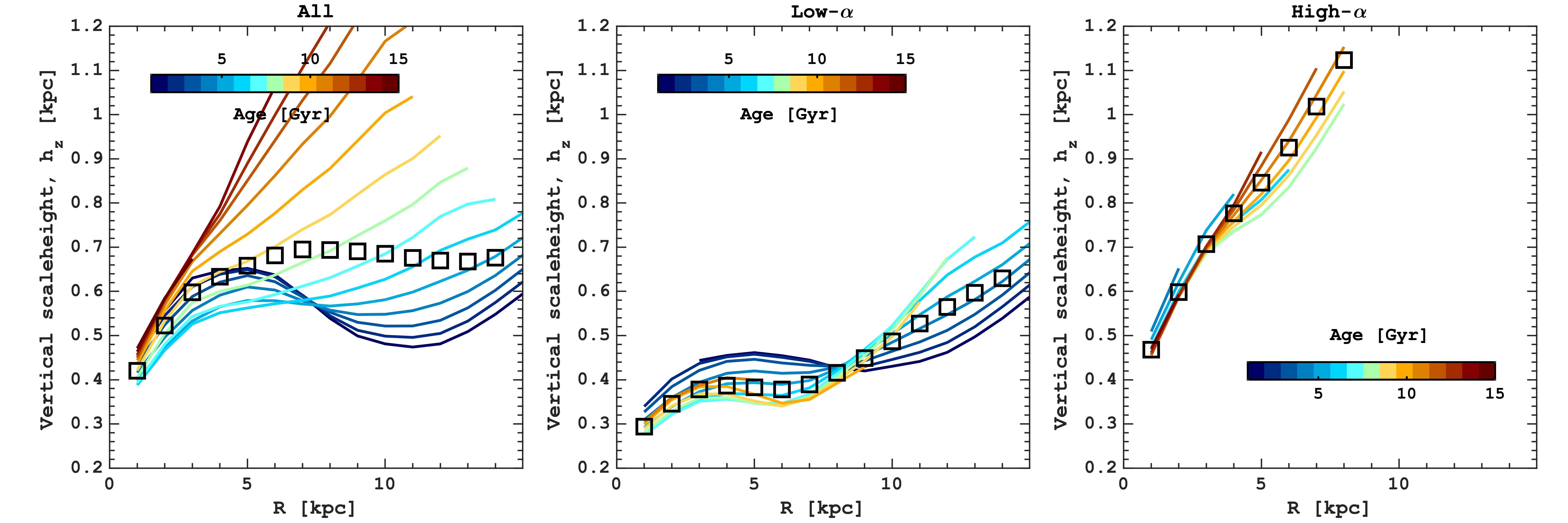}
    \caption{Flaring of the mono-age populations in the MW disc. The panels show the variation of disc scale height, $h_z$, with Galactocentric radius for all stars~(left), low-$\alpha$~(middle) and high-$\alpha$~(right) populations. Different colours correspond to mono-age populations, while the black squares show the trends for all stars together. Each line corresponds to the non-overlapping mono-ago population with the age bin size of $1$ Gyr. All mono-age populations show flaring, which is seen only outside the solar radius for the low-$\alpha$ populations. The strongest flaring is experienced by the high-$\alpha$ sequence stars.}
    \label{fig02::disk_flaring}
\end{figure*}

We find that the overall mass growth of the MW was quite rapid, with $\approx 40\%$ (or $\approx 2 \times 10^{10}~\Msun$) of its mass forming around $10$ Gyr ago. Such rapid in-situ stellar mass growth appears possible if this mass is concentrated in the discy component~\citep{2022MNRAS.516.2272S}, suggesting an early formation of the MW disc~\citep{2022MNRAS.514..689B,2024ApJ...962...84S,2024arXiv240918173S}. Recently, we demonstrated that this rapid evolution also supports the early formation of strong bars in the TNG50 galaxies~\citep{2024MNRAS.533.3975K}, with the emergence of a bar in the MW around $8$~Gyr ago, possibly shaping the end of the high-$\alpha$ sequence formation~\citep{2016A&A...589A..66H, 2024arXiv241021580B}. Alternatively, the impact of the last major merger has also been associated with the transition from high- to low-$\alpha$~\citep{2022MNRAS.515L..34L, 2024MNRAS.528L.122C}. However, on the contrary,  interactions of the MW with nearby galaxies are believed to result in bursts of the star formation~\citep[see, e.g.][]{2021MNRAS.506..531D,2024MNRAS.527.2426A}. This controversy can be addressed by considering the radial variations in the star formation history. For example, we can hypothesize that merger-driven quenching, followed by a subsequent rise of the SFR in the low-$\alpha$ regime, should proceed from the outside in, whereas in the bar-quenching scenario, the process should occur in the opposite direction. We discuss this in more detail in Section~\ref{sec2::results_AMR}.

The age distribution at different Galactocentric radii is presented in Fig.~\ref{fig02::age_radii_distibution} for all stars (left) and separately for the high-$\alpha$ and low-$\alpha$ populations (middle and right). It is important to note that, unlike the global mass-weighted age distribution, which can be considered as the star formation history, the age distribution at different radii does not allow such a direct interpretation. This is because stars can migrate over their lifetimes, leading to a mixing of star formation histories across different radii. The knowledge of stellar birth radii may allow us to resolve this problem, which is beyond the scope of this paper but addressed in~\citetalias{Mapping-sfh}. 

We observe that the peak of the age distribution for the low-$\alpha$ populations shifts towards the outer radii for younger stars, in agreement with the expected inside-out disc formation. Interestingly, the skewness of the distribution changes significantly, likely reflecting the effects of radial migration or related to SF episodes. If we ignore the effect of radial migration or assume that it primarily affects the shape of the distribution rather than the position of the peak, then the figure supports an inside-out formation scenario but predominantly in the inner regions. In this case, the low-$\alpha$ stars formed across the entire disc early on, but over time, star formation became more concentrated in the outer disc. In contrast, the high-$\alpha$ stars of different ages are well mixed in the inner disc, showing essentially no variation in their age distribution with Galactocentric distance. It is unclear whether this homogeneity results from radial migration~\citep{2002MNRAS.336..785S}, mixing caused by the last major merger~\citep{2012MNRAS.420..913B,2009MNRAS.397.1599Q} or if the high-$\alpha$ stars formed in an upside-down manner~\citep{2013ApJ...773...43B,2024arXiv241021377G}. It is reasonable to assume that all these processes contributed to the observed pattern, with their relative contributions yet to be determined.

\subsection{Geometric thick and thin discs, flaring}
In this subsection, we continue analyzing the spatial structure of the MW disc, aiming better to understand the geometrically defined thin and thick components and connect them to the chemically defined ones. An extensive discussion on the controversy between different definitions is presented in \cite{2013A&A...560A.109H}, which suggests that the inner thick disc is associated with the high-$\alpha$ populations. The metal-rich low-$\alpha$ populations comprise the inner thin disc, while the outer thin disc comprises the metal-poor tail of the low-$\alpha$ stars. This view is consistent with Fig.~\ref{fig02::disk_density_high_low}, where the high-$\alpha$ thick disc is prominently featured in the inner Galaxy. 

However, observations of external galaxies indicate that thick disc components, when measured structurally, are radially extended. This apparent contradiction was explained by \cite{2015ApJ...804L...9M} as the effect of disc flaring, where younger populations flare at progressively larger radii, thus populating an extended thick disc component. While radial migration can flare quiescent discs by increasing their scaleheight by a factor of $\sim2$ in $\sim4$ scale-lengths \citep{2012A&A...548A.127M}, when mergers are included this effect is a factor of $\sim10$ \citep{2008MNRAS.391.1806V, 2009ApJ...707L...1B}. The contribution of radial migration in the cosmological context is, thus, to reduce flaring because outward migrators arrive much cooler to the continuously extending strongly flared outer disc \citep{2014AA...572A..92M}.

A similar phenomenon is observed in the MW, where the outer disc is flared, contributing to the geometric thick component well beyond the compact high-$\alpha$ populations \citep{2011ApJ...735L..46B}. Therefore, it becomes evident that the geometric and chemical definitions of thick and thin discs are closely related in the inner galaxy where, however, the X-shaped bulge mixes them in a complex way. However, this correspondence breaks down when considering the MW as a whole and comparing it to external galaxies.

Since we have access to information about stellar populations across the entire MW, we can break this controversy and estimate the relative contribution of different components. In order to do this, we measured the density profiles of geometric thin and thick disc components by fitting the vertical stellar density distribution using a double-exponential law at different Galactocentric radii~(see Fig~\ref{fig02::thinthick_lowhigh_1D} in the Appendix). The obtained 2D density decomposition is presented in Fig.~\ref{fig02::disk_decomposition}~(left). Next, we convolved these maps with the density distribution of low- and high-$\alpha$ stars separately. This approach enables us to examine the contributions of the two $\alpha$-sequence populations~(see Fig.~\ref{fig02::disk_density_high_low}) to the geometrically defined MW discs, providing a clearer understanding of their spatial distribution and structural roles within the Galaxy.

We estimate the total mass of the geometric thick disc to be approximately $2 \times 10^{10} \Msun$, which is about $40\%$ of the MW stellar mass and it is surprisingly close to the mass of the high-$\alpha$ populations (see Fig.~\ref{fig02::age_distr0}). Interestingly, the thick disc is composed of equal proportions of high- and low-$\alpha$ stars. While the thin disc is predominantly made up of low-$\alpha$ populations, it still contains roughly $40\%$ high-$\alpha$ stars. Hence, each of the geometric components comprises a mixture of both $\alpha$ populations. We can conclude that the masses of the geometrically and chemically defined discs in the MW are very similar. It is hard to imagine a physical meaning for such a result; perhaps it is just a coincidence. However, it can be interesting to understand if simulations can reproduce such a picture and if this can be used as an extra criterion for selecting the MW analogues in models and among external components.

\begin{figure*}
    \centering
    \includegraphics[width=1\hsize]{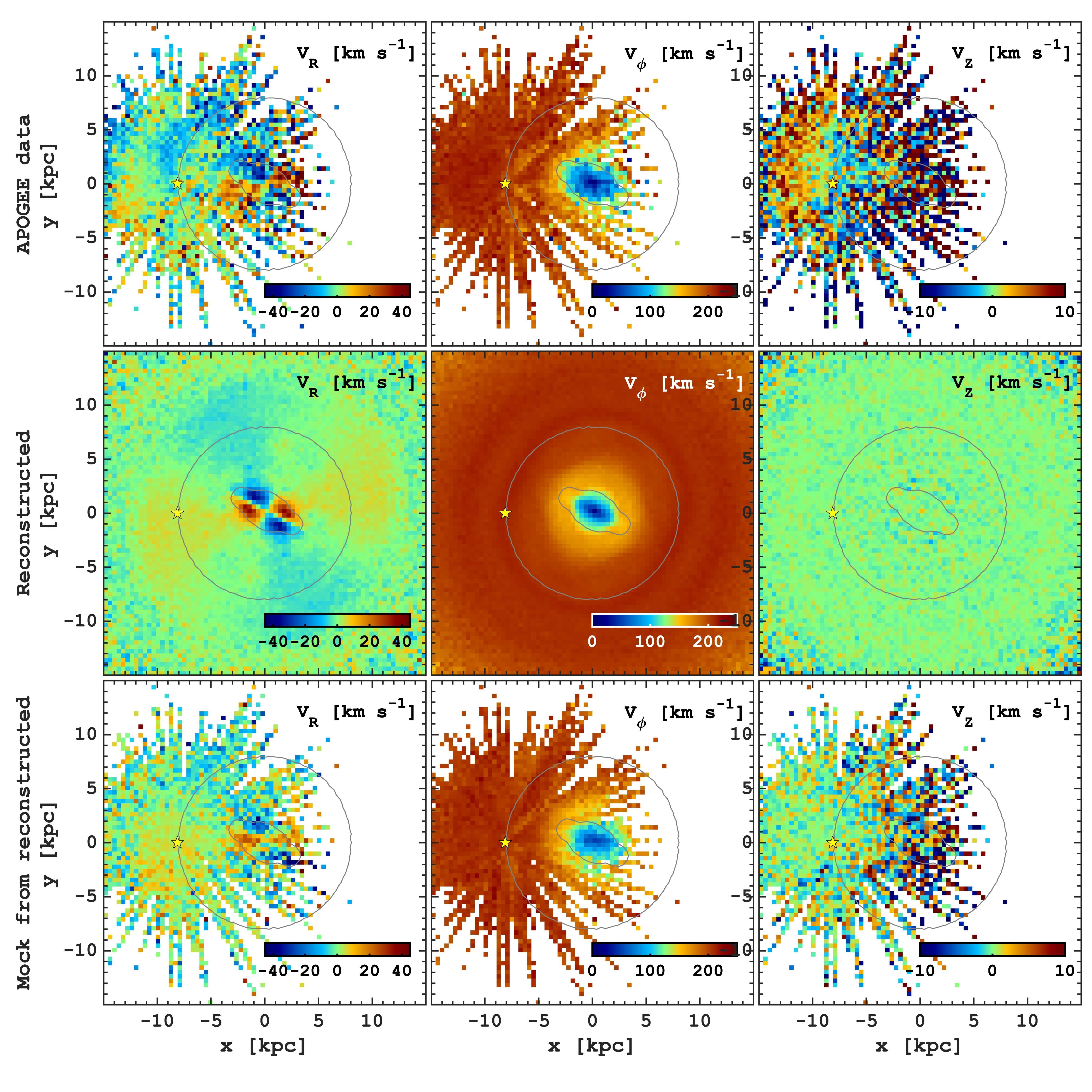}
    \caption{Face-on velocity component maps for the input APOGEE sample~(top) and mass-weighted orbit superposition result~(middle). The bottom row shows the orbit superposition result re-observed to mimic the APOGEE spatial footprint and take into the distance uncertainties~(see details in Sec.~\ref{sec2::results_kinematics_maps}). The contour lines show the isodensity levels highlighting the orientation of the bar of $27^\circ$. The position of the Sun~$(-8.12, 0)$ is marked by the yellow asterisk in all panels.}
    \label{fig02::face_on_velocities}
\end{figure*}

The composition of the thick component in Fig.~\ref{fig02::disk_decomposition} reveals another intriguing feature. In agreement with some previous studies, the MW thick disc overall has a roughly constant vertical scale height along the radius~(white lines in the left middle panel) of approximately $1$ kpc, while its low-$\alpha$ sub-population exhibits a very prominent flaring. To illustrate the effect of flaring, we decomposed the MW disc into mono-age (1 Gyr age bin) populations. The vertical density profiles of these mono-age populations are well-reproduced by a single exponential (see Fig.~\ref{fig02::disk_mono_age_density} in the Appendix) and the scaleheights~($h_z$) of these mono-age populations, measured at different radii, are shown in Fig.~\ref{fig02::disk_flaring} with different colours. The pattern observed in the figure closely reproduces the results of cosmological simulations presented in \cite{2015ApJ...804L...9M}, where the scale heights of coeval populations also increase with radius. However, the overall structure of the MW (considering all stars) does not show flaring, as illustrated by the black squares in the left panel. The increase in scale height within the inner region (less than 4-5 kpc) can be attributed to the presence of the boxy bulge. The middle and right panels of the figure reveal that both the low- and high-$\alpha$ populations exhibit flaring. Surprisingly, the high-$\alpha$ stars show a more pronounced flaring effect but its amplitude weakly changes with age. The latter, however, may be the result of relatively high age uncertainty for old high-$\alpha$ stars. Hence, we do not make strong conclusions regarding this similarity. The low-$\alpha$ populations demonstrate a monotonic increase of flaring amplitude with increasing age~\citep[see, e.g.][]{2014A&A...572A..92M}, which, however, is seen only outside the solar radius.

\section{Results: MW kinematics}\label{sec2::results_kinematics}

\subsection{In-plane velocity components}\label{sec2::results_kinematics_maps}
In this section, we focus on the kinematics of the MW stellar disc. In our previous paper, where we tested our approach on a simulated MW-like galaxy~\citetalias{Mapping-model}, we demonstrated that the APOGEE giant stars sample enables us to recover the full kinematic information about the disc stellar populations. Even more, one of the key advantages of our orbit superposition approach is that once weighted by stellar mass, it allows us to reveal the unbiased kinematic structure of the disk beyond the survey footprint, thereby correcting for the spatial selection incompleteness~(see Fig.~\ref{fig02::DFs}).

In Fig.~\ref{fig02::face_on_velocities}, we present the face-on projections of the three velocity components in Galactic coordinates: radial (\vr), azimuthal (\vp), and vertical (\vz). The top row of the figure displays the raw APOGEE data, showcasing several features extensively discussed in the literature. The radial velocity component exhibits a quadrupole pattern with at least two clearly visible positive and negative lobes. Such a pattern is expected in strongly barred galaxies like the MW. However, according to theory, this pattern should align with the bar's major axis, whereas we observe it oriented towards the Sun. This discrepancy, as mentioned in the introduction, results from distance and kinematic uncertainties known from early studies of the MW's morphology. Outside the bar region, a wave-like pattern is apparent, likely associated with spiral arms~\citep{ 2012MNRAS.425.2335S, 2014MNRAS.440.2564F, 2018A&A...616A..11G}. The rotational velocity shows a rising trend in the centre, with a nearly flat profile beyond 4 kpc. Interestingly, the APOGEE data do not display any bar-related signatures in the rotational velocity distribution~\citep[but see][based on the \Gaia data with similar results]{2023A&A...674A..37G}. The vertical velocity is zero on average within 10-11 kpc, without any notable features. In the outer regions, however, there is a systematic positive vertical velocity, highlighting the presence of the MW warp~\citep{1989ApJ...341L..13D, 2002A&A...394..883L, 2018MNRAS.481L..21P, 2022MNRAS.516.4988M}.

\begin{figure}
    \centering
    \includegraphics[width=1\hsize]{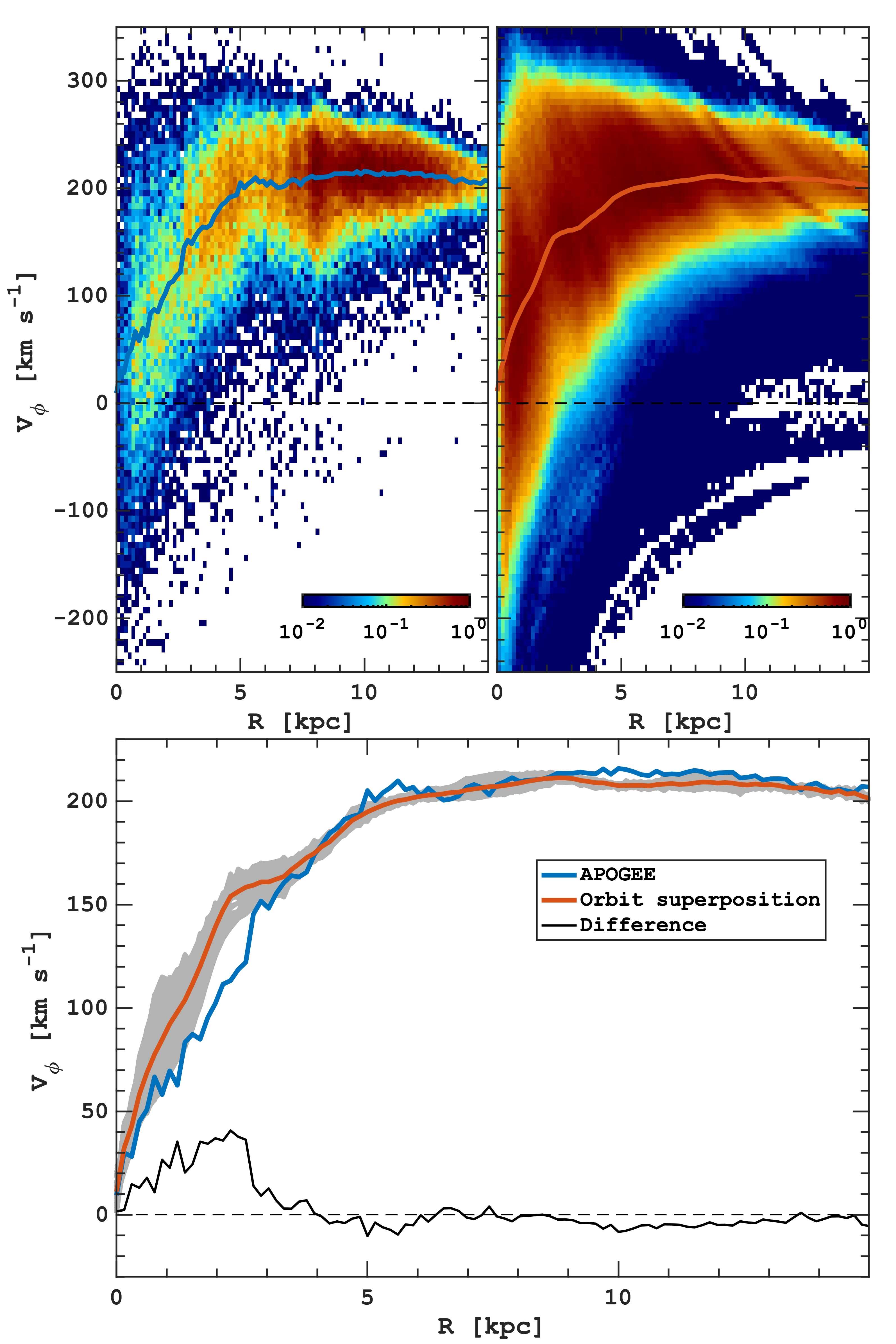}
    \caption{Azimuthal velocity~(\vp) distribution as a function of Galactocentric distance based on the APOGEE sample~(left) and the mass-weighted distribution orbit superposition~(right). The mean trends in each panel are shown by the blue and red lines, respectively. Their comparison is shown in the bottom panel, where the solid black line represents the difference. The variation of the mean rotational velocity with azimuth in the orbit superposition reconstruction output is marked by the grey colour. }
    \label{fig02::vrot}
\end{figure}

\begin{figure}
    \centering
    \includegraphics[width=1\hsize]{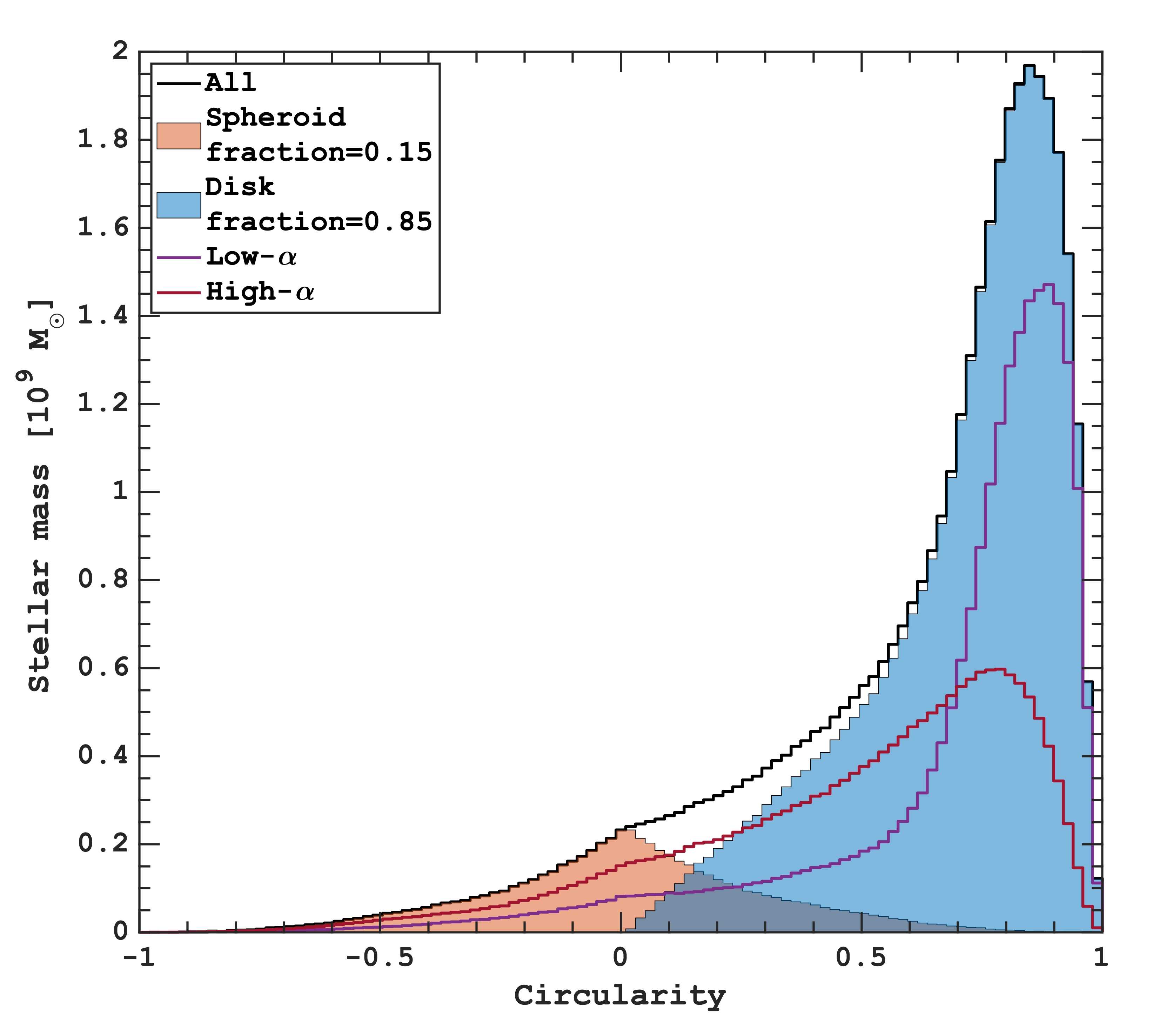}
    \caption{Stellar mass-weighted orbital circularity distribution for the whole MW~(black) and decomposed onto spheroid~(red) and disc-like kinematics~(blue). The spheroidal component is defined to be symmetric around zero circularity value; the remaining component is considered as the MW disc. The circularity distribution for high- and low-$\alpha$ populations are shown with dark red and purple colours, respectively.}
    \label{fig02::circ0}
\end{figure}

\begin{figure*}
    \centering
    \includegraphics[width=1\hsize]{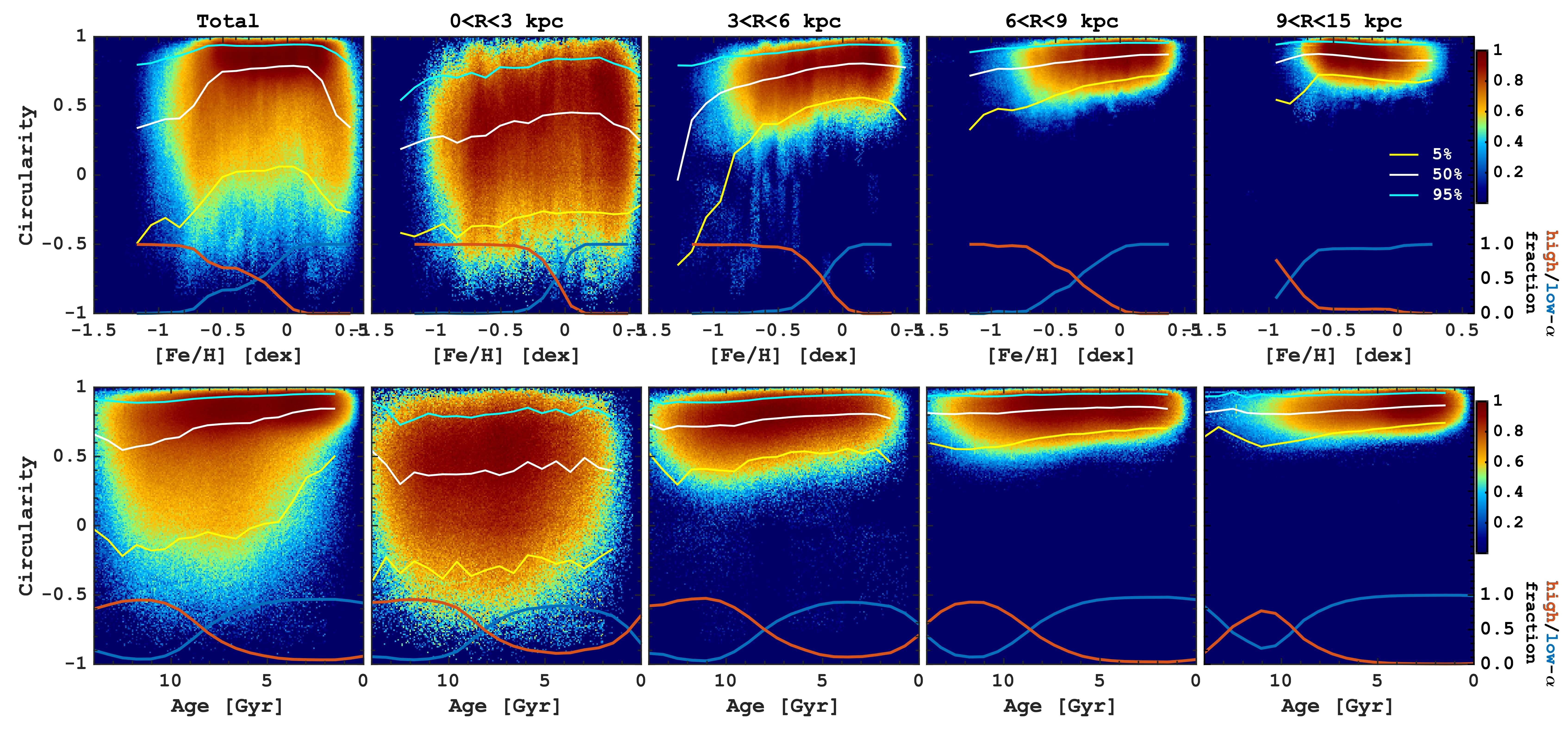}
    \caption{Stellar mass weighed distribution of circularity in different ranges of Galactocentric radii as a function of stellar metallicity~(top) and age~(bottom). The fraction of high- and low-$\alpha$ populations is shown by the red and blue lines in each panel. %{\bf We observe the difference because there is a radial metallicity gradient but not the age. The spinup is seen at the inner disc. At a given distance, the metal-poor stars are older  ... }
    }
    \label{fig02::feh_age_circularity_at_diff_radii}
\end{figure*}

The second row of Fig.~\ref{fig02::face_on_velocities} displays the reconstructed kinematics of the entire MW disc in the face-on projection. The radial velocity on the left shows the full quadrupole pattern correctly aligned along the bar, whose orientation~($27^\circ$ relative to the Sun - Galactic centre line) is highlighted by the isodensity contours. Outside the bar, there is a weak~($<5$~\kmps) pattern, predicted in the analytic calculations~\citep{2016MNRAS.461.3835M}. However, our model, by design, does not capture the effect of spiral arms seen in the APOGEE data. The rotational velocity, again contrary to the APOGEE, shows an ellipsoidal shape of the rising part, also aligned with the bar. The reconstructed vertical velocity is, on average, zero across the whole disc, as our orbit superposition model assumes a dynamical equilibrium, while the warp is clearly a non-equilibrium phenomenon~\citep{2020NatAs...4..590P,2024arXiv240509624H}. The recovered velocity maps show how the velocity field of the MW may look if our Galaxy was in dynamical equilibrium with no spiral arms. Of course, this is not the case in reality, as the Galaxy experiences different sorts of external interactions with LMC~\citep[see, e.g.][]{2018MNRAS.481..286L,2018MNRAS.473.1218L} and Sgr~\citep[see, e.g.][]{2018Natur.561..360A,2019MNRAS.485.3134L,2019MNRAS.490..797C} together with spiral waves passing across the disc and affecting the phase-space distribution of the disc stellar populations~\citep[see, e.g.][]{2012ApJ...750L..41W,2022A&A...663A..38K,2023A&A...674A..37G,2024MNRAS.52711393H}. 

Despite some expected limitations of our method, the kinematic solution we obtained using the orbit superposition approach can be used exactly to highlight various disequilibrium phenomena. Since our sample of APOGEE stars is rather small and quite noisy, as seen in the top row of Fig.~\ref{fig02::face_on_velocities}, we suggest that one can use the \Gaia DR3 data and subtract our orbit superposition solution to isolate the non-equilibrium phenomena, such as the kinematic imprint of spiral arms.
However, we stress that simply subtracting the reconstructed data from the APOGEE kinematics will highlight not only the effects not included in our model but also various biases present in the observational dataset. For instance, subtracting the top and middle radial velocity maps will primarily highlight the impact of distance uncertainty rather than real effects. Therefore, to make a proper comparison, we suggest generating a mock dataset from the reconstructed data and introducing kinematic and distance uncertainties first. 

This approach is illustrated in the bottom panels of Fig.~\ref{fig02::face_on_velocities}, where we ``re-observed'' our reconstructed MW stellar populations following the procedure described in \citetalias{Mapping-model} and used to create the APOGEE mock from a simulated galaxy. Here, we also include distance uncertainties as detailed in \cite{2024MNRAS.528.3576V}, assuming that the orbit superposition is ground truth. Because of such extra features,
as shown~(bottom left panel), the radial velocity quadrupole pattern is incorrectly oriented similarly to the APOGEE-based map~(top left), and the rotational velocity also does not align with the imposed bar orientation. At the same time our mock also reproduces well some artifacts seen in the rotational velocity map, such as straight lines with sharp variations of the mean velocity along the line-of-sight. The radial velocity amplitude appears to be a bit weaker compared to the APOGEE-based map, which can be partially explained by the lack of spirals in our mock, affecting the kinematics of stars at the edge of the bar once the spirals are connected to the bar~\citep{2024MNRAS.528.3576V}. Also, we do not include the kinematic uncertainties, which also affect the observed kinematics~\citep{2019MNRAS.490..797C}. The vertical velocity map appears noisy in the inner region with no systematic patterns. Nevertheless, a very close similarity of our mock with the APOGEE kinematics suggests that the reconstructed data closely reproduces the large-scale kinematics of the MW disc.

We suggest that a proper comparison of orbit superposition-based reconstruction with observational data requires thoughtful implementation of not only the distance uncertainties but also the radial velocity and proper motions uncertainties whose significance sharply increases with the distance from the Sun~\citep{2018A&A...616A..11G} which we leave for further investigation. Note that this is not a trivial exercise, as the uncertainties depend strongly on the position in the Galaxy in a complex way that is greatly related to extinction. Also, using a larger sample of input stars for the orbit superposition will be more beneficial compared to the limited sample used in this work, which is restricted by the available information on chemical abundances and stellar ages.

\subsection{Rotational velocity structure}\label{sec2::results_kinematics_vrot}
One of the key characteristics of disc galaxies, including the MW, is rotational velocity as it provides us with information regarding the global mass distribution but also small-scale effects related to non-axisymmetric structures. In Fig.~\ref{fig02::vrot}, we show the comparison of the APOGEE sample-based distribution of azimuthal velocity~(left) and the one we obtain using the orbit superposition approach~(right). Note that in the left panel, we show the number of stars in the $R-\vp$ plane, while in the right, the colour corresponds to the stellar density; however, for easier comparison, we scale both distributions by corresponding maximum values.
Both maps on the top show roughly the same behaviour; however, thanks to the full coverage of the MW disc and proper account for the stellar mass radial distribution in the right panel, we can observe many details which are obscured in the APOGEE-based \vp-distribution. On the right, we can see several diagonal overdensities, or ridges, which are seen well in larger observational datasets compared to the one we have here. Although such ridges are believed to be associated with various phenomena across the disc, e.g. Galactic spiral pattern~\citep[see, e.g.][]{2019MNRAS.490.1026H,2022A&A...663A..38K}, non-equilibrium phase-space features~\citep[see, e.g.][]{2019MNRAS.489.4962K, 2019MNRAS.488.3324F, 2019MNRAS.485.3134L} and so on, we can see that several of them can be very well recovered by our approach, even in the innermost region poorly covered by APOGEE. In this case, these features are naturally associated with the impact of the bar  resonances~\citep{2019A&A...626A..41M}.

The combination of different small-scale diagonal overdensities in the 
$\rm R-\vp$ plane can result in a wave-like pattern in the radial profile of the mean rotational velocity~\citep{2018MNRAS.479L.108K}, as illustrated by the red line in the right panel. This wave-like pattern is also observed in the APOGEE data; however, these velocity variations do not match precisely, as shown in the bottom panel of Fig.~\ref{fig02::vrot}. Outside $4$~kpc, the difference does not exceed $10$~\kmps. It is important to consider the limited coverage of the APOGEE footprint and the uneven number of stars in the azimuthal direction, which may affect the behaviour of the mean radial trend. The azimuthal variations of the rotational velocity here are highlighted by the grey shaded area, which shows the range of possible mean $\vp$ values in the azimuthal direction at a given distance from the centre. 

The biggest difference between APOGEE and the orbit superposition reconstruction kinematics is visible in the innermost $3$~kpc, where we notice an offset up to 40~\kmps. The explanation for such a difference is rather trivial. As we showed in the previous section, the distance uncertainties result in the velocity field ``stretching'' along the line of sight, affecting the distribution of individual components of the velocity vector.

\subsection{MW circularity distribution}\label{sec2::results_kinematics_circ}

One of the parameters widely used for the analysis of the kinematic structure of galactic discs is circularity, showing how much the angular momentum of a given tracer deviates from the maximum allowed angular momentum~(perfectly circular motion) at a given total energy~\citep{2003ApJ...591..499A}. It is often used in cosmological simulations not only to trace the settling of the galactic discs but also to do a kinematic decomposition into rotationally-supported disc and dispersion-dominated spheroidal components~\citep{2003ApJ...597...21A, 2014MNRAS.437.1750M}. 

In Fig.~\ref{fig02::circ0}, we show the stellar mass-weighted distribution of circularity. We remind the reader that neither the total energy nor the angular momentum are conserved in a non-axisymmetric potential. Hence, a single orbit can contribute to different parts of the distribution. The total circularity distribution peaks at the circularity value of about $0.85$ with a weak tail to the negative values, which unsurprisingly shows that the MW is a disc-dominated galaxy with no distinct spheroidal component, at least in terms of the kinematics. However, again, based on the circularity distribution, it is not a pure disc because there is a notable population of stars with circularity around zero and below.

To quantify a kinematically defined spheroid contribution, we assumed that the distribution of the circularity parameter of the spheroidal component is symmetric around zero, with the rest considered as the disc. The result of this decomposition is illustrated by the red and blue filled areas, showing that the total mass of the kinematically defined spheroidal component is about $15\%$. It is important to emphasize that this does not automatically imply that the MW has a classical bulge. Instead, it sets an upper limit on the mass of the spheroidal component, as several different stellar populations can exhibit such kinematics, such as proto-disc of the MW~(dubbed as Aurora~\citep{2020MNRAS.494.3880B}), heated high-$\alpha$ stars~(Splash/Plume~\citep{2020MNRAS.494.3880B,2019A&A...632A...4D}), ex-situ stars stripped during mergers~\citep{2018MNRAS.478..611B, 2018Natur.563...85H} and disrupted globular clusters~\citep{2017ApJ...849L..24M, 2023A&A...673A..44F}. An in-depth study of the MW bulge region is required to place tighter constraints on its inner spheroidal stellar component. However, when examining the circularity distribution for the low-$\alpha$ populations, we notice that about one-third of the kinematically defined spheroid comprises relatively young low-$\alpha$ stars~(see Fig.~\ref{fig02::age_radii_distibution}). This observation suggests that the possible mass of an old classical bulge could be as low as $10\%$ of the stellar mass of the MW, which aligns well with several independent estimations~\citep{2010ApJ...720L..72S, 2017MNRAS.469.1587D}. Nevertheless, without a more detailed understanding of the stellar populations in the bulge region, we cannot conclusively determine whether the MW hosts a massive classical bulge~(see more details in \citetalias{Mapping-bulge}).

We underline that the derived distribution of circularity shown in Fig.~\ref{fig02::circ0} can be used for the analysis of cosmological galaxy formation simulations and extragalactic observations~\citep{2018NatAs...2..233Z, 2022ApJ...930..153S} to identify the MW analogues, at least in terms of its kinematic composition. This distribution serves as a benchmark for comparing the MW's kinematic structure with those of other galaxies, aiding in the identification of similar galactic systems and enhancing our understanding of the MW's place in the broader context of galaxy formation and evolution.

The broad distribution of the circularity in Fig.~\ref{fig02::circ0} indicates the heating of the MW disc. To assess the details of this process in Fig.~\ref{fig02::feh_age_circularity_at_diff_radii} we present the distribution of circularity as a function of stellar metallicity~(top) and age~(bottom) where different panels correspond to different ranges of Galactocentric radii. The circularity distribution is quite complex in the inner 3 kpc, where the mean circularity increases with metallicity. However, such a trend is not very prominent as a function of age. This controversy can be related to either a complex relationship between age and metallicity~(see Section~\ref{sec2::AMRG_radii} below) of the populations mixed together by the bar/bulge structure, whose detailed investigation is presented in \citetalias{Mapping-bulge}. At the same time, this region is the most likely to contain pre-MW disc populations; the Aurora and Spin-up contributions can be recovered more robustly using more detailed chemical abundance information~\citep{2022MNRAS.514..689B, 2022ApJ...938...21M, 2024arXiv241022250K}; however, it is not clear whether these populations can be associated with a classical bulge component.

For the pure disc component at larger radii (3-9 kpc) in each panel of Fig.~\ref{fig02::feh_age_circularity_at_diff_radii}, the circularity gradually increases with age and metallicity, suggesting either the upside-down disc settling or the heating of stellar populations over time. In the outer disc, outside $9$~kpc, this behaviour is not so prominent, suggesting a modest impact of the dynamical heating, which might be explained by the lack of the heating factors, like spiral arms and massive molecular clouds scattering stars~\citep{2016MNRAS.462.1697A}. The most prominent feature here is seen in the $6-12$~kpc range~(third and fourth columns). The presence of the circularity tail down to $-0.5-0$ in the metal-poor part of the distributions indicates that stars with metallicity below $\approx -0.4$ and older than $\approx 8$ were heated up. This feature is similar to the Splash in-situ populations, whose kinematics are shaped by the Gaia-Sausage-Enceladus~(GSE, \cite{2018MNRAS.478..611B}, \cite{2018Natur.563...85H}) merger. As seen from these panels, the high-$\alpha$ populations are the most affected by such an event. This impact is not seen in the outer disc, likely because the high-$\alpha$ disc was and remained quite compact at the time of the merger. The figure shows that the GSE-caused heating process is mostly over by about 8 Gyr ago. There is, however, a similar feature -- a faint tail towards the low-circularity for stars with the age of $4-5$ Gyr. This feature may suggest that there was another significant disc heating event around that time, for instance, introduced by the first infall of the Sgr; despite the early phases of its orbit being highly uncertain, it is often involved in explaining all sorts of chrono-chemo-kinematic peculiarities seen for the low-$\alpha$ stellar populations. In this case, it is surprising that the effect of the heating at larger radii is not seen because an infalling on the Sgr-like orbit satellite would affect the disc outskirts earlier and more strongly, at least during the first passage~\citep{2024MNRAS.527.2426A}. The lack of such effects might imply that around 6 Gyr ago, the MW disc was not larger than $\approx 9$~kpc.

\subsection{Results: age-velocity dispersion relation}\label{sec2::results_AVR}
In the previous section, we discussed specific effects observed in the distribution of orbital circularity, particularly the smooth transition from hot to cold orbits as a function of Galactocentric distance and age, suggesting the dependence of the disc heating history on these parameters.  In general, the heating of stellar disc components has been argued to result from the impact of various mechanisms of stellar disc heating related to stochastic spiral patterns~\citep{1990MNRAS.245..305J, 2006MNRAS.368..623M}, bars \citep{2010ApJ...721.1878S}, molecular cloud relaxation~\citep{1951ApJ...114..385S, 1984MNRAS.208..687L, 2016MNRAS.462.1697A}, disc-halo interaction ~\citep{2011MNRAS.416.2802F}, infalling satellites~\citep{2004MNRAS.351.1215B,2016MNRAS.459.2905M}, and other processes~\citep{2011ARA&A..49..301V}. To summarize our findings regarding the MW disc heating quantitatively, we examine the well-known age-velocity dispersion relations~(AVR). 

\begin{figure}
    \centering
    \includegraphics[width=1\hsize]{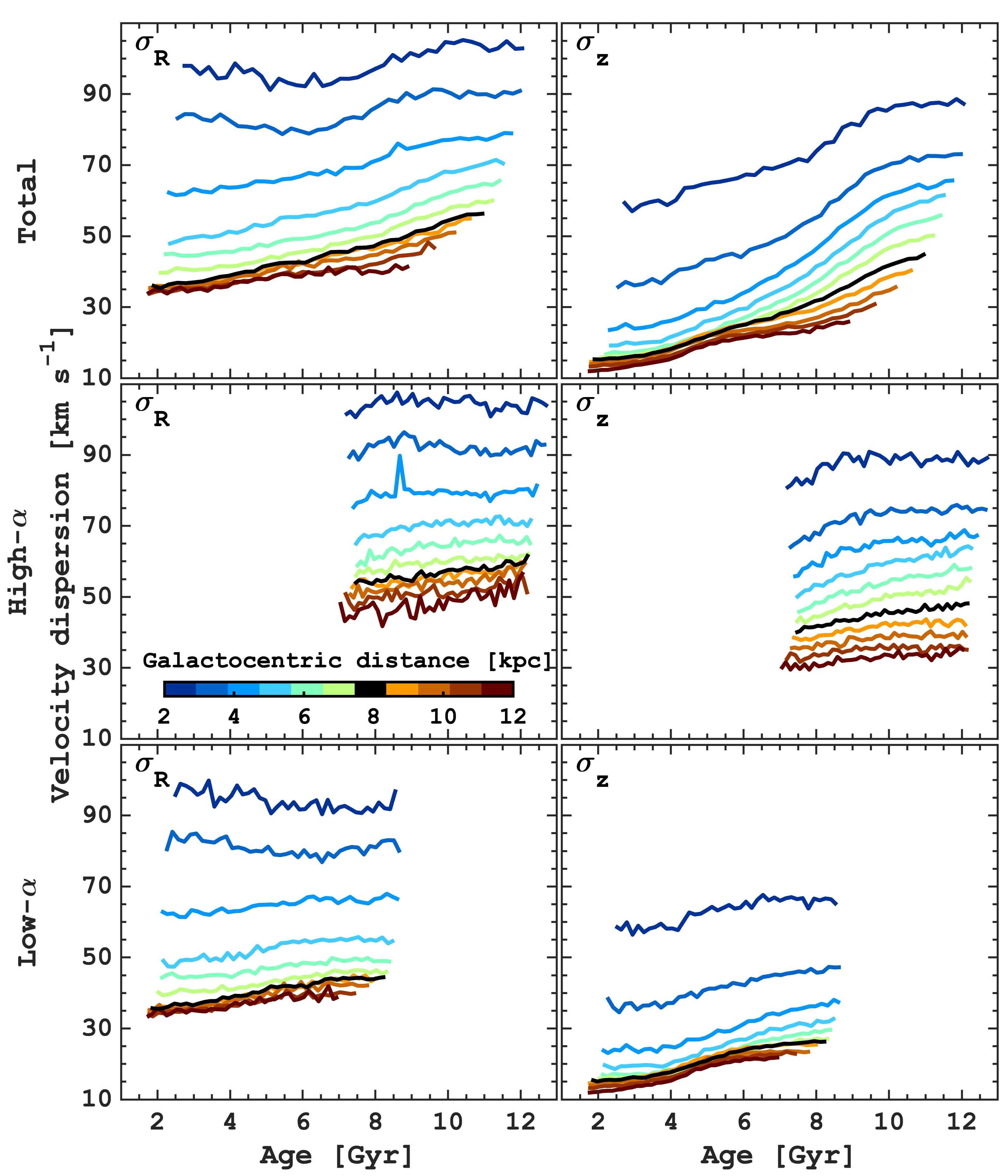}
    \caption{Stellar mass-weighted age-velocity dispersion components~(\sr, \sphi\ and \sz\ in different columns) for stars at different Galactocentric radii. The top, middle and bottom panels correspond to the all-stars, high- and low-$\alpha$ populations, respectively. The colour of lines corresponds to the relations for stars at different Galactocentric radii, where the black colour marks the solar radius~($\approx 8$~kpc). }
    \label{fig02::AVR}
\end{figure}

Figure~\ref{fig02::AVR} shows the AVR for stars selected in radial annuli of $1$~kpc width, which, thanks to the orbit superposition approach, we can explore across the entire Galaxy. The top row shows the \sr\ and \sz\ velocity dispersion components for all stars with age information, regardless of their chemical composition. At the solar radius, depicted in black, the velocity dispersion components increase monotonically from $(\sr,\sz)=(35, 15)$ for the youngest stars to $(55,45)$~\kmps for the oldest populations. At the innermost radii (<3-4 kpc), the radial velocity dispersion component (\sr) shows little to no dependence on age. This effect is evidently due to the presence of the bar. It suggests that stars formed within the MW bar region are born on quite eccentric orbits, rather than experiencing significant heating over time. This is further supported by the gradual increase in vertical velocity dispersion for stars in this region, indicating that stars formed 'hot' in a thin gaseous disc and were gradually heated primarily in the vertical direction. However, the exact mechanism for vertical heating by the bar remains unclear. In the case of a rigid bar rotating at a constant angular speed, significant heating is unlikely because positive and negative torques from the bar would cancel out along the orbit. This scenario changes if the bar parameters (strength, length, pattern speed) evolve over time, which likely results in bar-induced heating.

\begin{figure}
    \centering
    \includegraphics[width=1\hsize]{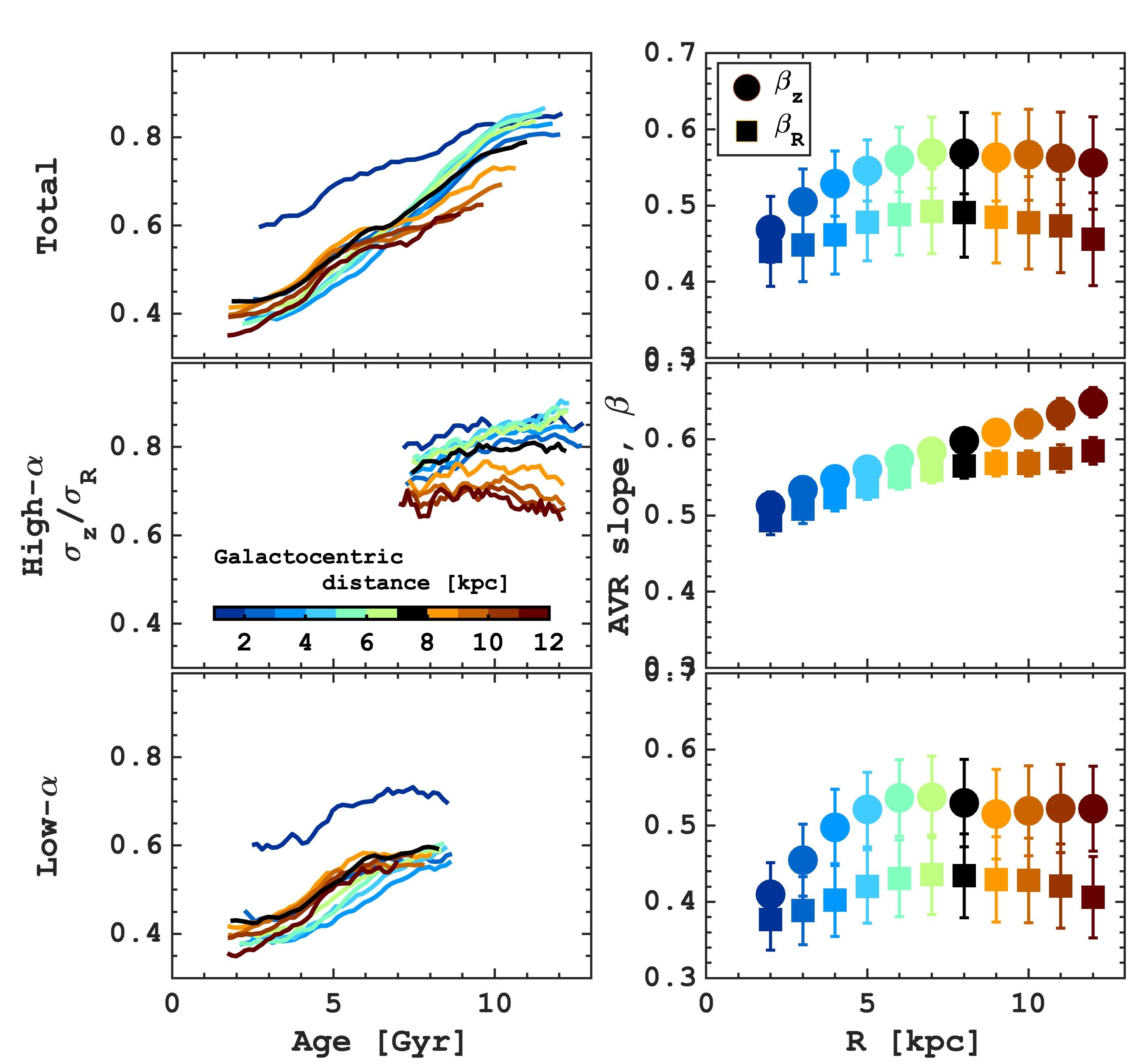}
    \caption{Parameters of the AVR. The left panels show the ratio between vertical and radial velocity dispersion components, while the right panels depict the power-law slope of the AVR, $\beta$. In both sets of panels, the colour corresponds to the Galactocentric distance. The solar radius values are marked with a black colour.}
    \label{fig02::AVR_params}
\end{figure}

At larger radii ($>4-5$~kpc) we observe a gradual increase of both velocity dispersion components in agreement with a number of studies of the data in the solar neighbourhood~\citep[see, e.g.][]{1977A&A....60..263W, 2009A&A...501..941H, 2001ASPC..230...87Q, 2009MNRAS.397.1286A, 2012ApJ...751..131B, 2019MNRAS.489..176M}. We only stress that the observed behaviour is usually interpreted as a gradual heating of stars formed on nearly circular orbits over time. However, some simulations demonstrate that the birth eccentricity of stars increases with age, suggesting a significant natal velocity dispersion originated from highly irregular motions in the ISM due to a bursty regime of star formation~\citep{2023MNRAS.519.2598G,2023MNRAS.523.6220Y}. Most likely, both the decline of the birth velocity dispersion and the follow-up heating take place in disc galaxies, but their relative importance is hard to estimate in the MW data alone. However, by analysing the AVRs for high and low-$\alpha$ populations, we can assess the importance of the physical conditions in the MW, which are believed to be different during the formation of these two chemically distinct components.

In the top-right panel of Fig.~\ref{fig02::AVR}, we observe notable changes in vertical velocity dispersion behaviour as a function of age and distance. Older stars ($>9$~Gyr) in the inner regions exhibit a nearly constant velocity dispersion, a trend confined to the innermost Galaxy within the bar length ($\approx 4$~kpc). For stars aged between 4 and 9 Gyr, the vertical velocity dispersion decreases across all radii, displaying a noticeable slowdown beyond the solar radius, characterized by a smooth step-like transition. In the inner Galaxy, the velocity dispersion decrease continues steadily toward the youngest stellar populations. The observed variations in the AVR trends seem to suggest a contribution of different factors affecting the kinematics of the MW stars in the inner and outer discs, while the solar radius seems to be unaffected by them~(see Section~\ref{sec2::discussion} for details). 

When we divide the MW disc stars into low- and high-$\alpha$ populations, a diverse picture emerges, as shown in Fig.~\ref{fig02::AVR}~(middle and bottom rows). There is a radical decline in velocity dispersion of the high-$\alpha$ stars. However, at any given radius, the velocity dispersion curves for the radial and vertical components remain nearly flat despite the stars spanning a broad range of ages. For the low-$\alpha$ stars alone (bottom row), we see the same trend observed for all stars (top row), but the heating of the low-$\alpha$ stars seems to be less effective. This implies that the AVR for all stars is largely shaped by the superposition of at least two different heating histories for the low- and high-$\alpha$ components.

These two histories can be read in Fig.~\ref{fig02::AVR_params} where we show the ratio between vertical and radial velocity dispersion components~(left), as well as the slope of the AVR at different Galactocentric radii~(right). The velocity dispersion ratio shows a clear increasing trend with age, suggesting a higher importance of vertical heating for older stars. The high-$\alpha$ sequence alone shows a monotonic decrease of the velocity dispersion ratio with increasing distance but no variations with age. The latter might be caused by the age uncertainty flattening particular trends~\citep{2014A&A...572A..92M}; however, since we observe the same flat distribution for the oldest and youngest stars, it suggests that these flat profiles, although might be obscured, are real. The high values of the velocity dispersion ratio~(\sr/\sz) for the high-$\alpha$ populations of $0.6-0.9$ are hard to explain by secular process~(spiral arms, bar, molecular clouds), which are able to explain the values of $\sr/\sz$ up to $\approx 0.4$~\citep{2017MNRAS.471.3057M}. The most likely explanation is that the high-$\alpha$ stars were heated up by a merger, where the most natural actor is GSE, accreted onto the MW about $8-11$ Gyr ago.

However, since the Splash/Plume component is rather faint~(see Fig.~\ref{fig02::feh_age_circularity_at_diff_radii} and Section.~\ref{sec2::results_kinematics_circ}), it seems that the GSE impact on the pre-existing MW disc was quite gentle. At the same time, it does not imply the GSE-progenitor was a low-mass galaxy. Such a limited impact is hard to understand if the existing disc was dynamically cold, but if it was already hot enough, then the effect of even a rather massive merger can be modest~\citep[see, e.g.][]{2023A&A...677A..91K}, as we observed in the high-$\alpha$ populations. Since there was very little time for the high-$\alpha$ stars to be heated secularly before the GSE infall, it is quite natural to assume that these populations were formed hot. Such hot populations at the early phases are seen in many simulations, essentially suggesting the upside-down formation of the early MW~\citep{2023MNRAS.523.6220Y,2024arXiv241021377G}. At the same time, we do not diminish the role of accreted stars in the heating of the in-situ populations, which might not happen due to direct tidal interaction. For instance, the kinematic misalignment of non-rotating accreted stars and in-situ disc populations can shape the velocity ellipsoids increasing $\sz/\sr$ up to $0.9-1$~\citep{2017A&A...597A.103K, 2021MNRAS.500.3870K} which we observe at the innermost radii for high-$\alpha$ populations.

\begin{figure*}
    \centering
    \includegraphics[width=1\hsize]{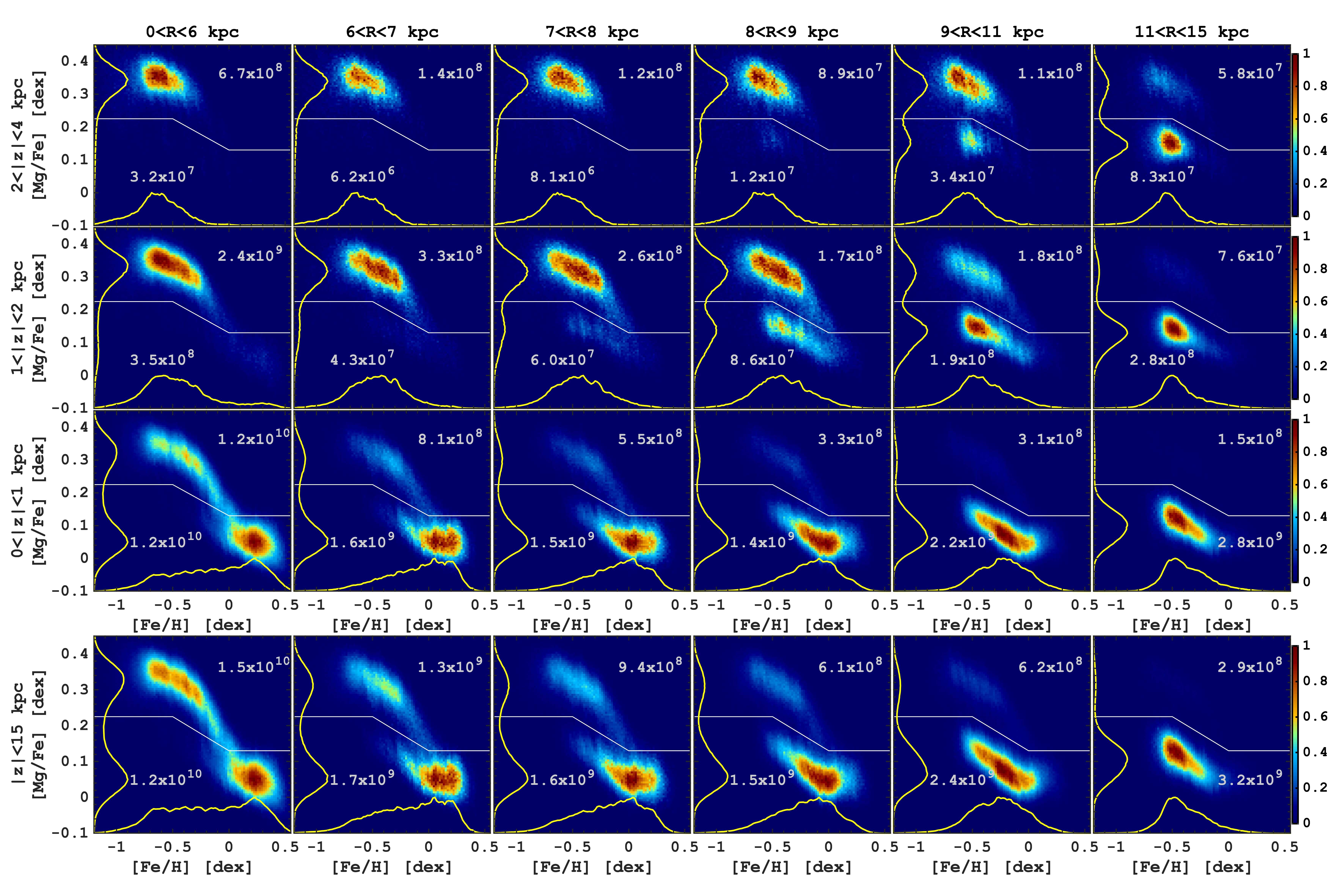}
    \caption{Distribution of stellar mass in the \MgFe\ vs. \FeH\ plane as a function of Galactocentric distance, R, and |z|. The bottom set of panels corresponds to the radial selection only, within 15~kpc from the mid-plane. In each panel, the yellow lines show the normalized \MgFe- and \FeH-distribution functions, and the white line shows a boundary used to separate high- and low-$\alpha$ populations. The masses of high- and low-$\alpha$ populations in a given radial and vertical bin marked with the grey numbers in \Msun\ units.}
    \label{fig02::bimodality_RZ}
\end{figure*}

\section{Results: MW disc chemical abundance composition}\label{sec2::results_abundances}
So far, we have explored the orbit superposition projection we obtained for the APOGEE giants in terms of structural and kinematic properties of stellar populations in the MW. In this section, we study in more detail the chemical abundance composition of the MW disc as a function of position inside the Galaxy and as a function of stellar age.

\subsection{Spatial variations of the disc chemical bimodality}
The  $\aFe$-bimodality is considered one of the most spectacular features of the MW disc. In Fig.~\ref{fig02::afe_feh_all}, we present the selection function corrected $\FeH-\MgFe$ plane. For the orbit superposition approach, it reveals the mass-weighted distribution for the entire Galaxy. A striking difference between the raw APOGEE data and the reconstructed density distribution is the dominance of metal-rich populations among the low-$\alpha$ stars, which weakens the low-metallicity ($<$-0.2 dex) tail of the low-$\alpha$ population. This correction reduces the apparent significance of the bimodal $\MgFe$ distribution, showing a dominant high-$\alpha$ sequence over the low-$\alpha$ sequence. However, it is important to note that the distribution in Fig.~\ref{fig02::afe_feh_all} is computed for the entire Galaxy, mixing populations from different Galactocentric radii.

To better understand the spatial variations of the stellar-mass weighted \FeH-\MgFe distribution, Fig.~\ref{fig02::bimodality_RZ} (top three rows) shows the classic variation of this relationship with Galactocentric radius and vertical height from the disc mid-plane. The bottom panels focus on the radial dependence for stars located within 15 kpc from the mid-plane. Each panel is divided by a white line to distinguish high- and low-$\alpha$ populations, with the stellar mass of these populations indicated for each spatial bin. Different radial bin sizes are used in the figure to highlight the key features of the $\FeH-\MgFe$ plane. Overall the figure shows the known trends~\citep{2014ApJ...796...38N, 2015ApJ...808..132H, 2018MNRAS.478.4513B, 2021MNRAS.506..150B, 2022ApJ...928...23E, 2024arXiv240919858B}.

The leftmost column displays data within $<6$~kpc, which roughly corresponds to the effective (or half-light) radius of the MW \citep{2024arXiv240605604L}. Notably, there is no evidence of disc $\alpha$-bimodality at a given metallicity inside this effective radius, a standard metric in extragalactic studies. This suggests that even if MW analogues possess an $\alpha$-bimodality, they will not demonstrate this feature within the effective radius. In fact, the bimodality is also not prominent in the radial range of $6-7$~kpc (second column). It starts to appear at $7-11$~kpc, but it is asymmetric and vastly dominated by the low-$\alpha$ sequence.

The radial variations of the $\FeH-\MgFe$ plane demonstrate an interesting behaviour of the low-$\alpha$ sequence. Inside the solar radius, it is essentially parallel to the \FeH-axis, but at larger radii, its lower metallicity tail shows an upward bend towards higher values of \MgFe. This may imply a different chemical evolution of the metal-rich and metal-poor parts~(relative $\approx -0.2$~dex) of the low-$\alpha$ sequence, spatially separated by $9-10$~kpc where the location of the bar outer Lindblad resonance~(OLR) should prevent mixing between inner and outer discs~\citep{2015A&A...578A..58H,2018A&A...616A..86H}.

\begin{figure*}
    \centering
    \includegraphics[width=1\hsize]{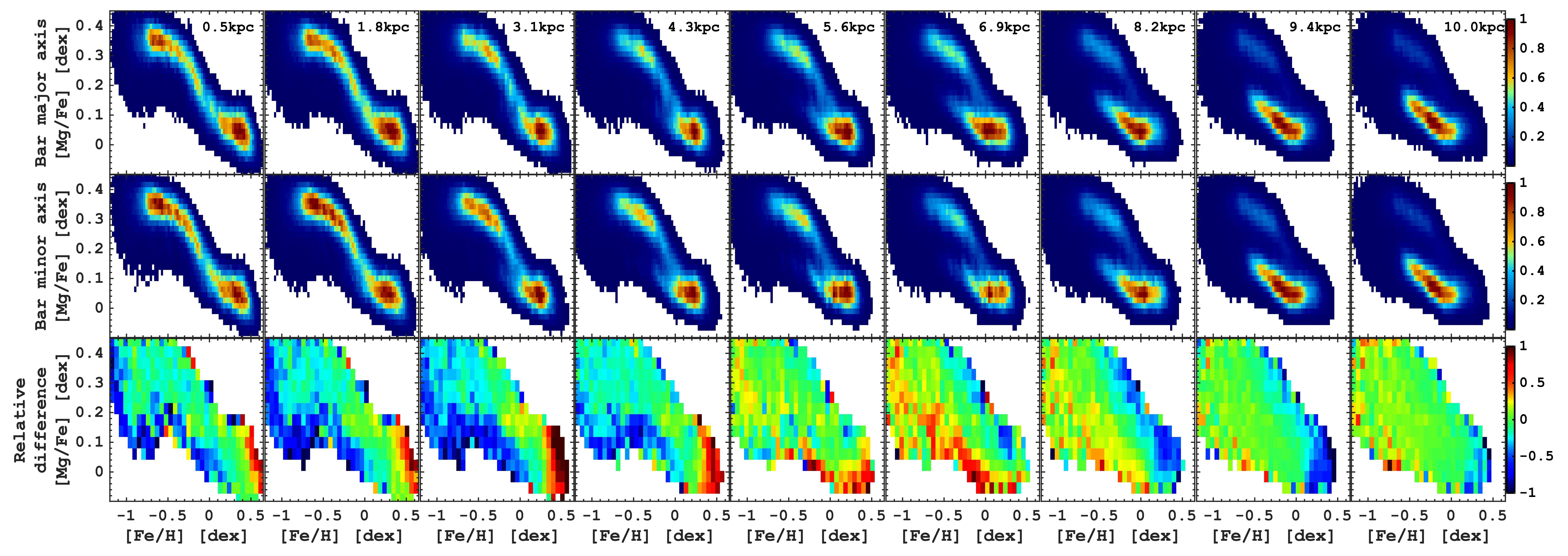}
    \caption{Distribution of stellar mass in the $\MgFe-\FeH$\ plane as a function of Galactocentric distance in $0.5$ kpc radially-thick and $15^\circ$-wide sectors along the bar major~(top) and minor~(middle) axes. The relative difference between two distributions at a given radius is shown in the bottom panels. The red~(blue) colour highlights the excess of stars along the major~(minor) axis.}
    \label{fig02::bimodality_azimuth}
\end{figure*}

The advantage of the orbit superposition approach is that it allows us to examine the \MgFe-\FeH plane as a function of azimuth. Since only the bar can potentially affect azimuthal abundance variations, Fig.~\ref{fig02::bimodality_azimuth} shows the density distributions along the bar major~(top) and minor axes~(middle). The relative difference between distributions along the major and minor axes is depicted in the bottom row, where red indicates an excess of stars along the major axis, and blue indicates an excess along the minor axis. Inside the bar (<4.3~kpc), we observe an excess of the most metal-rich stars along the major axis of the bar and a deficit of low-metallicity (mostly low-$\alpha$) populations. At intermediate radii (5.6-7~kpc), there is an excess of lower metallicity stars at a given \MgFe along the major axis. The following two radial bins (8.2-9.4~kpc) show an inverted pattern relative to the innermost regions. Beyond 10~kpc, the $\MgFe-\FeH$ distributions appear identical, which aligns with the expected scenario of bar-induced azimuthal variations. In this scenario, the azimuthal variations may arise due to differential mapping of stars with different kinematics. For instance, we notice that the high-$\alpha$ populations are weakly affected by the bar while the low-$\alpha$ stars are more sensitive to its presence. However, the low-$\alpha$ stars have different kinematics as a function of metallicity. In Fig.~\ref{fig02::AVR}, we showed that innermost stars, which are the most metal-rich, have lower velocity dispersion, and hence, they are more affected by the presence of the bar and more likely to be captured along its major axis. The opposite behaviour explains the deficit of metal-poor stars, which are hotter and less easily trapped along the bar major axis. Overall, this illustrates so-called kinematic fractionation phenomena~\citep{2017MNRAS.469.1587D}, shaping populations with different kinematics and resulting in the observed abundance variations. At the same time, it is not clear why we observe the inverted trend between corotation and the OLR of the bar.

\subsection{Mono-abundance populations: orbital parameters and spatial structure}\label{sec2::maps}

Viewing the Galactic stellar disc as composed of mono-abundance populations~(MAPs) has been advocated for providing an opportunity for a better understanding of the disc structure and evolution by allowing a more detailed analysis of the dynamics and chemical composition of distinct stellar populations. The utility of MAPs arises from the fact that in the presence of significant radial migration, chemical abundances are the only life-long characteristics that stars preserve~\citep{1979ApJ...229.1046T, 2002ARA&A..40..487F}. This makes them invaluable for isolating stellar populations without presuming a particular dynamical history and recovering the build-up of the entire Galactic disc.

\begin{figure*}
    \centering
    \includegraphics[width=1\hsize]{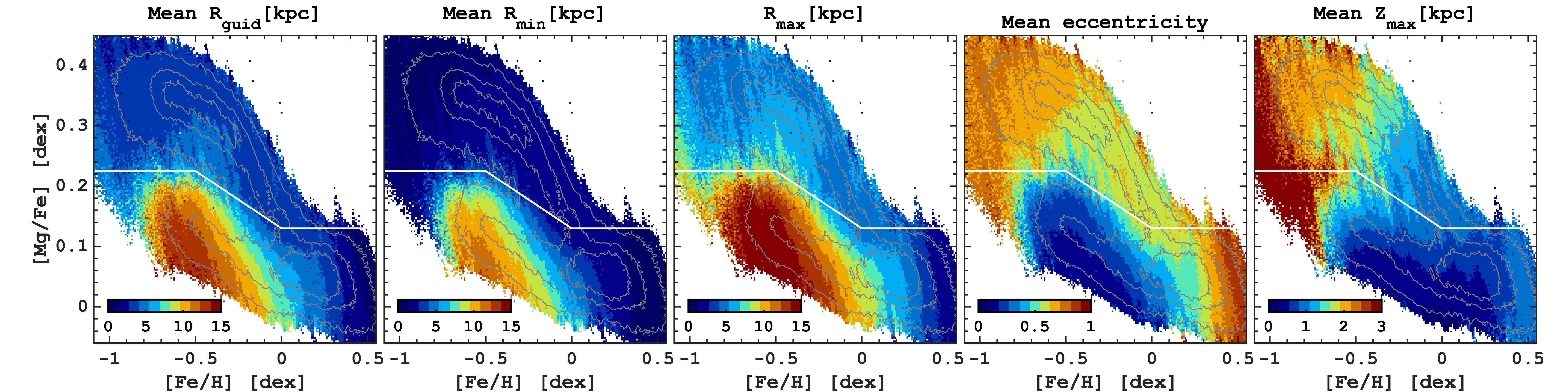}
    \caption{Mean orbital parameters values in the $\MgFe-\FeH$ plane for the entire MW disc. From left to right: guiding radius~($R_{guid}$), pericentre~($R_{min}$), apocentre~($R_{max}$), eccentricity and maximum vertical excursion~($Z_{max}$). The grey lines show the isodensity contours. The white lines correspond to the boundary used to define high- and low-$\alpha$ populations.}
    \label{fig02::MAPS_orb_params}
\end{figure*}

In Fig.~\ref{fig02::MAPS_orb_params} we show the $\MgFe-\FeH$ plane colour-codded by several orbital characteristics of enclosed stellar populations, from left to right: mean guiding radius, mean pericenter~($R_{min}$), mean apocentre~($R_{max}$), mean eccentricity~($(R_{max}-R_{min})/(R_{max}+R_{min})$) and mean vertical excursion~($Z_{max})$. We stress that these quantities were calculated for each orbit, and since they were obtained in a non-axisymmetric potential, they take into account the impact of the bar, often ignored in the orbital analysis of the MW stellar populations. 

The map of the mean guiding radius~(leftmost in Fig.~\ref{fig02::MAPS_orb_params}) sharpens the picture we observed in Fig.~\ref{fig02::bimodality_RZ}, where the high-$\alpha$ and metal-rich low-$\alpha$ stars are tightly confined in a limited range of Galactocentric radii, $<5$~kpc, thus representing the inner disc. The inner disc MAPs can be identified in the blue-colour region in the first two panels in Fig.~\ref{fig02::MAPS_orb_params}. Unsurprisingly, the MW bulge and the bar are predominantly made of these populations~\citep[see, e.g.][and references therein]{2016PASA...33...27D}. Although these stars are located in the inner disc, the low- and high-metallicity sub-populations show somewhat different behaviour manifested by a non-monotonic change of the mean eccentricity and $Z_{max}$ along metallicity. 

The subsolar metallicity high-$\alpha$ populations show a high radial extension with the mean $R_{max}\approx 5-6$~kpc and relatively high mean eccentricity $\approx 0.6-0.8$ and the mean $Z_{max}\approx 1-2$~kpc. The supersolar metallicity stars tend to be more compact with increasing metallicity, where the extremely-metal rich~(EMR) ones are trapped just inside $1-2$~kpc~\citep{2024arXiv240601706R} while being on extremely-radial orbits~(mean eccentricity is $>0.8$). The vertical excursion gradually decreases as a function of metallicity; only the EMR stars have slightly higher $Z_{max}$.

The remaining low-metallicity low-$\alpha$ populations, highlighted by the red colour in the two leftmost columns, constitute the outer disc since they have the mean guiding radius of $6-12$~kpc. Even more, by looking at the distributions of $R_{min}$ and $R_{max}$, we can conclude that the pericentres of these stars are located beyond $5-6$~kpc and their orbits never pass through the inner disc. This suggests that the low-metallicity low-$\alpha$ MAPs have a ring-like density distribution~\citep{2012ApJ...753..148B,2016ApJ...823...30B}, which is very thin in the vertical direction as seen from the $Z_{max}$ map in the rightmost panel. Nevertheless, these stars still show a negative radial metallicity gradient, which is seen as a decrease in the mean guiding radius as a function of metallicity.

Our results clearly demonstrate the dichotomy between high- and low-$\alpha$ populations at subsolar metallicities, which can be smeared out if the low-$\alpha$ populations are considered across the entire range of metallicities. This reinforces the idea about the presence of distinct inner and outer disc components, where the inner disc has a boundary at around $5-6$~kpc and is made of high- and low-$\alpha$ at subsolar and supersolar metallicities, respectively. The low-$\alpha$ populations with metallicities lower than solar represent the outer disc. 

\begin{figure*}
    \centering
    \includegraphics[width=0.499\hsize]{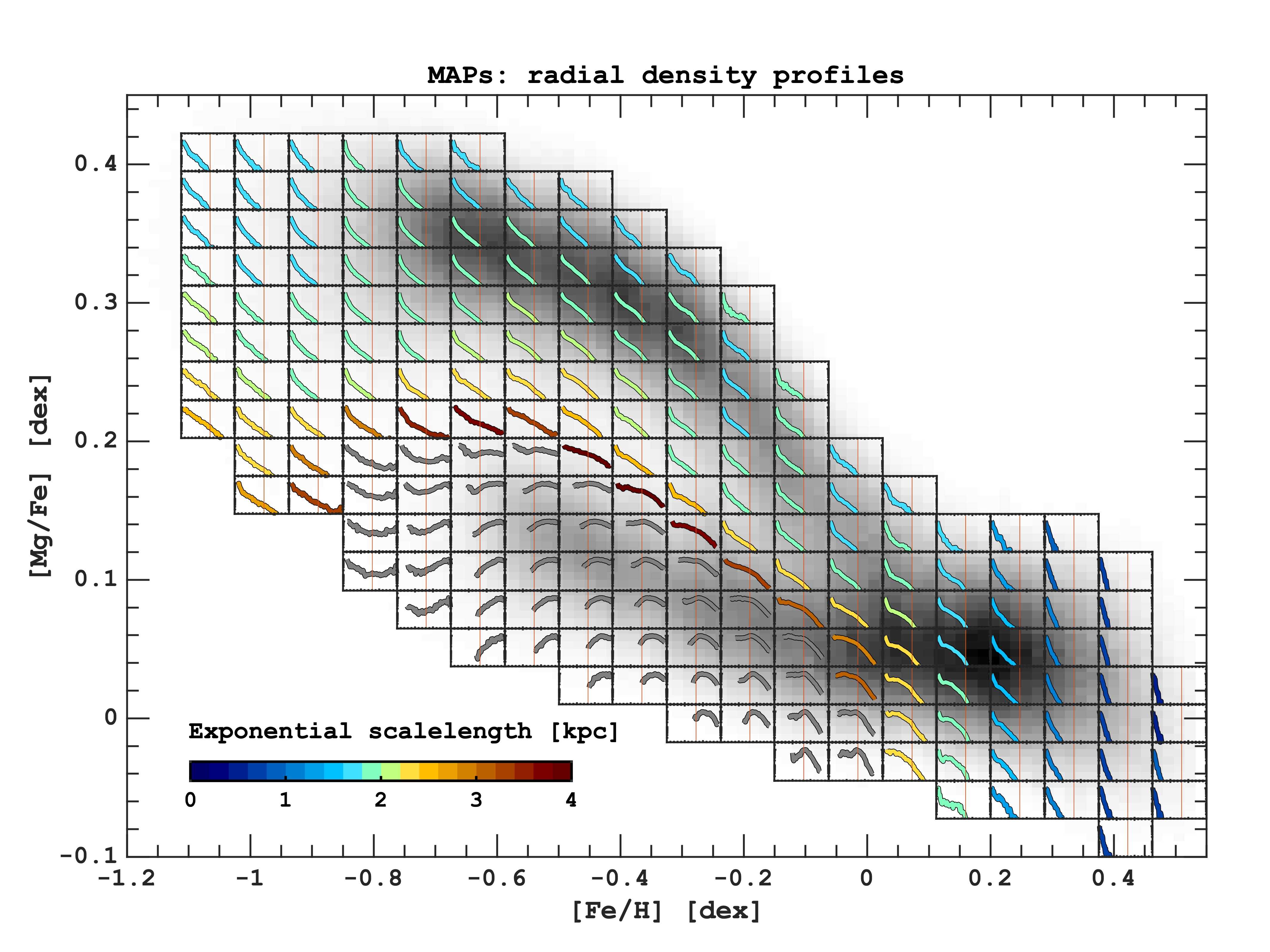}\includegraphics[width=0.499\hsize]{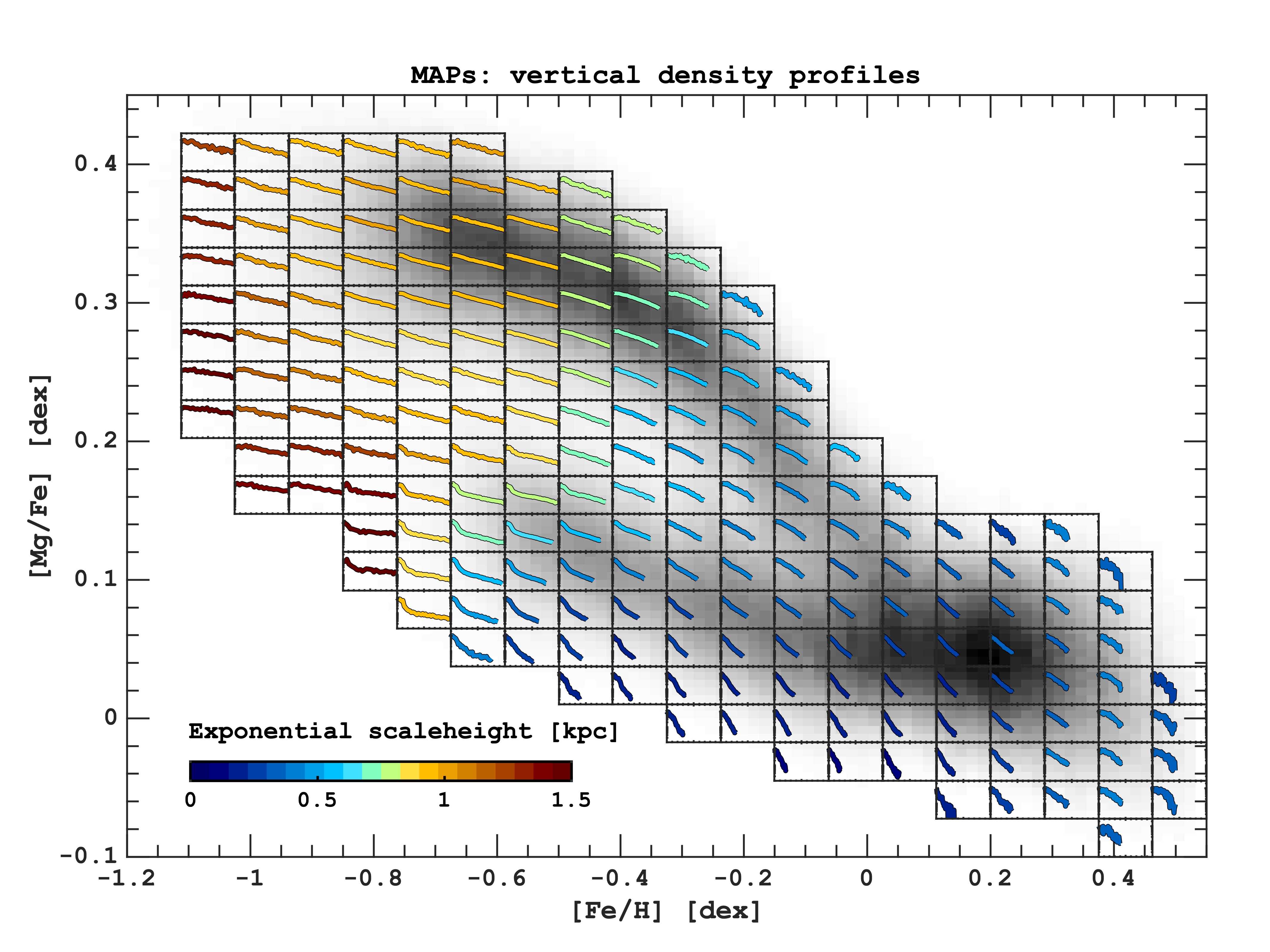}
    \caption{Radial and vertical structure of MAPs. The background map shows the stellar mass in the \MgFe\ vs. \FeH\ plane. Each subpanel shows the radial~(left) and vertical~(right) density profiles of MAPs selected in the abundance range from the background map. The colour of lines in the subpanels corresponds to the exponential scalelengths~(left) and scaleheights~(right) according to the colour bar. In the left subpanels, the low metallicity part of the low-$\alpha$ sequence does not show exponential profiles but rather a donut-like distribution, where the exponential fit fails to recover meaningful values of the scalelength. On the left, the location of the solar radius is marked by the red vertical lines.}
    \label{fig02::MAPs}
\end{figure*}

The connection between the inner and outer discs and thin and thick disc components requires a more detailed analysis of the spatial distributions of the MAPs. In Fig.~\ref{fig02::MAPs} we present the variations of the radial~(left) and vertical~(right) density profiles as a function of \FeH and \MgFe. The background maps show the stellar mass distribution in the \FeH-\MgFe plane while the subpanels show the density profiles of corresponding MAPs colour-codded by the exponential scalelength~(left) and scaleheight~(right). 

The trends observed in Fig.~\ref{fig02::MAPs} are strongly two-dimensional and can be described as follows. The radial scalelength for high-$\alpha$ populations is roughly constant, about $1.8-2.1$~kpc~\citep[see also]{2016ApJ...823...30B}. The low-$\alpha$ radial density profiles are more complex. The most metal-rich, or EMR, stars have a scalelength of less than $1$~kpc, which increases up to $3$~kpc at the solar metallicity. However, at subsolar metallicities, the radial density profiles can not be approximated with exponential profile, as they are more like a broken distribution with the peak around the solar radius at $\FeH\approx -0.1$ which shifts out to $\approx 15$~kpc at $\FeH\approx-0.7$. \cite{2017MNRAS.471.3057M} also showed that the low-$\alpha$ populations have a broken exponential profiles~\citep{2012ApJ...755..115B} which, however, we observe only at subsolar metallicities where the MAPs can be described as rings but not donuts, as they are very thin. Another difference between \cite{2017MNRAS.471.3057M} and our results is that we observe a more prominent shift of the MAPs density peak from $\approx 8$~kpc at $\FeH\approx-0.1$ up to $15$~kpc at $\FeH\approx-0.7$~dex, while \cite{2017MNRAS.471.3057M} found the density peak at $\approx 10$ for the lowest metallicity low-$\alpha$ MAPs~\citep[][]{2022MNRAS.513.4130L,2024A&A...683A.128C}. 

The decrease of the vertical scaleheight for the high-$\alpha$ stars can be interpreted as a steady up-side down settling of the MW disc with the increase of the metallicity. However, we need to stress that the presence of accreted stars at metallicities below $-0.5$~dex may affect this picture, as it can be suggested by the eccentricity and $Z_{max}$ panels in Fig.~\ref{fig02::MAPS_orb_params}. Depending on the definition of low- and high-$\alpha$ stars, the accreted populations most likely would contaminate both populations, which, for instance, will result in increasing the flaring of the outer low-$\alpha$ disc.

\begin{figure}
    \centering
\includegraphics[width=1\hsize]{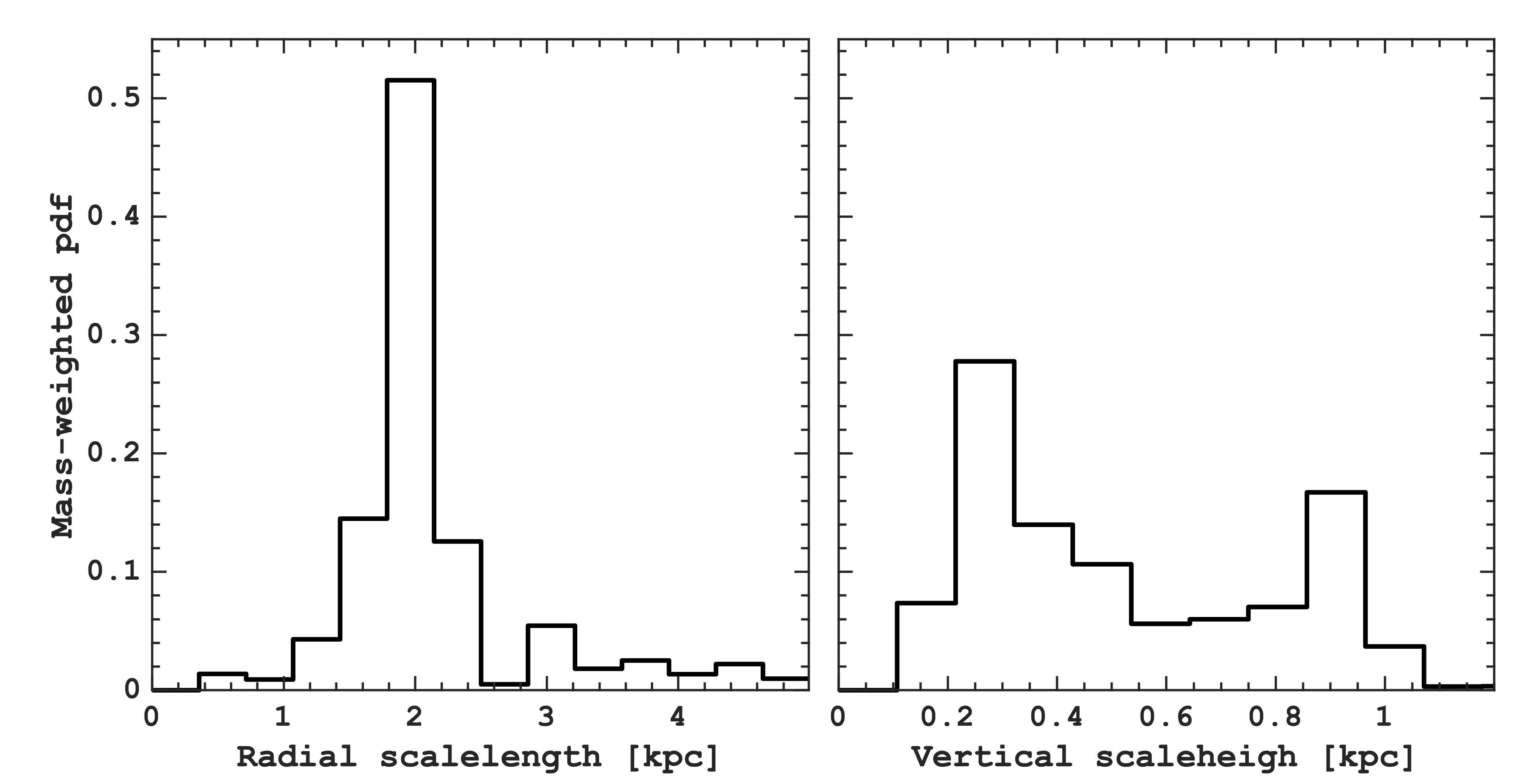}
    \caption{Mass-weighted radial~(left) and vertical~(right) scales of MAPs.}
    \label{fig02::MAPs_vscale_hist}
\end{figure}

The right panel of Fig.~\ref{fig02::MAPs} suggests that the scale height is monotonically decreasing with decreasing \aFe and increasing \FeH, implying no thin/thick disc dichotomy~\citep{2012ApJ...753..148B, 2017MNRAS.471.3057M}. However, we have already demonstrated the presence of distinct thin and thick disc components made of low- and high-$\alpha$ populations with different fractional contributions. Therefore, a smooth transition of the vertical scaleheight, i.e. discontinuity, for MAPs does not mean the lack of two distinct disc components, as different MAPs are mapped differently into geometric thin and thick discs. This is relatively easy to understand if we look at the background map in Fig.~\ref{fig02::MAPs}. The background density suggests that some MAPs have relatively small mass contributions, and once we weigh the scale height by the mass of MAPs, we can find the dominant spatial parameters of MAPs. In Fig.~\ref{fig02::MAPs_vscale_hist}, we show the stellar-mass weighted radial scale length and vertical scaleheight distributions. For the scalelength distribution, we do not account for MAPs with the broken exponential profiles, whose profiles are in grey in Fig.~\ref{fig02::MAPs}. Interestingly, we find the dominant scalelength of $2$~kpc, in agreement with earlier studies of MAPs~\cite {2012ApJ...753..148B, 2012ApJ...751..131B}. However, the vertical scaleheight distribution shows a prominent bimodal distribution with the peaks at $\approx 250-300$ and $\approx 900$~pc corresponding to the measurements of thin and thick MW discs~\citep{2016ARA&A..54..529B}.

Along the radial shift of the density peak, we observe a broadening of the low-$\alpha$ MAPs density profile towards lower metallicities. This can be attributed to the radial migration of stars, which is more evident for lower metallicity~(older) stars. However, the increase of the radial star-forming region seen in simulated galaxies is another mechanism~\citep{2024A&A...690A.352R}. Most likely, both processes shape the radial structure of the MAPs. We, therefore, confirm previous studies suggesting that the MW disc is composed of multiple subpopulations~(MAPs or mono-age) that smoothly span a range of properties \citep[e.g.,][]{1987ApJ...314L..39N, 1991PASP..103...95N, 2012ApJ...753..148B, 2012ApJ...751..131B, 2016ApJ...823...30B}; however, in terms of structural parameters, the disc is still characterized as the superposition of two distinct structures with different scale heights.

\begin{figure*}
    \centering
    \includegraphics[width=1\hsize]{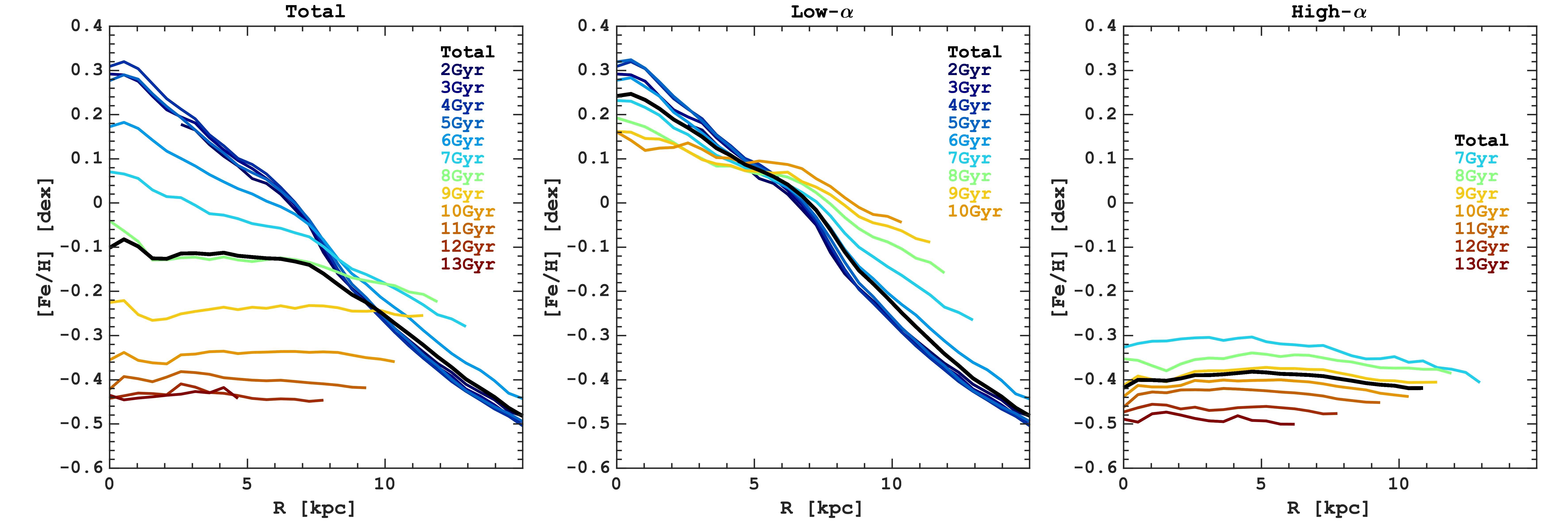}
    \caption{Mass-weighted mono-age profiles of the mean stellar metallicity for all MW stars~(left) and low-/high-$\alpha$ populations separately in middle/right panels. The lines of different colours correspond to different ages; the black lines show the averaged metallicity profile as a function of the Galactocentric distance of corresponding populations. }
    \label{fig02::feh_radial_monoage}
\end{figure*}

\begin{figure}
    \centering
    \includegraphics[width=1\hsize]{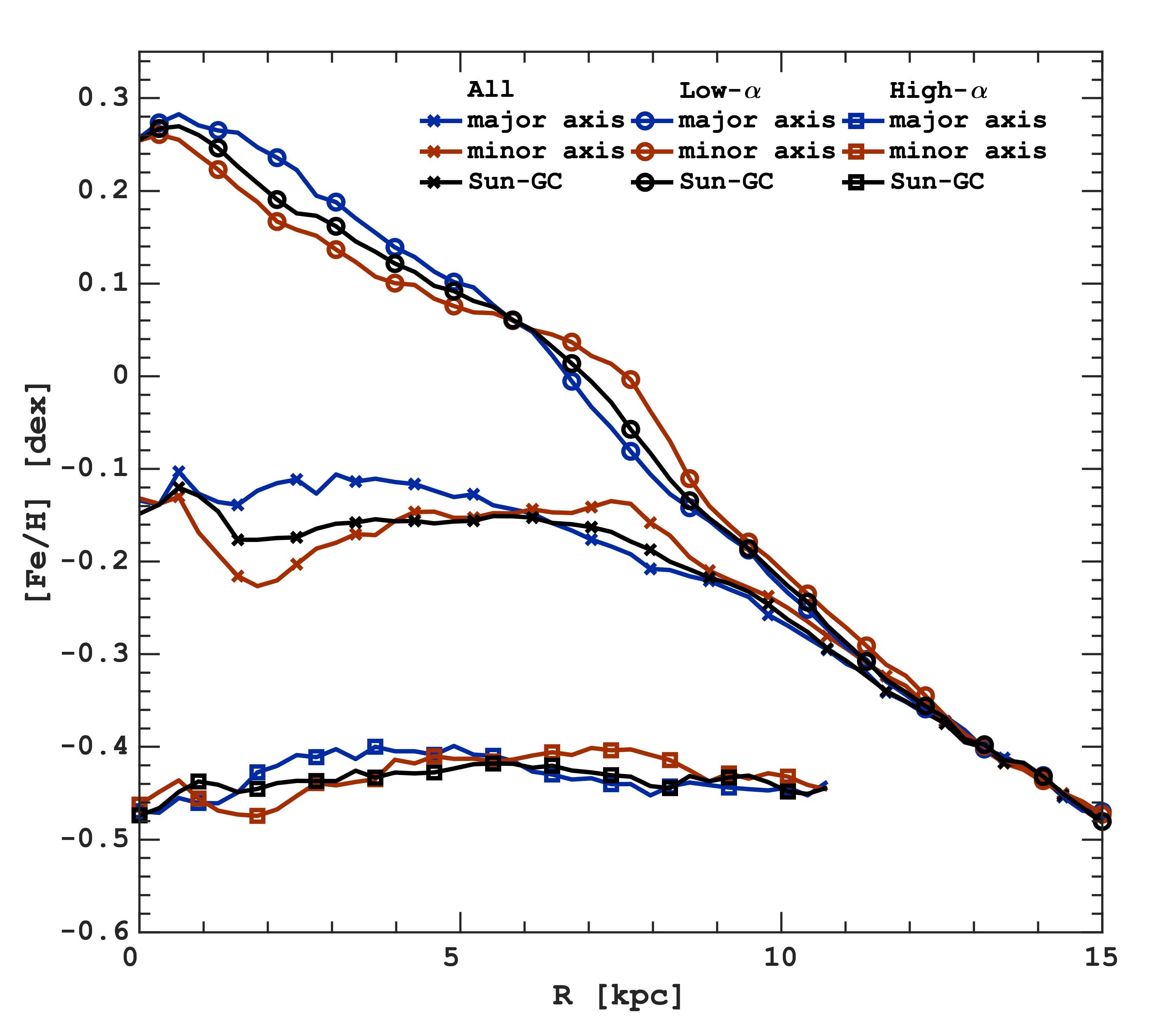}
    \caption{Stellar mass-weighted radial metallicity profiles in the MW disk. Different lines show the profiles along the line connecting the Sun with the Galactic centre and along the major and minor axes of the bar. From top to bottom, three groups of lines correspond to the low-$\alpha$, all and high-$\alpha$ stellar populations.}
    \label{fig02::feh_azimuthal_radial_gradient}
\end{figure}

\begin{figure*}
    \centering
    \includegraphics[width=1\hsize]{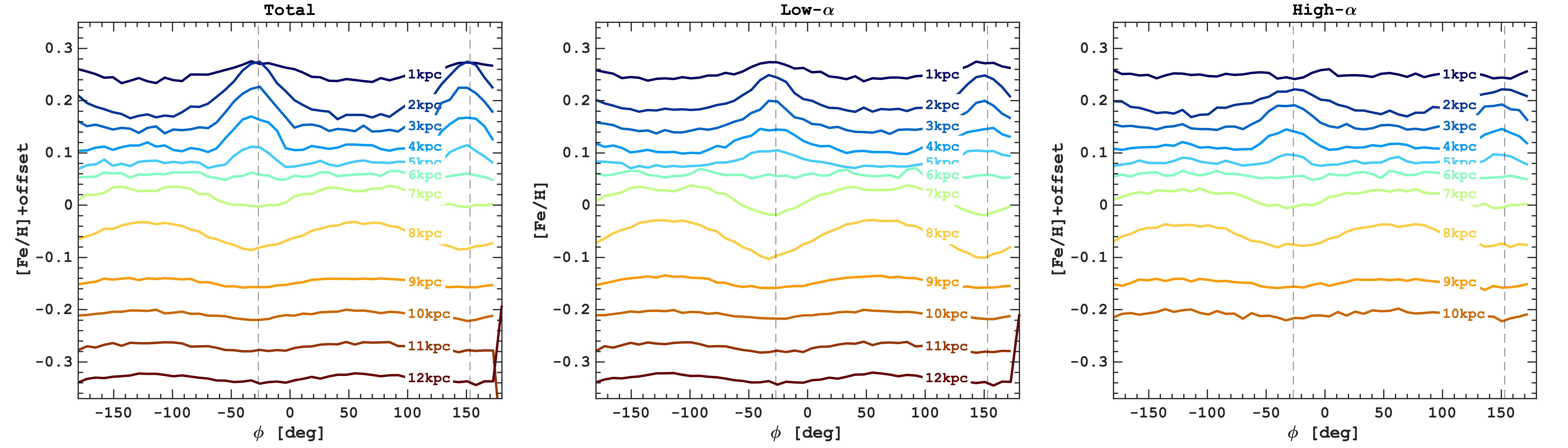}
    \caption{Mass-weighted azimuthal profiles of the mean stellar metallicity at different galactocentric radii for all stars~(left), low- and high-$\alpha$ populations. The profiles for the low-$\alpha$ stars are shown with the absolute values, while the profiles for all and high-$\alpha$ stars are shifted vertically for better visibility.}
    \label{fig02::feh_azimuthal_gradient}
\end{figure*}

\section{Results: metallicity profiles}\label{sec2::results_met_gradients}
\subsection{Radial metallicity gradients}

The orbit superposition reconstruction allows the distribution of abundances in the MW’s disc to be mapped in unprecedented detail. Abundance gradients have been observed in most disc galaxies and show that the abundances of metals decrease outwards. It has been well established the mean stellar metallicity decreases from the centre of the galaxy outwards~\citep{1992A&A...262..455G, 1989MNRAS.239..885M, 1995ApJ...444..721S, 1997ApJ...478..190A, 2010ApJ...721L..48K, 2014A&A...570A...6S, 2014A&A...563A..49S} which is a signature of the star formation profile evolution associated with an inside-out growth of galactic discs~\citep{1976MNRAS.176...31L, 1989MNRAS.239..885M, 2011ApJ...729...16K}. We note that the inside-out growth term is somewhat misleading as it may involve two interpretations. In one case, we can assume that the scalelength of subsequently formed stellar populations increases with time, resulting in more radially extended mono-age density distributions. At the same time, one can imagine the propagation of star formation from inner to outer regions of the galactic disc and concentrated in a small range of galactocentric radii. In both cases, the negative radial metallicity gradient is expected to be present, and it should provide valuable insights into the assembly history of the MW stellar disc.

In Fig.~\ref{fig02::feh_radial_monoage} the black lines present the radial metallicity profiles for all stars~(left panel) and low-/high-$\alpha$ populations~(middle and right panels) separately. The lines of different colours show the mean metallicity profiles for mono-age $1$~Gyr width sub-populations. In the analysis, we exclude the young high-$\alpha$ stars, which were rejuvenated by interactions or mass accretion~\citep{2015MNRAS.451.2230M}, as discussed above. The averaged metallicity profile~(black in the left panel) shows the two main behaviours: very flat inside the solar circle and a negative gradient of $\rm \approx -0.057\ dex\  kpc^{-1}$. The latter is in a good agreement with recent studies relying on different tracers:  $\rm -0.060\ dex\ kpc^{-1}$\citep{2014A&A...566A..37G}, $\rm -0.052\ dex\ kpc^{-1}$~\citep{2022A&A...659A.167R}, $\rm -0.064\ dex\ kpc^{-1}$~\citep{2022Univ....8...87S}, $\rm -0.056\ dex\ kpc^{-1}$\citep{2023A&A...674A..38G}, $\rm -0.053\ dex\ kpc^{-1}$~\citep{2024A&A...690A.147H}. The total inner flat metallicity gradient is the result of the superposition of flat profiles of stars older than $\approx 8$ Gyr and weakly negative gradients for younger stars. This suggests a rather sharp transition from the metallicity gradient behaviour around $8-9$~Gyr ago, or, since the flattening of the metallicity gradient is expected to be the result of radial migration~\citep{2012A&A...540A..56P, 2013MNRAS.436.1479K, 2012A&A...548A.126M, 2014A&A...572A..92M}, a rapid change in the efficiency of the migration. It is indeed quite remarkable how the flat metallicity profile at $8$~Gyr transformed to a negative gradient of $\rm \approx -0.025\ dex\ kpc^{-1}$ for $7$~Gyr old populations.

The low-$\alpha$ populations show the negative metallicity gradient for all ages younger $10$~Gyr, where the gradients flatten for older stars, as expected from the radial migration scenario. The high-$\alpha$ stars show no metallicity gradient. Although we remain cautious regarding the impact of age-uncertainly on the trends, we recover the fact that the mean metallicity increases with the mean stellar age, which suggests that the picture we observe is quite robust. We note that we obtain nearly identical flat metallicity gradients for the high-$\alpha$ populations if we adopt the recommended metallicity cut-off $-0.7$~dex for the age catalogue we use~\citep{2023ApJ...954..124I,2024AJ....167...73S}. This adjustment affects about $8\%$ of the total MW stellar disc mass, which will be missing in our age-related analysis of the orbit superposition model.

The radial metallicity profiles shown in Fig.~\ref{fig02::feh_radial_monoage} illustrate how a superposition of two chemically-distinct subpopulations~(low- and high-$\alpha$) results in mixed behaviour of the total gradient. Since the inner Galaxy seems to be dominated by the high-$\alpha$ stars, we observe the flat gradient, while the outer disc is made of only low-$\alpha$ populations with a negative gradient. The origin of such metallicity gradient dichotomy can be explained, for instance, if the high-$\alpha$ stars formed in an intense and rather short episode of star formation where the efficient mixing of the ISM prevents the development of the metallicity gradient. At the same time, stars formed on relatively hot orbits, and even without involving radial migration, release metals far away from their birth sites, additionally acting against the radial gradient development. Galaxy formation simulations do not provide a unique answer to the question of whether the (negative) early metallicity gradients exist or not~\citep{2022MNRAS.511.1667T,2019MNRAS.482.2208T,2016A&A...592A..93T}, highlighting the importance of mergers~\citep{2017MNRAS.471.3856T}, active galactic nuclei~\citep{2008ApJ...672L.107E,2017MNRAS.471.3856T} and subgrid physics prescription~\citep{2012A&A...540A..56P,2013A&A...554A..47G,2016A&A...587A..10M}.

In Fig.~\ref{fig02::disk_density_high_low}, we show that the density distribution of low- and high-$\alpha$ populations, although both trace the bar, show some diversity in their 3D distribution. Therefore, since these populations have different chemical compositions, one can expect to see the impact of the bar on the radial metallicity profiles at different angles relative to the bar. In Fig.~\ref{fig02::feh_azimuthal_radial_gradient} we present the radial metallicity profiles measured in $15$ deg sectors along major~(blue) and minor~(red) bar axes, as well along the line connecting the Sun with the Galactic centre~(black). As expected, although closely followed by each other, these profiles show a certain level of divergence. For the low-$\alpha$ populations, along the bar major axis, the mean metallicity is higher, and the radial gradient is steeper inside $\approx 6$~kpc compared to the bar minor axis direction. At larger radii, we observe the opposite trend up to the distance of $9$~kpc, and there is no difference between metallicity profiles further out. The high-$\alpha$ sequence shows negligible variations of the metallicity profile; however, we notice weak systematic variations in the mean metallicity across $6$~kpc. 

The superposition of low- and high-$\alpha$ populations give rise to rather complex metallicity profile variations for both populations considered together. Inside the solar circle, the metallicity gradient along the bar minor axis is slightly positive while it becomes negative along the bar major axis and progressively steepens with radius. In all the cases described above, the metallicity profiles along the Sun-Galactic centre line show the mean trends we observe for the directions of the major/minor bar axes. 

The reason behind the azimuthal variations of the radial metallicity gradient is the bar's influence on the distribution of stars with different kinematics~\citep{2013A&A...553A.102D,2023MNRAS.524..276F}. Stars on colder orbits are trapped by the bar more efficiently, and since they have relatively higher metallicity, their relative fraction is higher along the bar. Conversely, perpendicular to the bar, the fractional contribution of stars on hotter orbits increases, resulting in a lower mean metallicity along the bar's minor axis. This complex behaviour can explain the diversity of measurements of the radial metallicity gradient in the MW disc found in the literature.

\subsection{Azimuthal metallicity variations}
As demonstrated above, we observe not only changes in the radial metallicity gradient with azimuth but also variations in the mean metallicity. To emphasize this effect, Fig.~\ref{fig02::feh_azimuthal_gradient} shows the mean metallicity for all stars and high-/low-$\alpha$ populations as a function of azimuth. Different colours correspond to different Galactocentric radii. The orientation of the bar's major axis is highlighted by the black dashed lines.

We clearly observe periodic variations in the mean metallicity in all three panels within $6$~kpc. The metallicity profile peaks sharply at the angle corresponding to the orientation of the bar, effectively illustrating the bar itself. At $6$~kpc, the azimuthal metallicity variations disappear entirely due to the flat metallicity distribution. However, we observe unexpected metallicity azimuthal profiles at larger radii (greater than $7$~kpc). These variations are shifted by 90 degrees and exhibit a much less steep profile. Such a picture is observed in all three panels; however, for the high-$\alpha$ populations, which, as we showed above, have higher velocity dispersion, the strength of the metallicity variations is less pronounced. However, the fact that we still observe variations for old and hot populations suggests that the primary mechanism is the kinematic fractionation but not the local enrichment along the bar. Indeed, even if stars along the bar formed more metal-rich, they are phase-mixed on a very short time scale in the inner Galaxy, vanishing the azimuthal metallicity variations, as the rotational frequency of the bar is much higher compared to the stellar rotation. The fact that we do not observe any metallicity variations around 6~kpc strengthens this conclusion, as this radius corresponds to the bar corotation~\citep{2021MNRAS.508..728K, 2020A&A...634L...8K, 2022MNRAS.512.2171C}, where the effect of the local enrichment should be the strongest~\citep{2019A&A...628A..38S, 2023A&A...671A..56K}.   
By construction, our orbit superposition method is unable to capture the spiral structure directly, which is known to cause small-scale azimuthal variations~\citep{2018A&A...611L...2K, 2022A&A...666L...4P, 2023A&A...680A..85S}. Nevertheless, the reconstructed azimuthal metallicity variations can be used as a background to be subtracted from the data, making it possible to exclude the large-scale effects of the bar and highlight the spiral arms-related abundance patterns.

\subsection{Vertical metallicity gradients}
While the formation and evolution of radial abundance gradients in the MW, disc galaxies in general, and their simulated analogues are not completely understood but well-explored, the vertical metallicity gradients attract less attention. From a theoretical perspective, the vertical metallicity gradient can result from the upside-down formation process~\citep{2013ApJ...773...43B}. During the early collapse, the thickness of the star-forming region decreases over time, causing older, metal-poor populations to span larger distances from the midplane, while relatively younger, more metal-rich stars form closer to the midplane, or in a thinner layer. Consequently, a negative vertical metallicity gradient and a positive age gradient are expected in MW and other disc galaxies.

\begin{figure}
    \centering
    \includegraphics[width=1\hsize]{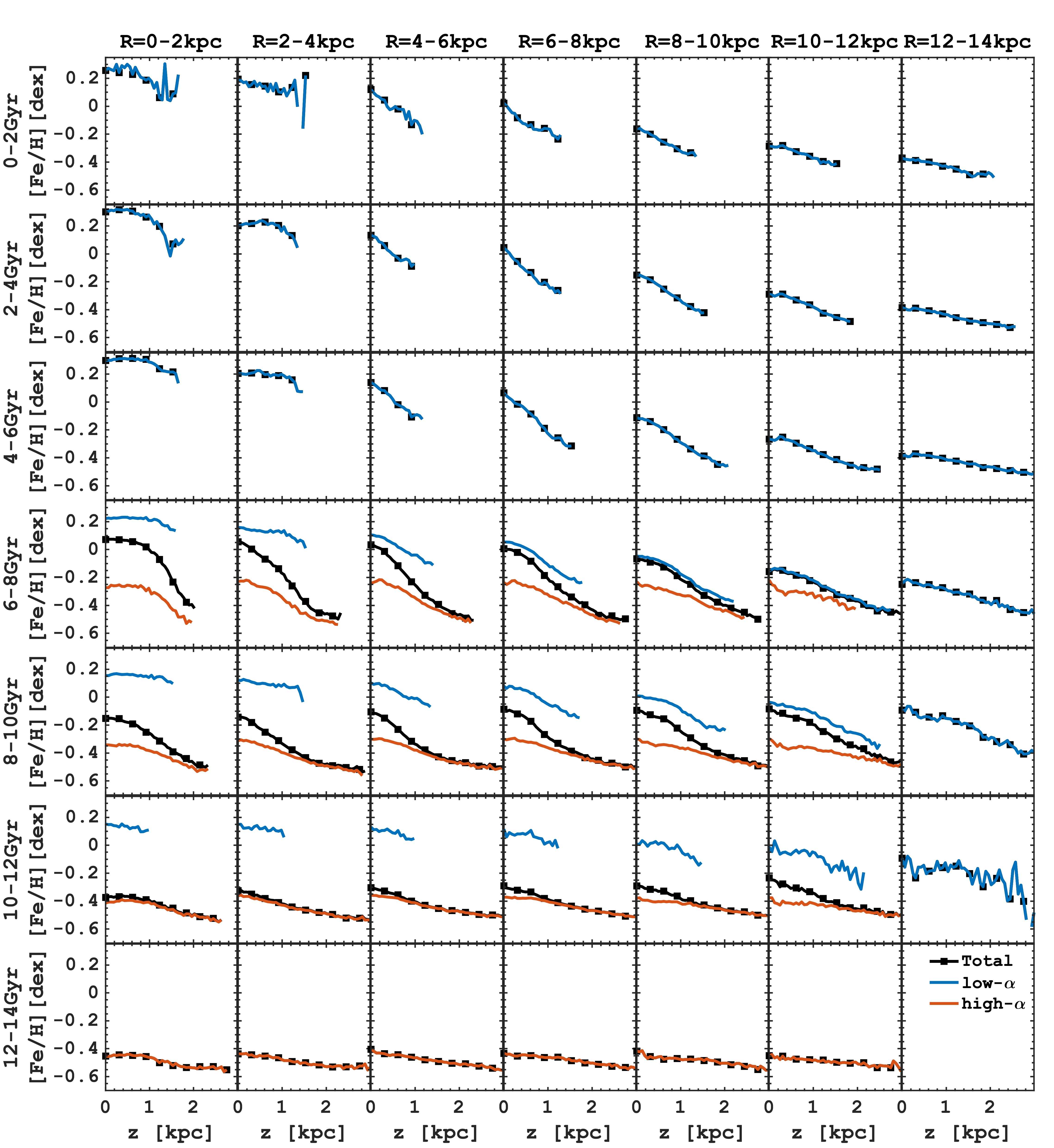}
    \caption{Vertical metallicity profiles as a function of galactocentric distance~(increasing from left to right) and stellar age~(increasing from top to bottom). In each panel, the metallicity profiles for low-, high-$\alpha$ and shown with blue and red colours, respectively. The total mass-weighted profiles are shown with the black lines.}
    \label{fig02::feh_vertical_gradient}
\end{figure}

\begin{figure}
    \centering
    \includegraphics[width=1\hsize]{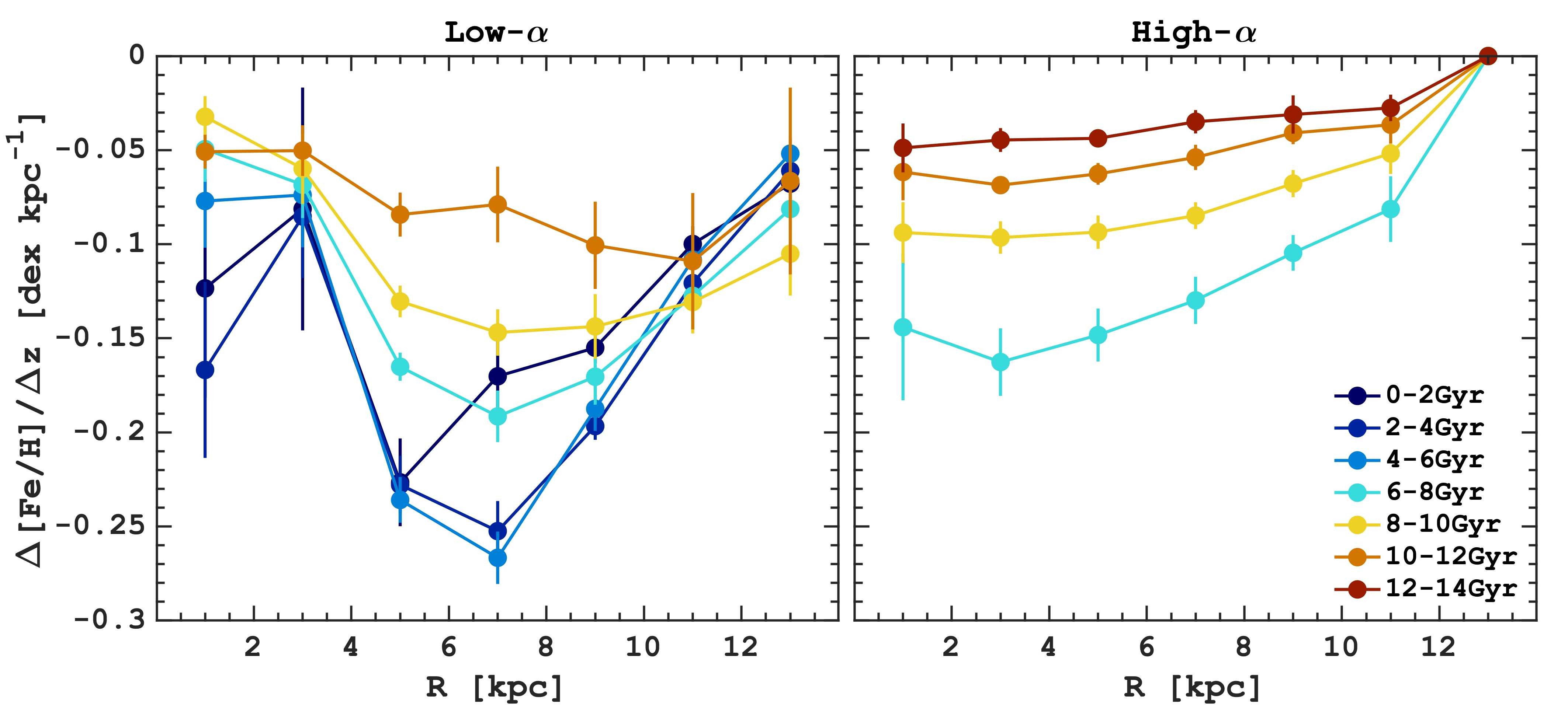}
    \caption{Vertical metallicity gradients for stars of different ages}
    \label{fig02::feh_vertical_gradient_values}
\end{figure}

\begin{figure*}
    \centering
    \includegraphics[width=1\hsize]{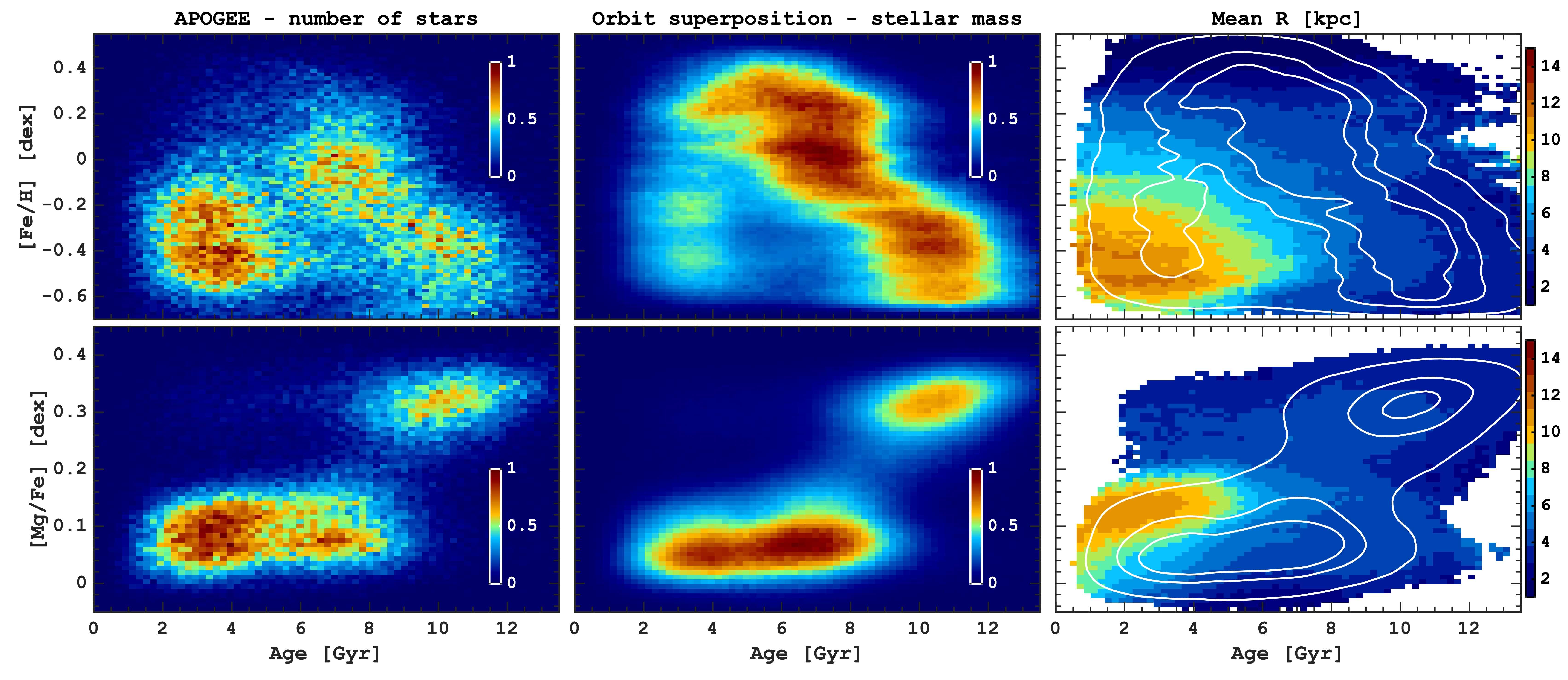}
    \caption{Age-\FeH~(top) and age-\MgFe~(bottom) relations. The left column shows the raw APOGEE number of stars while the middle column shows the stellar mass distribution from orbit superposition. The rightmost column shows the orbit superposition results colour-codded by the mean value of the guiding radius, where the white contours mark the density distribution in the corresponding coordinates.}
    \label{fig02::AMR_and_AMG_FE_weighted}
\end{figure*}

\begin{figure*}
    \centering
    \includegraphics[width=1\hsize]{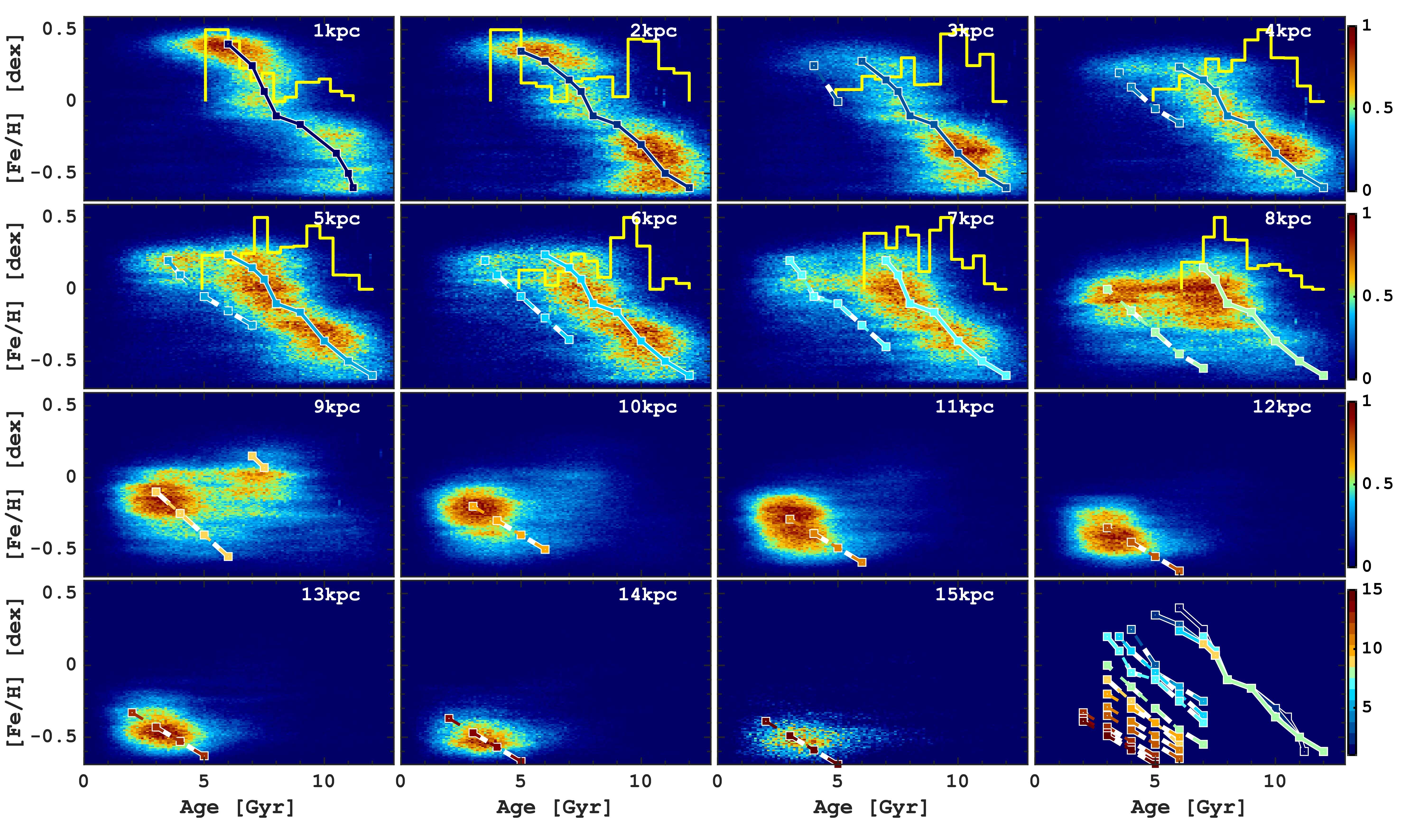}
    \caption{Stellar mass distribution in the age - \FeH plane. Each panel shows the normalized distribution for stars inside $1$~kpc-width guiding radii rings, as marked in the top right corner. The colour lines in each panel trace the time-dependence of abundance variation with stellar age obtained by optimization of age-\FeH~(this figure), age-\MgFe~(Fig.~\ref{fig02::AMG_radii}) and $\FeH-\MgFe$ distributions~(Fig.~\ref{fig02::AMRG_radii}). The rightmost bottom panel shows all the parametric curves for different guiding radii according to the colour bar. The yellow histograms show the normalized mass distribution along the age-\FeH curves for the old AMR sequence.}
    \label{fig02::AMR_radii}
\end{figure*}

\begin{figure*}
    \centering
    \includegraphics[width=1\hsize]{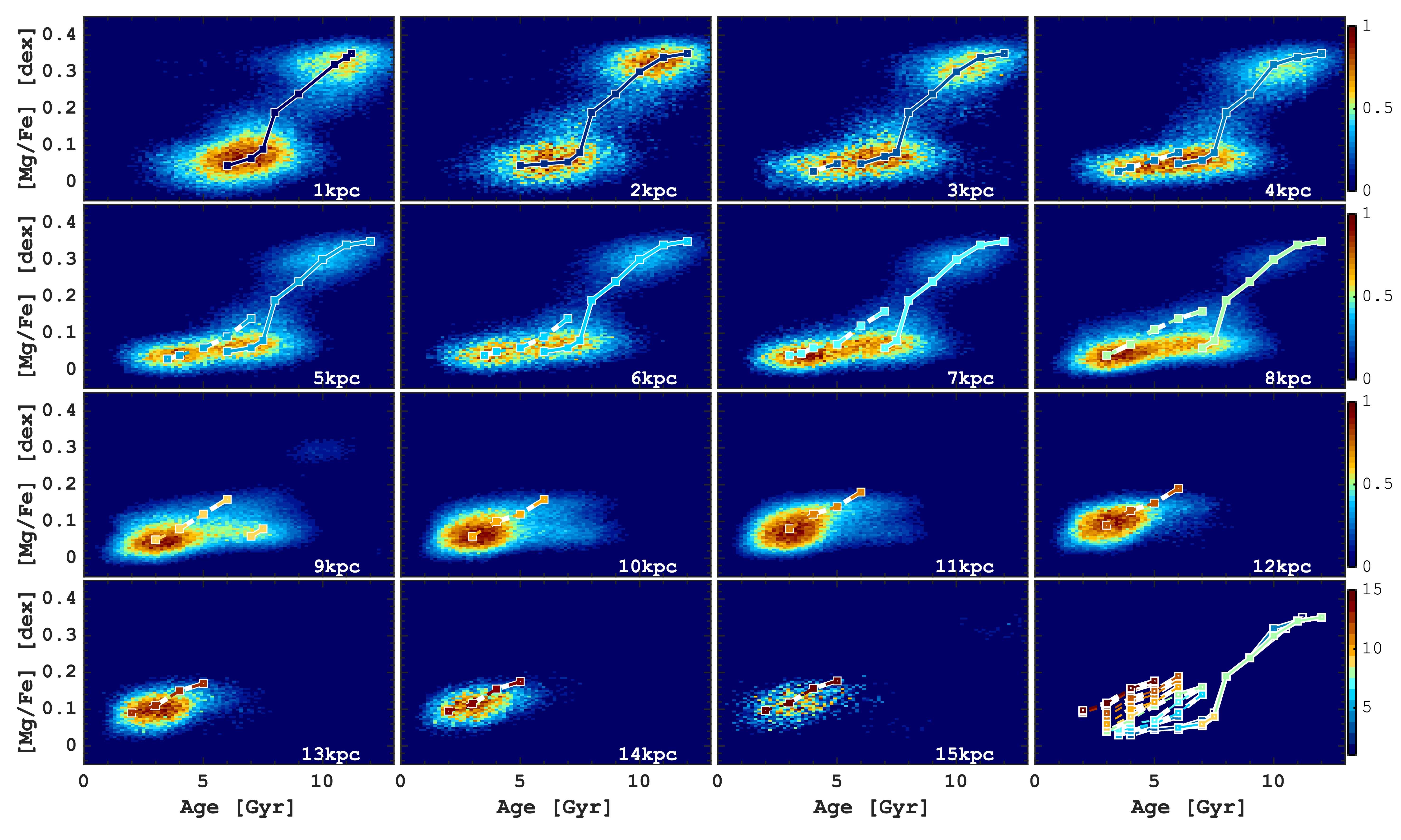}
    \caption{Same as in \ref{fig02::AMR_radii} but for the age - \MgFe relation. }
    \label{fig02::AMG_radii}
\end{figure*}

Since stars do not remain at their birthplaces but undergo radial migration~(churning) and vertical heating~(blurring)~\citep{2002MNRAS.336..785S, 2008ApJ...684L..79R, 2010ApJ...722..112M}, these processes can influence the present-day gradients of mono-age populations. \cite{2017MNRAS.464..702K} showed that outward migration leads to a positive vertical metallicity gradient, which can reverse its trend if the star-forming disc is flared. Additionally, \cite{2019MNRAS.487.3946M} demonstrated that even the superposition of flat metallicity profiles of mono-age populations can lead to a steep vertical gradient, also altered by the abundance uncertainties.

In Fig.~\ref{fig02::feh_vertical_gradient}, we present the vertical metallicity profiles of low-$\alpha$ (blue) and high-$\alpha$ (red) populations separately, as well as the total profiles (black), as a function of Galactocentric distance (from left to right) and age (from top to bottom). For the low-$\alpha$ stars, we observe mainly flat gradients within $R<3-4$~kpc. Variations in the gradient for the youngest populations (under 6 Gyr, top three rows) are likely caused by the X-shaped boxy bulge. Older populations, being kinematically hotter, do not show these features as they are less affected by the bulge emergence~\citep{2017MNRAS.469.1587D,2017A&A...606A..47F}. Beyond 4~kpc, outside the bar/bulge region, we find the steepest vertical gradients, which gradually flatten with increasing distance and age.

The presence of a vertical gradient for the youngest low-$\alpha$ stars may seem suspicious. However, given the $2$~Gyr age bins we use, it is likely due to the superposition of stars of different ages. Slightly older (and slightly metal-poor) stars tend to be hotter and contribute further away from the midplane compared to slightly younger (and slightly metal-rich) stars. Reducing the age bin size might seem like a solution, but age measurement uncertainties would likely dominate at some point and yield the same result. Additionally, this approach is impractical, as we can only select star particles within extremely narrow age ranges in simulations. Even in high-resolution simulations, the statistics would be very low, preventing accurate gradient measurement. A careful reader might notice a logical problem with this interpretation of the negative gradient, as the described mechanism would actually result in a positive gradient~\citep{2017MNRAS.464..702K}. However, as we showed already, the density distribution of low-$\alpha$ stars in small metallicity bins~(close to mono-age) is not exponential. Instead, it has a ring-like shape with decreasing density towards edges~(see Fig.~\ref{fig02::MAPs}, left). In this case, the radial heating can produce negative gradients. Therefore, the fact that we observe flattening of the vertical gradient with distance in the orbit superposition solution suggests a slower chemical enrichment when the stellar metallicity barely changes over time, as expected in the outer disc. The same behaviour with age is an illustration of the above-described mechanism, which flattens the initial negative gradient. 

The high-$\alpha$ populations generally follow the trends observed in the low-$\alpha$ stars but exhibit much less diversity in their metallicity profiles. At any age and Galactocentric distance, the vertical metallicity profile of the high-$\alpha$ stars is generally flatter. The combination of the low- and high-$\alpha$ vertical metallicity profiles results in strong overall metallicity gradients, as these populations have different median metallicities and density profiles at any given Galactocentric radius and age. 

The vertical metallicity profile behaviour is summarized in Fig.~\ref{fig02::feh_vertical_gradient_values}, which is in agreement with \cite{2015RAA....15.1209X}, who found that the vertical metallicity gradient flattens with age and Galactocentric distance outside the solar radius~\citep[see also][]{2019MNRAS.482.2189W}. Similarly, \cite{2014A&A...572A..33M} showed that the low-$\alpha$ disc has a negative vertical metallicity gradient, while the high-$\alpha$ sample shows a very shallow negative vertical metallicity gradient. The vertical gradients in the inner MW are less explored in the literature, also because the disc is dominated mainly by the bar/bulge structure whose 3D chemical abundance structure requires more detailed exploration, which is the topic of our follow-up work of the series~(\citetalias{Mapping-bulge}).

\section{Results: age-abundance relations and disc $\alpha$-dichotomy}\label{sec2::results_AMR}
\subsection{Global picture}
Age-abundance relationships, especially age-metallicity (or age-\FeH, AMR), offer crucial insights into the evolution of the MW, including its metals enrichment and star formation history. Fundamentally, the metallicity of newly formed stars increases over time, reflecting the accumulation of metals in the ISM due to pollution from previous generations of stars. Consequently, understanding the mean metallicity and its distribution across different ages allows us to constrain the chemical evolution of a galaxy, which is a play-off between star formation, which enriches the ISM, gas infall, which (usually) dilutes it, and outflows, which eject gas from the galaxy.

Interpreting the MW AMR based on limited datasets is challenging due to the influence of spatial coverage and selection function biases. For instance, the observed AMR can be distorted by the non-representative fractional contribution of stars from various Galactocentric radii. This artificial distortion is further complicated by radial migration and mixing processes that affect the current distribution of stars, placing them away from their birthplaces, thus shuffling enrichment histories of different Galactocentric radii with different ab initio unknown proportions. The importance of the radial dependence of the enrichment history is described as the inside-out process~\citep{1989MNRAS.239..885M, 2000MNRAS.313..338P, 2000A&A...362..921H} where the inner disc is thought to assemble first and form stars faster because of the high density of gas in the Galactic centre. 

Our orbit superposition approach mitigates the issue of selection function biases and provides stellar mass-weighted age-abundance relations. At the same time, the impact of radial migration is addressed in a follow-up work~(\citetalias{Mapping-sfh}). In Fig.~\ref{fig02::AMR_and_AMG_FE_weighted}, we present the age-\FeH(top) and age-\MgFe~(bottom) relations for the input APOGEE catalogue (left), the stellar mass-weighted distribution (middle), and the orbit decomposition-based mean Galactocentric map (right). We first observe the presence of two main overdensities in the age-\FeH maps~(top), which can be roughly separated by mean Galactocentric distance, revealing an apparent inner/outer disc dichotomy.  The existence of two AMR sequences~(and for several other elements) for the solar-type stars was shown by \cite{2020A&A...640A..81N}, who also noticed a small difference between the mean Galactocentric distance in the stellar orbits~\citep[see also][]{2021ApJ...920...23J}, where at a given radius, the locus of orbits of younger stars is located $\approx 0.5$~kpc further out compared to the older sequence. We stress, however, that this may simply reflect the age-velocity dispersion relation, as older stars experience more substantial asymmetric drift, so their guiding radii are close to the Galactic centre.

In our case, the younger group of stars, aged between 2-5 Gyr, is located in the outer disc (beyond $6-8$ kpc), while the inner-disc-dominated region shows a more extended and older age sequence. Although the AMR structure reveals these two prominent features, they appear differently in the APOGEE data compared to the stellar mass-weighted distribution. The young overdensity is less pronounced in the mass-weighted map, which is consistent with our earlier observations in Fig.~\ref{fig02::afe_feh_all}, where the low-$\alpha$ low-metallicity region of the \aFe-\FeH distribution is less prominent compared to the APOGEE data. The inner disc AMR sequence has also changed significantly. It is more pronounced for ages greater than 5 Gyr, and, more strikingly, its upper, super-solar metallicity~(>0.2~dex) part is highly prominent. The existence of such an AMR distribution is not surprising, as super-solar metallicity stars, especially in the Galactic centre~(see the top right panel) or the bulge region, have been known for some time~\citep{2006ApJ...636..821F, 2008A&A...486..177Z, 2024arXiv240601706R}; however, we find it to be very massive of $\approx 8.4\times 10^9\Msun$ for \FeH>0.2~dex. It seems that the most metal-rich stars are not the youngest, even in the MW centre; their presence shapes the global structure of the AMR. Hence, the natural assumption about the gradual increase of the metallicity over time or, at least, saturation~\citep{2018MNRAS.475.5487S}, in case of equilibrium chemical evolution~\citep[see, e.g.][]{2017ApJ...837..183W,2021MNRAS.508.4484J,2024arXiv241013256J}, is in tension with our mass-weighted AMR. The most natural explanation for such behaviour is the dilution of metallicity caused by the low-metallicity gas infall~\citep[see, e.g.][]{ 2015A&A...578A..87S,2019A&A...623A..60S}, in our case the dilution should be in order of $\approx 0.8-1$~dex on a very short time scale~($<1$~Gyr), hardly reproducible in galaxy-formation simulations; however, it is close to the prediction of the chemical evolution model by \cite{2019A&A...623A..60S}~\citep[see also,][]{2024arXiv240408059D}.
We recall, however, that the young age-\FeH sequence is dominated by stars located in the outer disc while the old one is made of the inner disc stars~(see the right panel in Fig.~\ref{fig02::AMR_and_AMG_FE_weighted}); hence if these populations are separated in space, there might be no need in the strong metallicity dilution~\citep{2024A&A...690A.352R, 2024arXiv241017326R}.

The age-\MgFe distribution in Fig.~\ref{fig02::AMR_and_AMG_FE_weighted}~(bottom) also shows the presence of two main overdensities, which are now related to high- and low-$\alpha$ sequences, equally seen for both APOGEE and mass-weighed distributions. The biggest difference here is that the lower part of the low-$\alpha$ sequence is more prominent compared to the APOGEE data, where there is more weight at younger ages. The middle bottom panel, in fact, shows three blobs: one is on top, corresponding to the high-$\alpha$, while the low-$\alpha$ sequence seems to have two overdensities, around $4$~Gyr and $7$~Gyr, corresponding to the relatively outer disc region and metal-rich older age-\FeH sequence. The similar structure of the low-$\alpha$ sequence was also suggested by \cite{2020A&A...640A..81N}~(see their Fig. 4). 

Fig.~\ref{fig02::AMR_and_AMG_FE_weighted} clearly shows that the AMR sequences and high-/low-$\alpha$ populations are not the same populations. As the old AMR sequence is made of both high- and low-$\alpha$ stars, while the young AMR sequence is a part of the low-$\alpha$ sequence~\citep[see also][]{2022MNRAS.510.4669S}. 

We need to stress that the picture we observe in Fig.~\ref{fig02::AMR_and_AMG_FE_weighted} is quite blurred; despite the large statistics, our approach in using the abundance uncertainties along the orbits of stars does not allow us to trace tiny details of the age-abundance relations. For instance, there is a visible vertical split of the low-$\alpha$ sequence in the age-\MgFe plane around $6-8$~Gyr in the APOGEE data~(see bottom left panel). We, however, attempt not to over-interpret the small-scale trends, as the data-driven and neural network methods may create artificial features in sparsely populated regions of parameter space in the training data. Therefore, we deliberately limit ourselves by the analysis of the global trends in the age-abundance plane, which, as we showed, have a tendency to vary with Galactocentric distance.

\subsection{Radial dependence}\label{sec2::AMRG_radii}
Studies on the radial dependence of the AMR have become feasible relatively recently due to advancements in age determination from both large-scale datasets and smaller, more precise asteroseismic data, as well as Gaia-based distance measurements. Using APOGEE DR 14, \cite{2019MNRAS.489.1742F} demonstrated that the metallicity-age relations (different from the AMR as \citealt{2019MNRAS.489.1742F} aimed to determine the averaged age for a given metallicity) are not monotonic but exhibit a notable turnover, shifting from high to low metallicities with increasing Galactocentric distance~\citep[see also][]{2022MNRAS.512.4697L,2022Natur.603..599X}. \cite{2019ApJ...871..181H} also observed a turnover in [C/N] at high metallicities, corresponding to a turnover in age.

\begin{figure*}
    \centering
    \includegraphics[width=1\hsize]{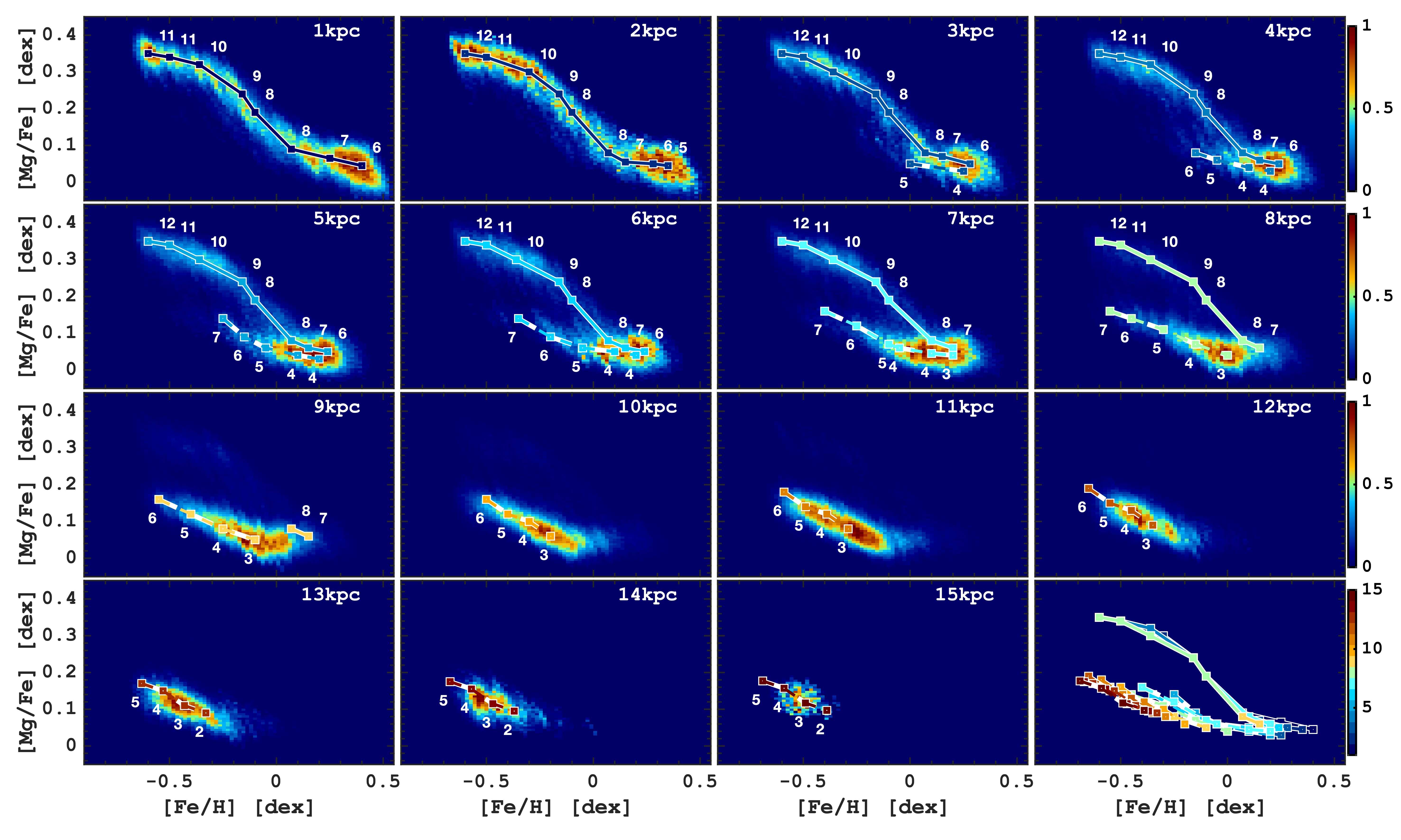}
    \caption{Same as in \ref{fig02::AMG_radii} but for the \FeH-\MgFe relation. The points along the lines mark the $1$~Gyr age interval. The numbers along the lines indicate the age (in Gyr) of stars.}
    \label{fig02::AMRG_radii}
\end{figure*}

The formation of the metallicity-age turnover can be readily understood within the framework of dual AMR sequences~\citep{2020A&A...640A..81N}, whose relative proportions vary with Galactocentric distance~\citep{2022MNRAS.511.5639L,2022MNRAS.510.4669S}. To investigate the origins of these two AMR sequences and their connection with high- and low-$\alpha$ populations, we present the age-\MgFe, age-\FeH, and \MgFe-\FeH relations in $\rm \pm 1 kpc$ radial bins at various Galactocentric radii from $1$ to $15$~kpc in Figs.~\ref{fig02::AMR_radii}, \ref{fig02::AMG_radii}, and \ref{fig02::AMRG_radii}. In each panel, the stellar density is normalized by the maximum value. Additionally, we indicate the location of the maximum density of the young and old AMR sequences identified in all three coordinate combinations simultaneously. For each Galactocentric distance, we attempt to determine the location of the AMR sequences within the $1$ Gyr age bin. We only draw the lines representing the sequences if we obtain a meaningful solution for at least two age bins. This, however, does not exclude small contamination from the tails of the AMR sequences at some radii.

{\bf In the inner two radial bins~($<3$~kpc)} in Fig.~\ref{fig02::AMR_radii}, we only see the presence of a single, old AMR sequence. At around 10 Gyr and $\FeH\approx -0.2$, we notice a flattening of the relation~\citep[see also][]{2022MNRAS.511.5639L} and since the \MgFe declines very rapidly, this suggests the lack of the CC-SNe enrichment, while the metallicity is still slowly increasing, likely mostly by the ejecta of previously formed stars, indicating the cessation of the SF, which was also seen in the stellar density distribution. The latter is illustrated by the yellow histograms showing the normalized~(by the twice maximum value, so the scale of the distributions is from 0 to 0.5) stellar mass distribution as a function of age. Here, we can see a drop in the mass distribution of stars younger than $\approx 10$~Gyr ago. We remain cautious in calling these histograms star formation histories, as they are based on relatively narrow one kpc rings, do not account for dead populations of stars and ignore possible mixing of stars formed at different radii. Nevertheless, we see a noticeable decline of stellar mass partially reflecting a decrease of the SFR found by \cite{2014ApJ...781L..31S,2015A&A...578A..87S} using chemical evolution models tuned to reconstruct the age-abundance structure of the MW, where stellar densities played no role. This episode was dubbed as quenching with the following cessation of the star formation in the inner Galaxy~\citep{2016A&A...589A..66H}. The reason for such decline of the stellar mass growth is still a matter of debate where competing scenarios include bar-induced~(more generally morphological~\citep{2009ApJ...707..250M}) quenching~\citep{2018A&A...618A..78H,2018A&A...609A..60K}, requiring the bar formation around $9-10$~Gyr~\citep{2024MNRAS.530.2972S,2024A&A...690A.147H}; or the impact of the GSE-progenitor infall~\citep{2019A&A...632A...4D,2024MNRAS.535..392L}. The GSE scenario, however, would imply a drop in metallicity~\citep{2018MNRAS.479.3381B,2024MNRAS.527.2426A} with an increase in \MgFe~\citep{2024MNRAS.528L.122C}, which we do not observe in the inner MW disc.  Hence, it is unlikely that the GSE merger is responsible for the SF quenching, at least inside $3$~kpc. Also, \cite{2018MNRAS.479.3381B} showed that wet significant~(1:10-1:3) mergers seem to enhance star formation~\citep[see also][]{2010ApJ...720L.149T,2017MNRAS.465.1934F,2022MNRAS.516.4922R}, but this picture, of course, depends on many parameters of the interaction~\citep{2008A&A...492...31D}, hence we can not rule out the GSE effect on the stellar mass growth rate around $10$~Gyr ago. In \cite{2021MNRAS.501.5176K}, we also showed that a rapid decline of the SFR can be caused by the stellar feedback alone, essentially removing star-forming gas from the disc and preventing its further infall, resulting in a transition from high- to low-$\alpha$ sequences.

{\bf At larger~(3-4~kpc) radii}, we observe essentially the same trends as in the innermost bins, but here the top part of the AMR, representing the metal-rich population of the bulge, fades away. The flattening of the AMR in the innermost radial bins is intriguing but, if real, its interpretation would require understanding how the presence of the bar and the bulge shape the chemical evolution in the central region, which, despite the progress in galaxy formation simulations~\citep{2018ApJ...861...88B, 2019MNRAS.485.5073D, 2020MNRAS.494.5936F}, remains illusive. From a different perspective, the lack of EMR stars just outside $\approx 3$~kpc suggests that they are not the subject of outward migration; hence, if we assume that stars formed in the inner disc were transferred outwards by the bar~\citep{2015A&A...578A..58H,2020A&A...638A.144K,2020ApJ...889...81W} then the EMR stars formed after the bar formation put a lower limit on the time formation of the MW bar to $7-8$~Gyr, when the very first EMR stars formed. We note that this is consistent with several studies of the MW bar time of formation~\citep{2024MNRAS.530.2972S, 2024A&A...690A.147H}.

{\bf The intermediate radii~(5-9~kpc)} show the cohabitation of both AMR sequences~(see Fig.~\ref{fig02::AMR_radii}), which, however, barely overlap in age at a given radius where we need to keep in mind the average a conservative age uncertainty of $1-1.5$~Gyr. Since we are not certain about the origin of these two AMR sequences, either they are created by the migration~\citep{2021MNRAS.507.5882S,2023MNRAS.523.3791C} or a rapid and strong~($\approx 0.5$~dex) metallicity dilution~\citep{2019A&A...623A..60S,2023A&A...670A.109S}, we do not connect them, for now considering their distinct origin. The maximum metallicity value decreases outwards for the sequences, reflecting the radial metallicity gradient for both AMR sequences. The chemical abundance diversity between two AMR sequences is seen as well in the age-\MgFe relation. The difference between two AMR sequences becomes the most evident in \FeH-\MgFe plane~(see Fig.~\ref{fig02::AMRG_radii}) where the old and young AMRs represent high- and low-$\alpha$ sequences, respectively, converging at metallicity $\approx 0.2$.  

Regarding the $\MgFe-\FeH$ plane, the old AMR sequence barely changes with radius inside $\approx 8$~kpc and represents the upper branch of the high-$\alpha$ sequence with age $>10$~Gyr and younger the most metal-rich populations of the low-$\alpha$ sequence. The fact that there is not much radial dependence of the high-$\alpha$, seen almost identical in age-\FeH and age-\MgFe (Fig.~\ref{fig02::AMR_radii} and ~\ref{fig02::AMG_radii}), may suggest that these populations formed in a well-mixed environment~\citep{2016A&A...589A..66H,2018A&A...618A..78H} with a weak or negligible radial abundance gradient. At the same time, even if there was a radial gradient, its importance in driving chemical enrichment is not evident, as early formed stars~($>10-11$~Gyr ago or $z>2-3$) would have high random motions~\citep{2023MNRAS.525.2241H,2023MNRAS.523.6220Y}, and thus they would release metals into the ISM away from their birthplaces smearing out the existing radial abundance structure on a very short time scale, $<<1$~Gyr. Therefore, on top of the ISM mixing driven by the turbulent motions, we need to consider its non-local~(far from birth radii) pollution by newly formed stellar populations at high redshift. The latter is difficult to implement directly in the Galactic chemical evolution models. However, the non-local enrichment aligns well with the instantaneous ISM mixing assumption, making different models successful in reproducing the enrichment history of the old AMR sequence.

{\bf The outer radii~($>9$~kpc)} show the young AMR sequence only, which corresponds to the young tail of the low-$\alpha$ population~(see Fig.~\ref{fig02::AMG_radii}). Hence, a notable feature at these distances is the absence of stars of the old AMR sequence, suggesting the lack of mixing between the inner and outer disc outside $8-9$~kpc. Despite the widely accepted idea regarding radial migration shaping the structure of the MW disc, it seems to be inefficient in moving out stars from the inner disc just outside the solar circle. Although the inner disc is older and its stars have had more time to migrate, the outer disc shows literally no signature of the high-$\alpha$ stars but also the metal-rich low-$\alpha$ stars. 

Even more, the outward inner disc migrators also seem to be locked inside $\approx10$~kpc, which roughly corresponds~(we need to keep in mind that the OLR is not infinitely thin but it is made of stars located in a broad range of radii~\citep{2007MNRAS.379.1155C}) to the present-day location of the bar OLR, abruptly limiting migration further out~\citep{2015A&A...578A..58H, 2018A&A...616A..86H, 2020A&A...638A.144K, 2024A&A...690A.147H}. We, however, should not disregard another perspective. Since migration is a process believed to be mainly driven by transient spirals~\citep{2002MNRAS.336..785S, 2008ApJ...684L..79R, 2012MNRAS.422.1363S, 2014ApJ...794..173V}, we may speculate that the absence of migration across $8-9$~kpc suggests that the spiral structure inside this radius is more regular than needed for efficient blurring. This assumption might not be so extreme in the presence of the bar, as it likely regulates the appearance of the spiral density waves~\citep{2022ARA&A..60...73S}. This does not mean that bar-regulated spiral density waves are not transient, but they more likely will appear/disappear with more determined frequency, and the disc stars gain/lose angular momentum without changing their time-averaged guiding radius. In such a case, stars can preferentially migrate due to the long-term evolution~\cite[slowdown, see, e.g.][]{2020A&A...638A.144K,2024A&A...690A.147H} and short time-scale variations of the bar parameters~\citep{2020MNRAS.497..933H,2024MNRAS.528.3576V}. The most vigorous mixing is therefore expected in between the bar resonances $\approx 5-10$~kpc, where stars can be captured by the resonances and released from them at different radii if the bar strength declines~(Marques et al., in prep). This picture we observe in Fig.~\ref{fig02::AMG_radii} where two AMR sequences spatially overlap only in the $5-9$~kpc region. 

\begin{figure}
    \centering
    \includegraphics[width=1\hsize]{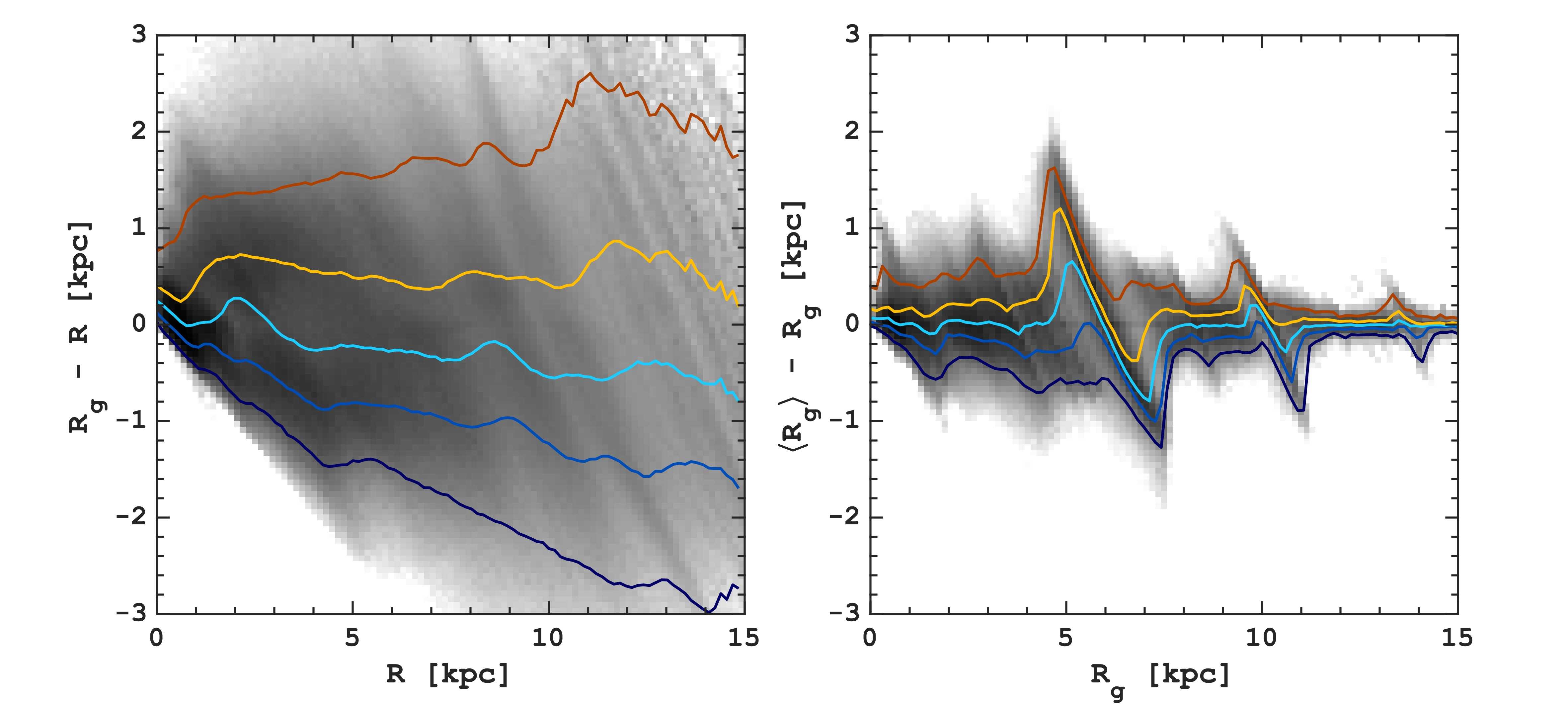}
    \caption{Illustration of the mixing caused by the bar. Left: normalized stellar density distribution in guiding~($R_g$) - instantaneous~($R$) plane. Right: normalized stellar density distribution in guiding~($R_g$) - mean guiding ~($\langle R_g\rangle$, orbit averaged) plane. In the right panel, the effect of the main bar resonances CR, OLR, and 4:3  is seen as diagonal overdensities. We stress that these features are not the effect of radial migration~(churning), as we consider a constant bar pattern speed and ignore the presence of spiral arms in the orbit superposition modelling.}
    \label{fig02::mixing_r_rg}
\end{figure}

To illustrate the mixing~(not migration) introduced by the bar in Fig.~\ref{fig02::mixing_r_rg}~(left) we show the difference between guiding radius~($R_g$) and instantaneous Galactocentric distance~($R$) as a function of the latter. The difference between $R$ and $R_g$ suggests that stars are oscillating around their guiding radii with an amplitude up to $1.5-3$~kpc. However, we need to keep in mind that for most stars, guiding radii also periodically change with time in a barred potential~\citep{2008gady.book.....B}. The strength of this libration was estimated as about $0.5$~kpc in an MW-type simulated galaxy~\citep{2024A&A...690A.147H}. In Fig.~\ref{fig02::mixing_r_rg}~(right) we quantify it for the MW by showing the difference between the mean~(orbit-averaged) and the time-dependent~(along the orbit) guiding radius. The picture we observe here very closely resembles classic migration-caused angular momentum change in the vicinity of resonances~\citep{2010ApJ...722..112M, 2011A&A...527A.147M, 2012A&A...548A.126M}; however, the process whose manifestation we observe here is different. The figure suggests that the guiding radii change a lot around the main bar resonances introducing additional mixing, on top of blurring. The strength of the guiding radii libration of about $0.5-1.5$~kpc may represent a substantial fraction of migration strength of $\approx 1.4-2.4$~kpc obtained in different simulations and the MW data analysis~\citep[][]{2008ApJ...684L..79R, 2019ApJ...884...99F, 2020ApJ...896...15F, 2021MNRAS.501.5176K, 2021MNRAS.502..260B}. However, the strength of the libration varies strongly and non-monotonically with radius, possibly contributing to the skewness of the MDF at different radii~\citep{2016ApJ...818L...6L}. The right panel of Fig.~\ref{fig02::mixing_r_rg} shows several peaks around $6$, $11$, and $13$~kpc, corresponding to the locations of the CR, OLR, and 4:3 resonance, respectively. These radii coincide with the regions where we observe a mixture of the two AMR sequences, between $5$ and $9$~kpc~(Fig.~\ref{fig02::AMR_radii}). Thus, the mixing rate introduced by the bar resonances may appear to be sufficient to explain the observed pattern; however, more quantitative analysis involving stellar birth radii estimation is needed to distinguish the orbits libration from the genuine churning~\citep[see, e.g.][]{2018MNRAS.481.1645M, 2022MNRAS.511.5639L, 2023MNRAS.523.2126P, 2024MNRAS.527..321W, 2024MNRAS.528.3464R}.

This confirms that the extended solar vicinity is a transition zone between the inner~(old AMR sequence) and outer disc~(young AMR sequence), as anticipated in \cite{ 2013A&A...560A.109H,2016A&A...589A..66H, 2017A&A...608L...1H}. Note that the young AMR sequence alone has a dip in the centre, as it was known for individual MAPs~\citep{2012ApJ...753..148B,2016ApJ...823...30B} and presented in the previous analysis~(see Fig.~\ref{fig02::MAPs} and Section~\ref{sec2::maps}). For instance, in the inner $\approx 6$~kpc, we observe only the most metal-rich part of the young AMR sequence, which suggests that the inward migration~(churning) is relatively modest and does not exceed $2-3$~kpc~\citep[see, e.g.][]{2024arXiv241017326R}. Even more, from this 'migration efficiency', we need to subtract the radial mixing~(blurring) caused by the bar, and, in particular, the libration of stellar orbits in a non-axisymmetric potential~\citep{2008gady.book.....B}, which can not be eradicated by using guiding radii, as they periodically change over time~\citep[see, e.g.][]{2007MNRAS.379.1155C, 2015A&A...578A..58H}. 

\section{Discussion}\label{sec2::discussion}
\subsection{MW disc build-up}
In the previous section, we discussed the age-abundance structure of the MW disc we recovered using orbit superposition. A striking feature here, known since recently, is the presence of two prominent age-\FeH sequences which do not match high- and low-$\alpha$ sequences, at least in their classic definition. We characterize these two sequences by optimizing the stellar density in three age-abundance coordinates, as illustrated in Figs.~\ref{fig02::AMR_radii}-\ref{fig02::AMRG_radii}. Since the evolutionary link between these two sequences is not obvious, and the roles of radial migration and, more generally, mixing are not trivial to assess without more specific analysis, we do not connect the age-abundance sequences in these figures. However, one natural possibility to connect the formation histories of the AMR sequences is to assume a metallicity dilution at the end of the old AMR sequence formation caused by the fresh gas infall, which, despite being able to explain the main features, fails to reproduce the details of the abundance-age relations quantitatively~\citep{2020A&A...640A..81N}.

\cite{2013A&A...560A.109H,2019A&A...625A.105H} described a scenario in which the high-$\alpha$ populations started to form in a turbulent gaseous disc with a chemical evolution well explained by the closed-box model inside $\approx 6-7$~kpc~\citep[][]{ 2014ApJ...781L..31S, 2015A&A...578A..87S, 2018A&A...618A..78H}. After the quenching of star formation, about $10$~Gyr ago~\citep{2016A&A...589A..66H,2020MNRAS.497.3557L}, gas previously expelled from the inner disc, mixed with the surrounding low-$\FeH$ gas in the halo, started to return to the outer regions~($R>6-7$~kpc). This mixture of pre-enriched gas gives rise to the outer disc or subsolar part of the low-$\alpha$ sequence. In parallel, the inner disc resumed to grow after the quenching, producing the most metal-rich stars until $\approx 5$~Gyr ago.

\begin{figure}
    \centering
    \includegraphics[width=1\hsize]{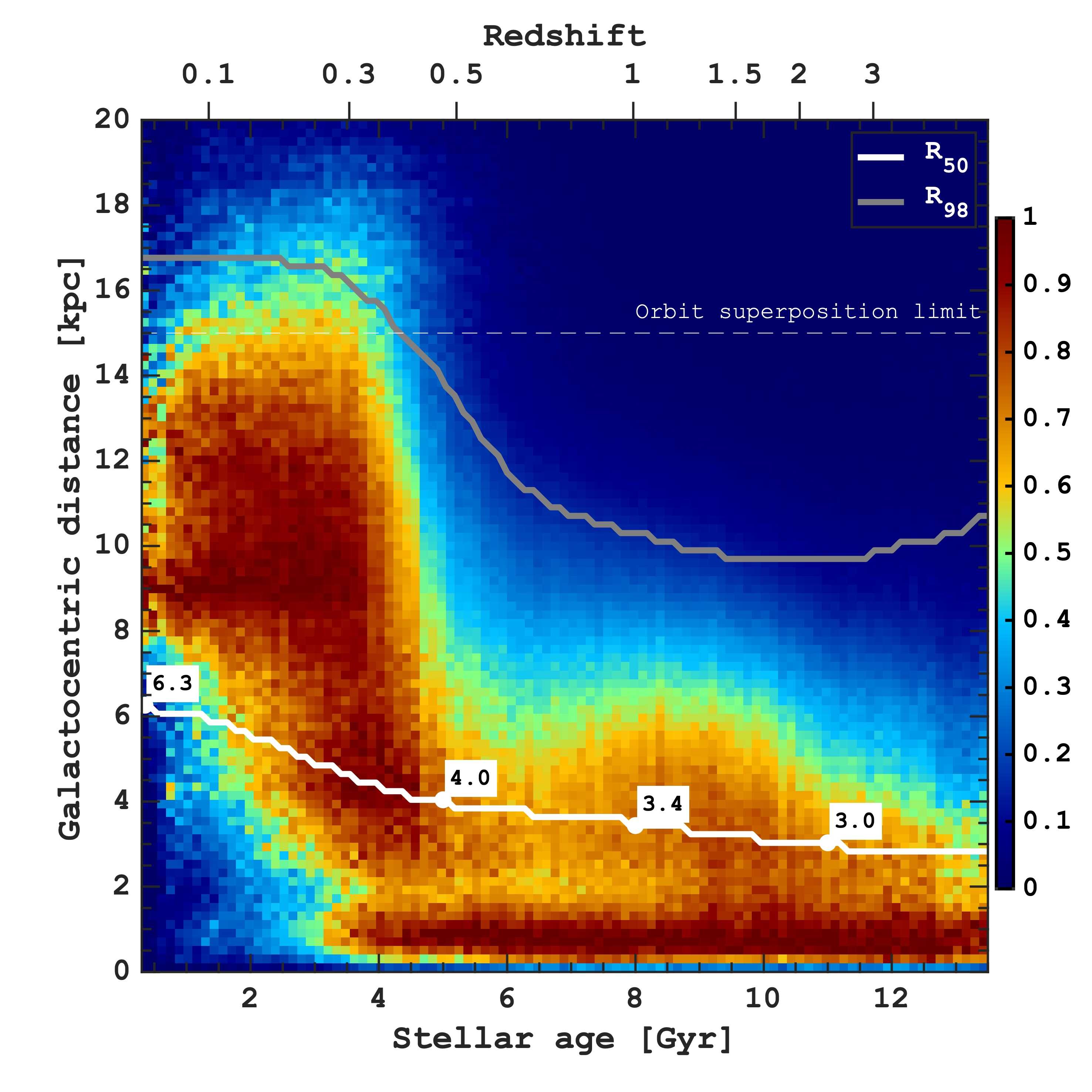}
    \caption{Radial distribution of stellar mass for mono-age populations. For each age bin, the distributions are normalized by the maximum value. The white line shows the effective radius evolution, assuming that the radial mass distribution does not change with time for mono-age populations. Several values of the effective radius at different ages are given along the while line. The grey line shows the radius of $98\%$ of the enclosed mass, $R_{98}$. We note that $R_{98}$ can be underestimated for ages earlier than $\approx 6$~Gyr, as our orbit superposition model is designed to fit the stellar density within $15$~kpc. Hence, it can be considered the lower limit in this age range.}
    \label{fig02::size_evolution}
\end{figure}

Below, we elaborate more on the scenario introduced by \cite{2019A&A...625A.105H} and discuss some details which we can speculate on using the orbit superposition-based results we presented throughout the paper. As we already discussed in Sec.~\ref{sec2::AMRG_radii}, the GSE scenario can not be ruled out, but it is less preferable compared to the bar formation to explain the SF quenching in the inner disc. In \cite{2018A&A...609A..60K}, we demonstrated that bars increase gas velocity dispersion inside their corotation radius and, as a result, suppress the star formation without removing gas from the galaxy. However, additionally, bars are known to be responsible for transporting gas towards the inner kiloparsec by generating torques that put gas clouds on intersecting trajectories. This leads to shocks, in which the gas loses angular momentum and funnels inwards~\citep{1992MNRAS.259..345A, 2005A&A...441.1011G, 2013ApJ...771....8L}. The gas reaching the very centre can feed the central black hole, trigger active galactic nuclei feedback ~\citep{2005AJ....130.1472P, 2014ApJ...792..101D, 2021A&A...656A..60A}, which can, in turn, rapidly suppress SF and cause quenching~\citep{2019A&A...624A..81M, 2021MNRAS.500.4004D, 2024arXiv241021580B}.

After the quenching, star formation restarted in the innermost region. However, since the metallicity continues to increase monotonically, the most likely source of this star formation is the remaining gas. If fresh gas had been (re)accreted into the central region, we would have observed a dilution of metallicity, which is not the case. Therefore, the bar-quenching scenario is plausible, as it does not significantly affect the gas content and its parameters. At this time~($8-10$~Gyr ago), the physical conditions in the Galactic centre are different from the rest of the disc~\citep{2010MNRAS.407.2091G, 2021A&A...656A.133Q, 2023A&A...673A.147P}. Due to high pressure deep in the potential well, metals released by newly formed stars remain locked in the inner kpc, allowing efficient enrichment, producing EMR stars~\citep{2024A&A...690A.352R} which we observe confined closely to the Galactic centre~\citep{2024arXiv240601706R}. The fact we observe the EMR stars stopped to form around $5$~Gyr ago suggests the exhaustion of gas not only in the Galactic centre but also inside the corotation radius, typically seen in barred galaxies~\citep[so-called star formation desert; ][]{2018MNRAS.474.3101J, 2020A&A...644A..79G, 2024A&A...687A.255S, 2024MNRAS.533.3975K} and simulations~\citep{2018A&A...609A..60K,2024arXiv241021580B}. In agreement with this picture, we observe a decline, but not the complete absence, of young stars($<4$~Gyr old) in the inner MW~(see Fig.~\ref{fig02::age_radii_distibution}), which most likely formed outside the corotation and migrated to the inner galaxy or trapped by the bar on highly-eccentric orbits~(see Fig.~\ref{fig02::mixing_r_rg}). The latter is also confirmed by a slight increase of the radial velocity dispersion of young stars in the inner $<3$~kpc~(see Fig.~\ref{fig02::AVR}).

In the picture we described, the inner MW disc~(or old AMR sequence) formed in a sort of isolation, without strong influence from outside. We can also notice a continuous formation of the entire inner disc from the metal-poor~(at least $\FeH\approx -0.7$~dex, as the ages for more metal-poor stars are not valid in the DistMass catalogue~\citep{2024AJ....167...73S}) up to the most metal-rich~($\FeH \approx 0.5$~dex) stars. This is seen in the density profiles of MAPs in Fig.~\ref{fig02::MAPs}. The scalelength of the inner disc MAPs is $\approx 2$~kpc and its nearly constant as a function of metallicity~(and \MgFe) up to \FeH$\approx 0.2$. Only at higher metallicities~(and youngest ages) does it rapidly drop. The latter, however, we interpret as the influence of the bar, essentially preventing the SF outside the nuclei, again dating the bar formation by $~\approx 7-8$~Gyr ago. The lack of the radial scalelength evolution with age is somehow in tension with the inside-out formation, suggesting an increase of the scale length with time~\citep{2013ApJ...773...43B,2021MNRAS.503.5826A}. It also looks like there is not much place for the effect of migration, as it tends to increase the scalelength for older populations~\citep{2006ApJ...645..209D,2008ApJ...675L..65R,2012A&A...548A.126M} but again we do not observe this either. This conclusion might not be very strong if we would observe a constant scale length for old stars only, but the age range of the old AMR sequence is from $\approx 5$ to $\approx 13$~Gyr. We argue that the scaleheight of MAPs of old AMR sequence monotonically decreases from $1$~kpc at the metal-poor end to $0.1-0.2$~kpc for the most metal-rich populations, illustrating upside-down formation but not heating. One can assume that mergers can result in strong heating of older~(high-$\alpha$) populations at early times; however, it is hard to imagine the impact of mergers only in the vertical direction and no impact on the radial extension of populations. We conclude that the old AMR sequence or the inner disc~($<6-7$~kpc) was formed gradually in an upside-down manner on a time-scale of $13$ to $5$~Gyr ago with negligible influence by external factors. Its formation was over due to gas exhaustion. 

\begin{figure}
    \centering
    \includegraphics[width=1\hsize]{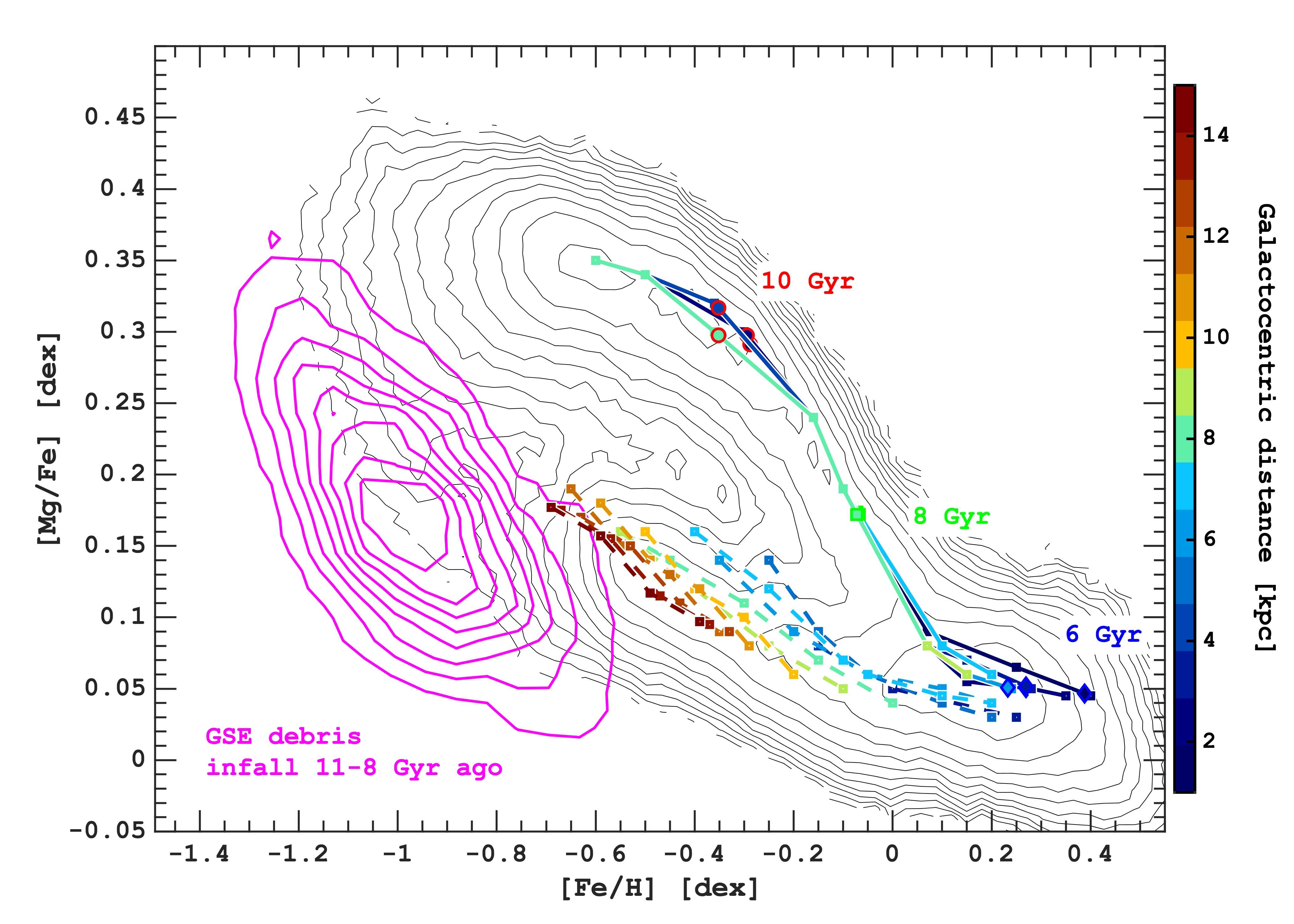}
    \caption{Possible origin of outer MW disc. The lines of different colours represent the \MgFe-\FeH fits at different radii, where several age points are marked by different markers. The black background contours show the overall distribution in the \MgFe-\FeH plane, while the magenta contours show the location of the accreted~(GSE).}  \label{fig02::MW_evolution}
\end{figure}

The inner disc evolution we described above implies a rather compact disc of the MW until $z \sim 0.5-1$. In contrast, the outer disc, vastly dominated by the young AMR sequence, started to form around $\approx 6-7$~Gyr ago~($z\sim 0.6-0.7$) at the latest. We illustrate this in Fig.~\ref{fig02::size_evolution}, showing the stellar mass density distribution as a function of age and Galactocentric distance. The white line marks the 'evolution' of the effective radius~($R_{50}$) calculated as the half-mass distance for all stars older than a given age. The grey line shows the radius of $98\%$ enclosed mass~($R_{98}$). We emphasize that Fig.~\ref{fig02::size_evolution} displays the current radial distribution of mono-age populations; however, their birth distribution may have been different~\citep{2018MNRAS.481.1645M, 2024MNRAS.535..392L}. The significance of this discrepancy is likely to grow as these populations age with a tendency for older populations to be more compact, but, as we discussed above, we do not expect this effect to be significant.

If we count the enclosed mass of the old AMR sequence, we can get the effective~(half-mass, $\rm R_{50}$) radius of the MW of $\approx 3.4$~kpc. Although, as we discussed above, the MAP structure does not indicate a strong scalelength evolution, this value can be considered as an upper limit for the MW size at $z \approx 0.5-2$. The present-day~(counting stars of all ages) effective radius is $\approx 6.3$~kpc~\citep[][found a half-light radius of $5.75\pm0.38$~kpc]{2024arXiv240605604L}. This suggests that since the end of the old AMR sequence formation, the MW effective size increased by $\approx 85\%$ till now. This size increase is close in between $20-50\%$~(since $z \approx 1$) found using extragalactic systems~\citep[see e.g.][]{2014ApJ...788...28V, 2019ApJ...880...57M, 2021ApJ...921...38K} and theoretical expectations, suggesting the evolution of nearly a factor of $\sim 2$ of the size of disc galaxies since $z \sim 1$~\citep{1998MNRAS.295..319M}. We also need to mention the results by \cite{2024A&A...682A.110B}, who using more physically-motivated galaxy size measurements~\citep[see also][]{2020MNRAS.493...87T,2022A&A...667A..87C}, demonstrated a dramatic evolution of disc galaxies since $z\sim 1$, with an average radial growth rate of the MW-like discs of $\rm \approx 1.5\ kpc\ Gyr^{-1}$. This is very much in line with the MW size evolution if we consider the 'evolution' of the outer radius enclosing $98\%$ of the mass, shown by the grey curve in Fig.~\ref{fig02::size_evolution}. The outer size of the MW remains nearly constant, around $10$~kpc, implying that there was no star formation and chemical enrichment outside this distance possible. However, starting from $7$~Gyr ago, the MW acquired the bulk of its outer disc during the following $3-4$~Gyr.

\begin{figure}
    \centering
    \includegraphics[width=1\hsize]{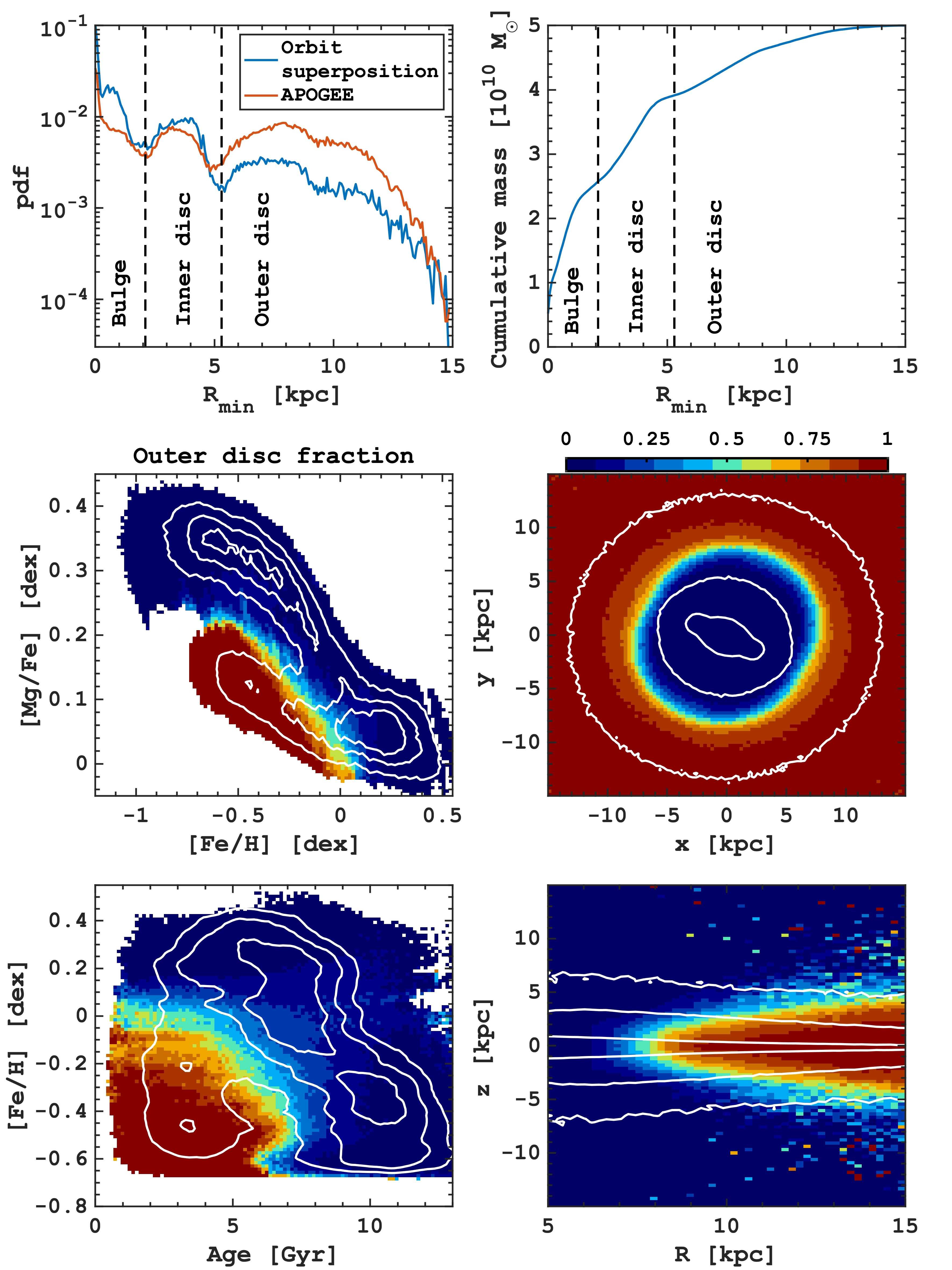}
    \caption{Kinematic definition of the MW outer disc. The top row shows the pdf~(left) and cumulative mass distribution~(right) as a function of pericentre values, $R_{min}$. The vertical lines mark the local minima of the pdf on the left, setting up the boundaries between the bulge~($R_{min}<2.1$), inner~($2.1<R_{min}<5.5$~kpc) and outer disc~($R_{min}>5.3$~kpc). The group of the bottom panels illustrate the outer disc selection in $\MgFe-\FeH$~(middle left), $x-y$~(middle right), age-\FeH~(bottom left) and $R-z$~(bottom right) coordinates. The maps are colour-coded by the stellar mass fraction of the outer disc relative to the total stellar mass in corresponding coordinates.}
    \label{fig02::rmin_selection}
\end{figure}

What is the origin of the outer MW disc? A plausible explanation is that it was made of a mixture of gas expelled during the inner disc formation~\citep{2013A&A...560A.109H, 2019A&A...625A.105H} combined with external gas in some proportion. To better understand the nature of the ex-situ gas component, we can assume that its chemical composition, combined with the chemical composition of the high-$\alpha$ sequence from around $7-10$ Gyr ago, represents the chemical composition of old outer disc stars. Hence, if we assume that these two gas sources contributed nearly equally to the outer disc build-up, we can use sort of 'abundance distance'~\citep{2023MNRAS.523.2126P} to find the chemical abundances required from the ex-situ gas. In Fig.~\ref{fig02::MW_evolution}, we show the \FeH-\MgFe plane where the background contours reflect the density distribution of the entire MW disc and the lines of different colours corresponding to the age-abundance relations we obtained using Figs.~\ref{fig02::AMR_radii}-\ref{fig02::AMRG_radii} at different radii. If we connect the high-$\alpha$ sequence chemical composition $7-10$ Gyr ago with the old outer disc and continue these lines towards lower metallicities and lower \MgFe abundances we end up in the \FeH-\MgFe region occupied by the debris of the GSE~\citep[see, e.g.][]{2010A&A...511L..10N, 2011A&A...530A..15N, 2012ApJ...761..161S, 2014MNRAS.444..515R, 2018ApJ...863..113H, 2018Natur.563...85H}, whose density distribution of the latter is marked by the magenta contours obtained from the GMM-based chemo-kinematic selection of the raw APOGEE data from \cite{2023arXiv231005287K}. From this very descriptive scheme, we can propose that the outer disc was formed from the mixture of solar or slightly sub-solar metallicity high-$\alpha$ and the gas content stripped from the GSE.

Using the VINTERGATAN zoom cosmological simulations of a MW-type galaxy~\citep{2021MNRAS.503.5826A}, \cite{2021MNRAS.503.5868R} presented a scenario in which the formation of the inner disc follows the scheme we described earlier. At the time of the last major merger, their galactic disc spans approximately $5$~kpc and is characterized by the old AMR sequence. The delayed formation of the outer disc is stimulated by the passage of this last major merger, which triggers star formation in a misaligned gaseous disc surrounding the inner stellar component. Once both the inner and outer discs settle into the same plane, they enrich each other, ultimately appearing as a chemically discontinuous single (geometrically thin) disc component. This scenario suggests that the low-\aFe sequence is populated in situ through two formation channels: one in the inner galaxy and one in the outer galaxy, each with distinct metallicities. This aligns with our analysis, where the low-$\alpha$ sequence comprises both old (inner) and young (outer) AMR sequences. However, in the VINTERGATAN scenario, it remains unclear why the \aFe abundance of the tilted disc, which did not host significant star formation, reaches the levels of the chemically evolved inner disc. A qualitatively similar picture is observed in the NIHAO simulations by \cite{2020MNRAS.491.5435B}, who demonstrated that the major merger introduces fresh, metal-poor gas that dilutes the ISM metallicity without affecting the \aFe. In this case, it is not straightforward to envision a rapid settling of the ex-situ gas from the accreted galaxy into a thin outer disc. Here, the gas expelled during the inner disc formation might play a crucial stabilizing role, particularly if it retains angular momentum. 

The scenario of the MW disc formation we presented above is likely best illustrated by a combination of the VINTERGATAN~\citep{2021MNRAS.503.5868R} and NIHAO~\citep{2020MNRAS.491.5435B} simulations. An important insight from this picture is that the distinction of the MW into low- and high-$\alpha$ components, despite its prominence in the chemical abundance space, is less genetically distinct compared to the decomposition of the MW onto inner and outer discs. This idea has been circulated in several works~\citep{2013A&A...560A.109H, 2021MNRAS.503.2814C, 2017A&A...608L...1H, 2019A&A...625A.105H}, which, thanks to our orbit superposition approach we can assess without contamination of the APOGEE selection function. In Fig.~\ref{fig02::rmin_selection}, we provide a relatively simple selection, allowing us to distinguish inner from outer discs without introducing sharp spatial, kinematic or chemical abundance cuts. The top panels present the mass-weighted distribution of pericentres or $R_{min}$ obtained in the barred potential. Quite surprisingly, the pdf~(top left) reveals three distinct components: $R_{min}<2$, $2<R_{min}<5.5$ and $R_{min}>5.5$~kpc, corresponding to the bulge, inner disc and outer discs, respectively. Masses of these components are: $2.5$, $1.5$ and $1\times 10^{10}$~\Msun~(see top right panel). 
Although the spatial distribution of the outer disc stars is pretty trivial and presented in the middle and bottom right panels of Fig.~\ref{fig02::rmin_selection}, the $R_{min}$ selection naturally results in a rather clean sample of the subsolar metallicity part of the low-$\alpha$ sequence~(middle left) and young AMR sequence~(bottom left).

\section{Summary}\label{sec2::summary}

In this study, leveraging data from the APOGEE spectroscopic survey, we present the first application of the orbit superposition method to the MW. Our novel approach demonstrates the capability to correct the observed distribution function of stellar populations for the selection function and mitigate the effects caused by the spatial footprint while simultaneously reconstructing the complete structural, kinematic, chemical, and age composition of the MW's disc. Using this panoramic view of the main MW stellar component, we found the following.

\begin{itemize}

    \item After the weighting stellar populations according to their mass from orbit superposition, we do observe the bimodal distribution of $\aFe$ at subsolar metallicities. Compared to the raw APOGEE data, the low-$\alpha$ sequence in this metallicity regime is less prominent. We also observe a substantial increase in the contribution of the metal-rich populations~($\FeH>0.25$) dominating the innermost MW. We detect mild variations in the \MgFe-\FeH distribution with azimuth, which does not imply azimuthally-non-homogeneous chemical enrichment but rather the redistribution of stellar populations in the MW barred potential depending on the kinematics.
    
    \item Thanks to the mass reconstruction, we are able to break the degeneracy between chemical~(Galactic studies) and geometrical~(extragalactic studies) definitions of thin and thick discs. In agreement with several independent studies, we showed that the total mass of the high-$\alpha$ sequence is $\approx 44\%$ of the MW disc mass. Curiously, the geometric thick disc has the same mass fraction, but it is made of a mixture of high- and low-$\alpha$ populations. These measurements provide an opportunity for a more direct comparison of the thin/thick disc properties of MW with those of extragalactic systems.
    
    \item We confirm that, while the averaged metallicity profile is flattened in the inner $<4-5$~kpc region, the mass-weighted radial metallicity gradient is negative, making the MW a typical disc galaxy. At the same time, we found the azimuthal variations of the gradient inside the solar circle caused by the bar. This, in the term, results in the azimuthal variations of the mean metallicity, which should be taken into account while measuring the chemical abundance variations associated with spiral arms, which our approach cannot detect directly.

    \item We provide the complete description of the mono-abundance~(in the \FeH-\MgFe plane) populations. We confirm the monotonic decline of the vertical scale height with increasing metallicity and decreasing \MgFe. This, however, does not imply discontinuity between thin and thick disc components, as the mass-weighed scale-height distribution shows two prominent peaks around 300 and 900~pc, corresponding to the geometric thin and thick disc populations. The radial scalelength shows a similar behaviour as the scale height, except for the subsolar metallicity low-$\alpha$ populations, which, in fact, correspond to the distinct outer MW disc.

    \item Our results reinforce the recent finding by showing that the AMR relation of MW breaks up into two prominent sequences, which, thanks to our orbit superposition approach, we associate with the inner and outer discs. According to this picture, we argue that inner/outer MW disc copulations are more distinct regarding their enrichment histories and orbital parameters compared to the high-/low-$\alpha$ terminology. The old age-metallicity sequence formed inside $\approx 6$~kpc with overall declining SFR with a period of a cessation around $8$~Gyr ago. The young AMR sequence started to form around $6$ Gyr ago outside the solar radius. This process results in the rapid increase of the MW disc size and the effective radius since redshift $\approx 0.7$. 
        
\end{itemize}

In this work, we present a comprehensive quantitative analysis of the current state of the MW stellar disc. Using the orbit superposition approach, we demonstrate its effectiveness in bridging gaps in our observational data. However, we argue that refining this methodology by including direct kinematic constraints combined with more precise and larger spectroscopic datasets from upcoming surveys like 4MOST~\citep{2019Msngr.175....3D}, MOONS~\citep{2020Msngr.180...18G} and SDSS-V~\citep{2019BAAS...51g.274K} and more precise determination of stellar ages will allow us to place even tighter constraints on the MW past. The panoramic insights provided by the orbit superposition method open the door to exploring the MW as an extragalactic analogue, a step that is crucial for understanding whether the Galactic evolution is universal or if the MW-mass systems experience very diverse evolutionary paths. This broader perspective will be vital for constraining galaxy formation models and deepening our understanding of our place in the cosmic landscape.

\begin{acknowledgements}
We thank Alexander Stone-Martinez for making the DistMass catalogue publicly available\footnote{\url{https://data.sdss.org/sas/dr17/env/APOGEE_DISTMASS/}}. SK thanks Sofia Feltzing for the discussion about spinning galaxies in October 2024.
\\
\thanksleiden 
\\
\thanksmiapb
\\
\thankssdss
\\
\thanksgaia
\end{acknowledgements}

\bibliographystyle{aa}
\bibliography{refs}

\begin{appendix}
\section{Extra figures}
\begin{figure*}
    \centering
    \includegraphics[width=1\hsize]{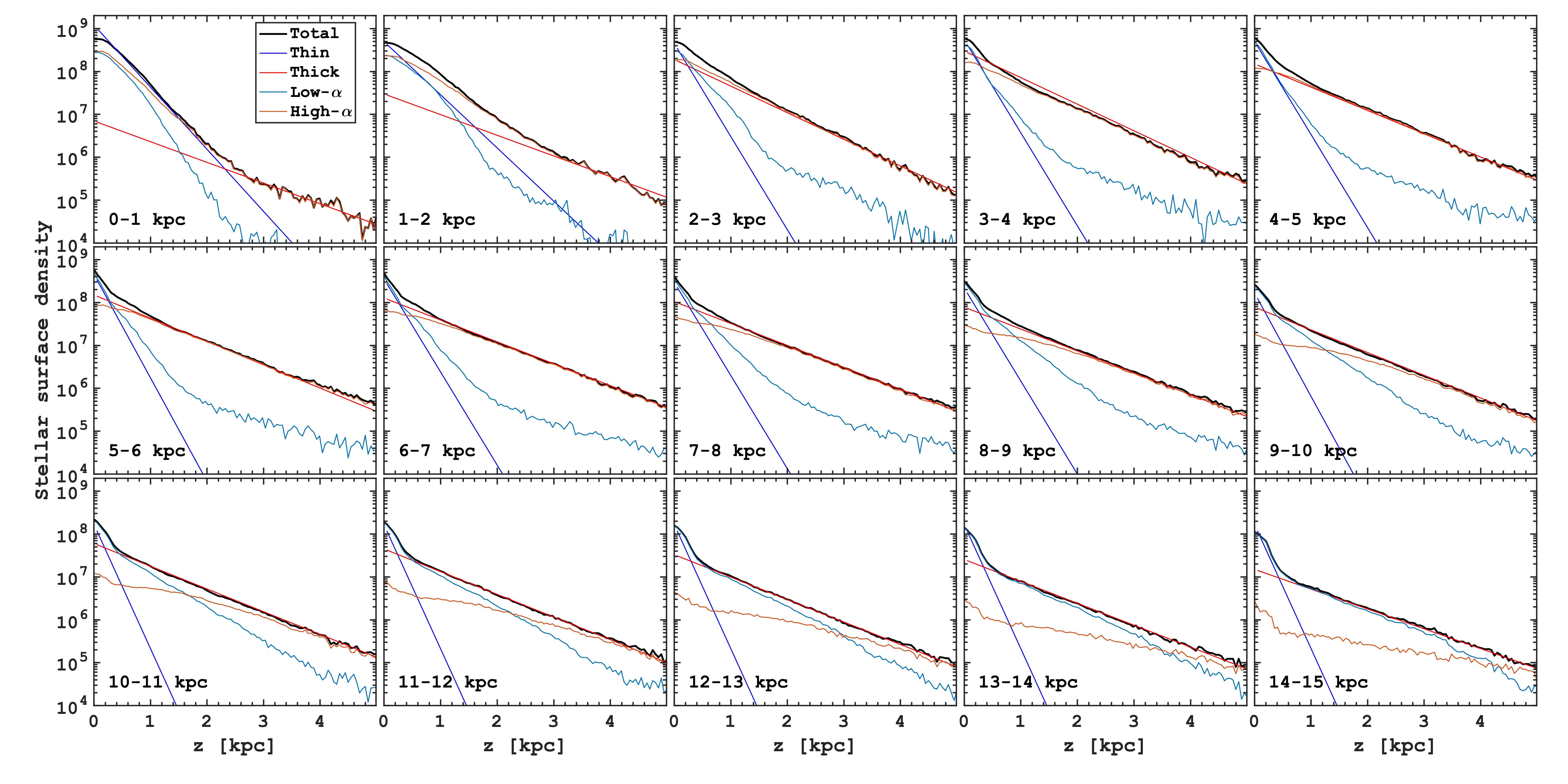}
    \caption{Decomposition of the MW stellar density profiles at different Galactocentric radii.}
    \label{fig02::thinthick_lowhigh_1D}
\end{figure*}

\begin{figure*}
    \centering
    \includegraphics[width=1\hsize]{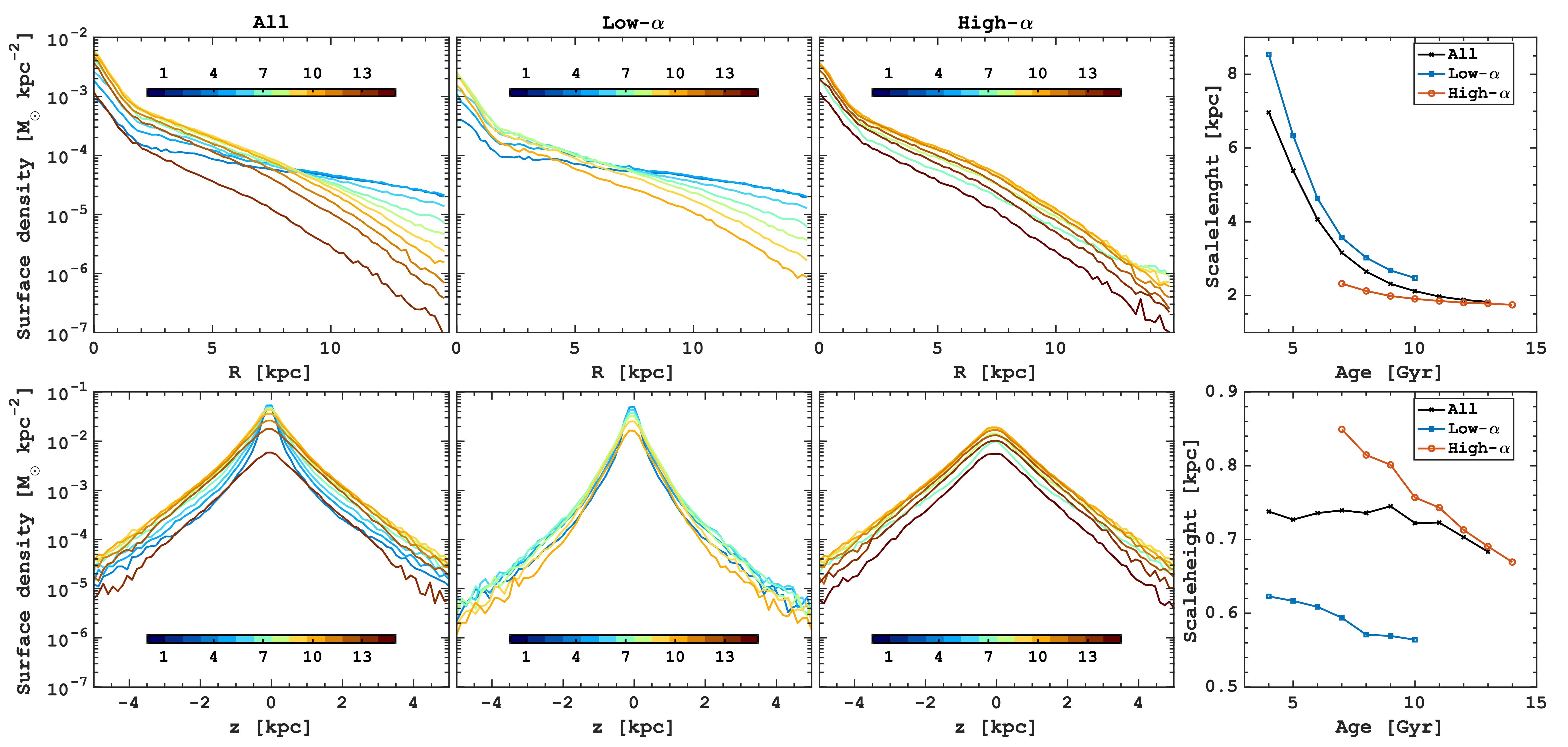}
    \caption{Surface stellar density profiles of mono-age populations in radial~(top) and vertical~(bottom) directions for all stars~(left), low- and high-$\alpha$ populations separately. The rightmost panels show the exponential scalelength~(top) and scaleheigh~(bottom) for mono-age stellar populations.}
    \label{fig02::disk_mono_age_density}
\end{figure*}

% \begin{figure*}
%     \centering
%     \includegraphics[width=1\hsize]{figs02/disk_radial_met_gradients_z_APOGEE-Dr17_0000.jpg}
%     \caption{From left to right: stellar mass-weighted radial metallicity profiles in the MW disc at different distances from the mid-plane for all~(left), low-~(middle) and high-$\alpha$~(right) stellar populations. In each panel, the vertically-averaged profiles are shown by the black lines, while contributions from different slices away from the midplane are shown by different colours, as marked in each panel. }
%     \label{fig02::feh_radial_gradient_z}
% \end{figure*}

\end{appendix}
\end{document}